\documentclass[
aps,prd,
showpacs,twocolumn,notitlepage,
amssymb,amsmath,amsfonts,mathrsfs,
nofootinbib,superscriptaddress,
floats,floatfix
]{revtex4-2}

\usepackage{graphicx}
\usepackage{xcolor}
\usepackage[colorlinks=true]{hyperref}
\hypersetup{citecolor=cyan,linkcolor=magenta}
\usepackage{multirow,array}

\usepackage{amsmath}
\usepackage{graphics}
\usepackage{epstopdf}
\usepackage{enumerate}

\usepackage{savesym}
\savesymbol{tablenum}
\usepackage{siunitx}
\restoresymbol{SIX}{tablenum}

\usepackage{hyperref}

\newcommand{\del}{\partial}

\newcommand{\mnras}{Mon. Not. R. Astron. Soc.}
\newcommand{\pasj}{Publ. Astron. Soc. Jpn.}
\newcommand{\apjs}{Astrophys. J. Suppl. Ser.}
\newcommand{\apjl}{Astrophys. J. Lett.}

\newcommand{\beq}{\begin{equation}}
\newcommand{\eeq}{\end{equation}}
\newcommand{\beqn}{\begin{eqnarray}}
\newcommand{\eeqn}{\end{eqnarray}}
\newcommand{\pa}{\partial}

\begin{document}


\title{Outflow energy and black-hole spin evolution in collapsar scenarios}

\author{Masaru Shibata}
\affiliation{Max-Planck-Institut f\"ur Gravitationsphysik (Albert-Einstein-Institut), Am M\"uhlenberg 1, D-14476 Potsdam-Golm, Germany}
\affiliation{Center for Gravitational Physics and Quantum Information, Yukawa Institute for Theoretical Physics, Kyoto University, Kyoto, 606-8502, Japan}

\author{Sho Fujibayashi}
\affiliation{Max-Planck-Institut f\"ur Gravitationsphysik (Albert-Einstein-Institut), Am M\"uhlenberg 1, D-14476 Potsdam-Golm, Germany}

\author{Alan Tsz-Lok Lam}
\affiliation{Max-Planck-Institut f\"ur Gravitationsphysik (Albert-Einstein-Institut), Am M\"uhlenberg 1, D-14476 Potsdam-Golm, Germany}

\author{Kunihito Ioka}
\affiliation{Center for Gravitational Physics and Quantum Information, Yukawa Institute for Theoretical Physics, Kyoto University, Kyoto, 606-8502, Japan}

\author{Yuichiro Sekiguchi}
\affiliation{Department of Physics, Toho University, Funabashi, Chiba 274-8510, Japan}

\date{\today}


\begin{abstract}
We explore the collapsar scenario for long gamma-ray bursts by performing axisymmetric neutrino-radiation magnetohydrodynamics simulations in full general relativity for the first time. In this paper, we pay particular attention to the outflow energy and the evolution of the black-hole spin. We show that for a strong magnetic field with an aligned field configuration initially given, a jet is launched by magnetohydrodynamical effects before the formation of a disk and a torus, and after the jet launch, the matter accretion onto the black hole is halted by the strong magnetic pressure, leading to the spin-down of the black hole due to the Blandford-Znajek mechanism. The spin-down timescale depends strongly on the magnetic-field strength initially given because the magnetic-field strength on the black-hole horizon, which is determined by the mass infall rate at the jet launch, depends strongly on the initial condition, although the total jet-outflow energy appears to be huge $> 10^{53}$\,erg depending only weakly on the initial field strength and configuration.  
For the models in which the magnetic-field configuration is not suitable for quick jet launch, a torus is formed and after a long-term magnetic-field amplification, a jet can be launched. For this case, the matter accretion onto the black hole continues even after the jet launch and black-hole spin-down is not found. We also find that the jet launch is often accompanied with the powerful explosion of the entire star with the explosion energy of order $10^{52}$\,erg by magnetohydrodynamical effects. We discuss an issue of the overproduced energy for the early-jet-launch models. 

\end{abstract} 


\maketitle

\section{Introduction}\label{sec:intro}

The collapsar model~\cite{Woosley1993,Macfadyen1999} is the widely accepted model for explaining the central engine of long gamma-ray bursts. In this model, one supposes a massive, rotating, and magnetised progenitor star that collapses into a black hole. After the formation of a spinning black hole, one assumes that the black hole is penetrated by a poloidal magnetic field with a sufficiently high field strength, with which the Poynting luminosity by the Blandford-Znajek mechanism~\cite{Blandford1977} is sufficiently high. Motivated by this idea, a number of general relativistic magnetohydrodynamics simulations (in the fixed black-hole spacetime) have been performed in the last two decades (e.g., Refs.~\cite{Komissarov:2005wj,McKinney:2006tf,2008MNRAS.385L..28B,Komissarov:2009dn,Tchekhovskoy:2011zx,Bromberg:2015wra,Gottlieb:2021srg}), and indicated that jets are indeed launched in the presence of strong poloidal magnetic fields that penetrate a spinning black hole, which are hypothetically assumed.  

In the force-free approximation, the Poynting luminosity associated with the Blandford-Znajek mechanism is approximately written as (e.g., Ref.~\cite{McKinney:2004ka})
\beqn
    {dE \over dt} \approx {4 \over 3}(B^r)^2 M_{\rm BH}^4 \hat r_+^2(\hat r_+ + 2) \omega (\Omega_\mathrm{BH}-\omega), \label{eq0}
\eeqn
where $B^r$ is the typical value of the (lab-frame) radial magnetic-field strength on the black-hole horizon, $M_{\rm BH}$ is the black-hole mass, $\hat r_+=1 + \sqrt{1-\chi^2}$ with $\chi$ being the black-hole spin and $M_{\rm BH}\hat r_+$ being the radius of the black-hole horizon in the Boyer-Lindquist coordinates (e.g., Ref.~\cite{Bardeen:1972fi}), $\omega$ is the angular velocity of the magnetic field lines, and $\Omega_\mathrm{BH}$ is the angular velocity of the black hole written as (e.g., Refs.~\cite{Bardeen:1972fi,Thorne1986})
\beq
\Omega_\mathrm{BH}={\chi \over 2M_{\rm BH} \hat r_+}. \label{eq01}
\eeq
To derive Eq.~(\ref{eq0}) we assume that $B^r$ and $\omega$ are constant on the black-hole horizon. Throughout this paper we use the geometrical units in which $c=1=G$ where $c$ and $G$ are the speed of light and gravitational constant, respectively. 
We note that $dM_\mathrm{BH}/dt=-dE/dt$ in the absence of matter accretion onto the black hole, and that the source of the Poynting luminosity is the rotational kinetic energy of the black hole. We also note that Eq.~(\ref{eq0}) is valid only when the poloidal magnetic field penetrates the entire surface of the black-hole horizon. If the poloidal magnetic field penetrates a part of the surface, the luminosity is lower. 

Although $\omega$ is a function of spatial coordinates determined by the detailed magnetic-field profile, we assume it as a constant for simplicity and set it as $\omega=f \Omega_\mathrm{BH}$ where $f$ is assumed to be a constant as well because previous numerical studies often showed that $f$ is $\sim 1/2$ (see, e.g., Ref.~\cite{McKinney:2004ka}). Then, Eq.~(\ref{eq0}) is written as
\beqn
    {dM_\mathrm{BH} \over dt} &\approx& -{f(1-f) \over 3}\left(B^r M_{\rm BH}\chi\right)^2
    (\hat r_+ +2) \nonumber \\
    &\approx & -1.1 \times 10^{50}\, f_{1/2} {1-f \over 1/2}
    \left({M_{\rm BH} \over 10M_\odot}\right)^2 \nonumber \\
    && \times \left({B^r \over 10^{14}\,{\rm G}}\right)^2
    \left({\chi \over 0.7}\right)^2
    \left({\hat r_+ +2 \over 4}\right)\,{\rm erg/s}, \label{eq1}
\eeqn
where $f_{1/2}=f/(1/2)$. 
In the following we suppose the typical values of $B^r$, $M_\mathrm{BH}$, and $\chi$ as $10^{14}$\,G, $\sim 10M_\odot$, and $\agt 0.5$ because with these values, the typical luminosity of long gamma-ray bursts can be reproduced assuming that the conversion efficiency of the Poynting luminosity to gamma-ray luminosity is of order $10\%$ and an opening angle of the jet is $5^\circ$--$10^\circ$. 

Associated with the energy extraction, the angular momentum of the black hole is also extracted with the rate~(e.g., Ref.~\cite{McKinney:2004ka})
\beqn
{dJ_\mathrm{BH} \over dt}&\approx& -{4 \over 3}(B^r)^2 M_{\rm BH}^4 \hat r_+^2 
(\hat r_+ +2)(\Omega_\mathrm{BH}-\omega) \nonumber \\
&=&-{2(1-f) \over 3} (B^r)^2 (\hat r_+ +2) \hat r_+ M_{\rm BH}^3\chi.
\label{eq2}
\eeqn

Before proceeding, a relation between the loss of the angular momentum and mass of the black hole is derived. From Eqs.~(\ref{eq1}) and (\ref{eq2}), we obtain
\beq
J_{\rm BH} {dJ_{\rm BH} \over dt}
={2 M_{\rm BH}^3 \hat r_+ \over f} {dM_{\rm BH} \over dt}, \label{eq4}
\eeq
where we used $J_\mathrm{BH}=M_\mathrm{BH}^2\chi$. 
Because $\hat r_+$ depends only weakly on $M_{\rm BH}$ and $J_{\rm BH}$ for moderately spinning black holes, we here approximate that it is constant and integrate
Eq.~(\ref{eq4}) in time, giving
\beq \Delta J_{\rm BH}^2 \approx {\hat r_+
  \over f} \Delta M_{\rm BH}^4,
\eeq
where $\Delta J_{\rm BH}^2$ and $\Delta M_{\rm BH}^4$ are the total changes of $J_{\rm BH}^2$ and $M_{\rm BH}^4$ during the black-hole evolution by the Blandford-Znajek mechanism.  Setting $M_{\rm BH}=M_0 + \Delta M$ where $M_0$ is the initial black-hole mass and $M_0 \gg |\Delta M|$ with $\Delta M<0$ is always satisfied, we obtain $\Delta M_{\rm BH}^4\approx 4M_0^3 \Delta M$. 

If a substantial amount of the angular momentum is extracted from the black hole,  $\Delta J_{\rm BH}^2$ may be approximated by $J_0^2$ where $J_0$ is the initial value of $J_{\rm BH}$. Then, we obtain 
\beqn
|\Delta M| &\approx& {f  \over 4 \hat r_+}\left({J_0 \over M_0^2}\right)^2M_0
\nonumber \\
&\approx &5.5\times 10^{53} \,f_{1/2} \left({\hat r_+ \over 2}\right)^{-1}
\left({\chi_0 \over 0.7}\right)^2 
\left({M_0 \over 10M_\odot}\right)\,{\rm erg},\nonumber \\
\label{eq7}
\eeqn
where $\chi_0=J_0/M_0^2$. Thus, the total energy budget for the spinning black holes with  typical mass of $M_0 \agt 4M_\odot$ is larger than $10^{53}$\,erg for $\chi_0 \agt 0.5$ if $f_{1/2}\sim 1$. The total energy of gamma-ray bursts (including the afterglow and associated supernova) are less than $10^{53}$\,erg (typically $\alt 10^{52}$\,erg) for the majority~\cite{2008ApJ...675..528L}, so that the spin angular momentum of the black hole should not be entirely transferred to the matter surrounding the black hole during the stages of the prompt gamma-ray emission, its  afterglow, and associated supernova (unless the factor $f$ is extremely small); otherwise, they had to be extremely bright.

From the spin angular momentum of the black hole, $J_{\rm BH}=M_{\rm BH}^2\chi$,
and the angular-momentum extraction rate of Eq.~(\ref{eq2}), 
we can estimate the timescale of the spin-down as
\beqn
\tau&:=&{J_{\rm BH} \over |dJ_{\rm BH}/dt|}={3 \over 2(1-f)(B^r)^2 \hat r_+(\hat r_+ + 2)M_{\rm BH}} \nonumber \\
&\approx& 1.0 \times 10^4 \,\left( {1-f \over 1/2} \right)^{-1}
\left({B^r \over 10^{14}\,{\rm G}}\right)^{-2}
\nonumber \\
&&~~~\times\left({M_{\rm BH} \over 10M_\odot}\right)^{-1}
\left({\hat r_+ (\hat r_+ + 2) \over 8}\right)^{-1} \,{\rm s}.\label{eq9}
\eeqn

For the duration of a gamma-ray burst of $\Delta t$, the spin angular momentum of the
black hole decreases to $J_0\exp(-\Delta t/\tau)$, and thus, for $\Delta t \ll \tau$, 
\beq
|\Delta J_{\rm BH}^2|=J_0^2 \left[1 - \exp(-2\Delta t/\tau)\right] \approx 2 J_0^2 (\Delta t/\tau),\label{eq10}
\eeq
where we assumed that $\tau$ is approximately constant. 
Hence, 
\beqn
|\Delta M| &\approx& {f_{1/2} \Delta t  \over 4 \hat r_+ \tau}\left({J_0 \over M_0^2}\right)^2M_0
\nonumber \\
&\approx &1.1 \times 10^{52} \,f_{1/2}
\left({1-f \over 1/2}\right)
\left({\Delta t \over 10^2\,{\rm s}}\right)
\left({B^r \over 10^{14}\,{\rm G}}\right)^2
\nonumber \\
&&~\times
\left({\hat r_+ + 2 \over 4}\right)
\left({\chi_0 \over 0.7}\right)^2 
\left({M_0 \over 10M_\odot}\right)^2\,{\rm erg},\label{eq11}
\eeqn
yielding the typically required magnitude for the long gamma-burst energy with the typical duration of $\Delta t=10$--$100$\,s. This analysis suggests that if a few percent (i.e., $\Delta t/\tau$) of the rotation kinetic energy of a spinning black hole is liberated, the total energy of long gamma-ray bursts can be explained assuming that the Blandford-Znajek mechanism is the primary mechanism of the central engine. 

In recent papers~\cite{Gottlieb:2023cgm,Jacquemin-Ide:2023aax}, the authors suggest that the black hole may spin down significantly within a timescale of order 10\,s in the context of the collapsar scenario.  
However, as we illustrated above, if the black hole is formed with an appreciable spin magnitude of $\chi\agt 0.5$, the total rotational kinetic energy available for the extraction by the Blandford-Znajek mechanism is $\agt 10^{53}$\,erg, which is too large to explain the observed energy of long gamma-ray bursts (and afterglows) with the typical luminosity of $dE/dt \sim 10^{50}$\,erg/s. 
Our analysis suggests that only a fraction of the rotational kinetic energy and angular momentum of a black hole should be extracted to reproduce typical long gamma-ray bursts and afterglows.

To examine how much spin angular momentum is extracted from spinning black holes during stellar collapse, we perform a neutrino-radiation magnetohydrodynamics simulation in full general relativity. For the magnetohydrodynamics simulations we employ an axisymmetric numerical-relativity code developed in Refs.~\cite{Shibata:2021bbj,Shibata2021b} with a modification by which the spin angular momentum of black holes is better resolved (see Appendix B of Ref.~\cite{Fujibayashi2023BH}). For the initial condition, we employ a model from stellar evolution, which  results in a rapidly rotating progenitor star of Ref.~\cite{Aguilera-Dena2020oct}, and construct an initial data composed of a spinning black hole and infalling matter with weak poloidal magnetic fields by using the method developed in our previous paper \cite{Fujibayashi2023BH}. We will show that only when the initial magnetic-field strength is high in the vicinity of black holes and the field is aligned well with the spin axis of the black hole, the timescale of the black-hole spin-down becomes very short with $\leq 100$\,s, while for a reasonable choice of the initial field strength, the spin-down timescale is much longer than the typical duration of long gamma-ray bursts or the spin-up by the matter accretion onto the black hole overcomes the spin-down by the Blandford-Znajek mechanism. We also show that the magnetic-field strength on the horizon at the jet launch is determined by the mass infall rate (i.e., the ram pressure) at the launch time, and thus, for the later jet-launch models, the magnetic-field strength on the black-hole horizon is lower and the spin-down timescale becomes longer. 

The paper is organized as follows. In Sec.~\ref{sec2}, we summarize the setup in the present numerical simulation. In Sec.~\ref{sec3}, numerical results are presented focusing on the mechanism of the jet launch in the present setting, on the outflow energy, and on the evolution of the black-hole spin by the Blandford-Znajek mechanism. Section~\ref{sec4} is devoted to a summary and discussion, in particular on the problem of the overproduced energy by the Blandford-Znajek mechanism. Throughout this paper, $k_\mathrm{B}$ denotes Boltzmann's constant. 

\section{Simulation setup}\label{sec2}

We employ the same formulation and simulation code as in Refs.~\cite{Shibata:2021bbj,Shibata2021b} for the present neutrino-radiation magnetohydrodynamics study. Specifically, we numerically solve neutrino-radiation resistive magnetohydrodynamics equations in full general relativity in this code. A  tabulated equation of state referred to as DD2~\cite{banik2014a} is employed, 
with the extension of the table down to low-density ($\rho\approx\SI{0.17}{g/cm^3}$) and low-temperature ($k_\mathrm{B}T=10^{-3}$\,MeV) region; see Ref.~\cite{Hayashi:2021oxy} for the procedure. In this paper, we take the ideal magnetohydrodynamics limit by setting a high conductivity $\sigma_c$ with which the resistive dissipation timescale is much longer than the simulation time ($\gg 10$\,s). 

In the present work, the key ingredient is to accurately evolve the mass and angular momentum of black holes. For this purpose, we have modified the treatment inside black-hole horizons for our Einstein's equation solver (a test result for evolving a vacuum black hole with a dimensionless spin parameter of $0.8$ is presented in Appendix~B of Ref.~\cite{Fujibayashi2023BH}). Specifically, in the current setting (grid spacing $\Delta x\leq 0.016M_\mathrm{BH}$; see below), the numerical error for the mass and dimensionless spin is within $1.5$\% and 0.5\%, respectively, for the time evolution of $5$\,s. For the dimensionless spin, the error size is much smaller than the spin-down fraction shown in Sec.~\ref{SecIIId}.


Following our recent work~\cite{Fujibayashi2023BH}, we prepare a system of a spinning black hole with matter infalling to the central region instead of using the original progenitor star model. This is partly motivated to save computational costs but is mainly from the physical consideration. As described in Eq.~(\ref{eq1}), the typical magnetic-field strength required on the horizon is $B\sim 10^{14}$\,G for the long gamma-ray burst models. The magnetic pressure for such a field strength is $B^2/8\pi=O(10^{26})$\,${\rm dyn/cm}^2$. On the other hand, the ram pressure of the infalling matter for given values of the rest-mass density $\rho$ and the infall velocity $v_\mathrm{infall}$ is
\beqn
\rho v_\mathrm{infall}^2
&\approx &2.2 \times 10^{26}\left({\rho \over 10^6\,{\rm g/cm^3}}\right)
\left({v_\mathrm{infall} \over c/2}\right)^2\,{\rm dyn/cm^2}.\nonumber \\
\label{eq:ram1}
\eeqn
This suggests that until the density of the infalling matter decreases below $\sim 10^6\,{\rm g/cm^3}$, the magnetic pressure cannot overcome the ram pressure to launch a jet or outflow by the Blandford-Znajek mechanism (hereafter, we refer to a small-opening angle  outflow along the $z$-axis as a jet even if it is not very relativistic inside the star). In the early stage of the stellar core collapse and black-hole evolution, the rest-mass density near the black hole is much higher than $10^6\,{\rm g/cm^3}$. For this reason, we start the simulations from a black hole and infalling matter. We note that a jet could be launched earlier in the presence of an extremely strong fossil magnetic field, but we do not consider this possibility in this paper. 


To obtain the initial data we first take the progenitor models from a stellar evolution calculation of Ref.~\cite{Aguilera-Dena2020oct} for which the black hole is likely to be formed in a short timescale after core bounce and be evolved simply by the accretion from the outer region without forming an accretion disk in an early stage~\cite{Fujibayashi2023BH}. We then construct the initial data by solving constraint equations of general relativity in the hypothesis that in the early stage of the black-hole evolution, the system is composed of a spinning black hole and nearly free-falling matter. In this paper, we employ the model for which the zero-age main-sequence mass of the progenitor is $M_\mathrm{ZAMS}=35M_\odot$~\cite{Aguilera-Dena2020oct} (i.e., \texttt{AD35} model of Ref.~\cite{Fujibayashi2023BH}). This progenitor star is very compact at the onset of the collapse, and hence, it is reasonable to assume that a black hole is formed in a short timescale after the onset of the collapse~\cite{Oconnor2011apr}. We set up an initial data at a stage just prior to the formation of a disk. For such a choice, the mass and dimensionless spin of the black hole are $M_\mathrm{BH,0}=15M_\odot$ and $\chi_0=0.66$ (see Ref.~\cite{Fujibayashi2023BH} for a detail), and the rest mass and angular momentum of the matter outside the black hole is $M_\mathrm{mat}=10.5M_\odot$ and $J_\mathrm{mat}=4.32J_\mathrm{BH,0}$ at the initial stage. Here, $J_\mathrm{BH,0}=M_\mathrm{BH,0}^2\chi_0$, and the mass and dimensionless spin of the black hole are determined by analyzing the equatorial and polar circumferential radii, $C_e$ and $C_p$, respectively, of apparent horizons~(e.g., see Ref.~\cite{Shibata2016a}). Specifically, the mass is determined by the relation of 
\beq
M_\mathrm{BH}= {C_e \over 4\pi},
\eeq
and the dimensionless spin is determined from $C_p/C_e$, which is a monotonic function of the dimensionless spin, $\chi$, for Kerr black holes and can be used to identify the value of $\chi$. We also check that the mass and spin obtained by them satisfy the relation of the area, $A_\mathrm{AH}=8\pi M_\mathrm{BH}^2(1 + \sqrt{1-\chi^2})$, with high accuracy (the error is less than 0.1\%). 
We cut out the matter outside $10^5$\,km because the computational domain in our simulation is $10^5\times10^5$\,km for $\varpi$ and $z$ where $\varpi$ is the cylindrical coordinate. As shown in our paper~\cite{Fujibayashi2023BH}, the matter infall onto the black hole with no disk formation proceeds for the first $\approx 2$\,s for this model, illustrating that our assumption is valid. 

We also performed several simulations employing the $M_\mathrm{ZAMS}=20M_\odot$ model of Ref.~\cite{Aguilera-Dena2020oct} and found that the results are qualitatively very similar to those for the $M_\mathrm{ZAMS}=35M_\odot$ model. For the $M_\mathrm{ZAMS}=20M_\odot$ model, the matter infall rate is lower than that for $M_\mathrm{ZAMS}=35M_\odot$, and hence, a jet can be launched with a lower magnetic-field strength. Besides this quantitative difference, we do not find a significant modification about the conclusion of this paper. 

\begin{table}[t]
    \centering
    \caption{List of initial setting: model name, maximum magnetic-field strength and the type of the magnetic field configuration initially given, and the value of $\varpi_0$ in units of $10^3$\,km ($\varpi_{0,3}$). The last two columns show whether a jet launch is found or not and whether the spin down of the black hole is found or not in the simulation time, typically, of $\sim 10$\,s.
    }
    \begin{tabular}{cccccc}
    \hline\hline
        Model & ~$B_\mathrm{max}$\,(G)~ &  Config & $~~\varpi_{0,3}$~~ & ~~Jet?~~ & Spin down?\\
        \hline
        B11.5 & $3\times 10^{11}$ & Eq.~(\ref{eqBz}) &1& Yes & Yes \\
        B11.3 & $2\times 10^{11}$ & Eq.~(\ref{eqBz}) &1& Yes & Yes \\
        B11.0 & $1\times 10^{11}$ & Eq.~(\ref{eqBz}) &1& Yes & Yes \\
        B10.5 & $3\times 10^{10}$ & Eq.~(\ref{eqBz}) &1& Yes & No \\
        B10.0 & $1\times 10^{10}$ & Eq.~(\ref{eqBz}) &1& No &  No\\
        Br11.0 & $1\times 10^{11}$ & Eq.~(\ref{eqB}) &1& Yes & Yes \\
        Br10.5 & $3\times 10^{10}$ & Eq.~(\ref{eqB}) &1& No & No \\
        Bq12.5 & $3 \times 10^{12}$ & Eq.~(\ref{eqB2}) &1& Yes & No\\
        Bq12.0 & $1 \times 10^{12}$ & Eq.~(\ref{eqB2}) &1& No & No\\
        Bq11.0 & $1 \times 10^{11}$ & Eq.~(\ref{eqB2}) &1& No & No \\
        Bq11.0b & $1 \times 10^{11}$ & Eq.~(\ref{eqB2}) &5& Yes  & No \\
        Bq11.0c & $1 \times 10^{11}$ & Eq.~(\ref{eqB2}) &10& Yes & Yes \\
        \hline
    \end{tabular}\\
    \label{tab:model}
\end{table}

We superimpose a poloidal magnetic field, with which the electromagnetic energy density is initially much smaller than the rest-mass density, to the spinning black hole and infalling matter. Because it is not clear what kind of the magnetic-field profile in the infalling matter around a massive black hole is developed in the stellar core collapses, we choose a rather ad hoc poloidal field configuration in the present numerical experiment, although it is a strong assumption to initially give an aligned poloidal field. We primarily prepare the magnetic field only of the $z$ component and $\sqrt{\gamma}B^z$ is a function only of $\varpi$ where $\gamma$ is the determinant of the three metric $\gamma_{ij}$. Here, we set that $(B^z)^2(\varpi)$ is approximately proportional to the pressure on the equatorial plane, which results in
\beq
B^z = {B_0 \over \varpi\sqrt{\gamma}}{d \over d\varpi} 
\left(\sqrt{{\varpi_0^2 \over \varpi^2 + \varpi_0^2}} \varpi^2\right),\label{eqBz}
\eeq
where $\varpi_0$ is a constant with a fiducial value of $10^3$\,km, which is $\approx 45M_\mathrm{BH,0}$, and $B_0$ is a constant which determines the magnetic-field strength. With this setting, the divergence-free condition of the magnetic field is automatically satisfied. Because the magnetic-field lines are aligned with the spin axis of the black hole and the magnetic-field strength does not decrease with $z$, this setting is quite favorable for launching a jet along the spin axis; we intentionally choose this setting to study a jet launch, subsequent spin-down of black holes, dependence of the spin-down rate on the initial magnetic-field strength, and Poynting luminosity by the Blandford-Znajek mechanism.

For several models, we also choose
\beqn
B^\varpi&=&-{B_0 \over \varpi\sqrt{\gamma}}
{\pa \over \pa z}
\left(\sqrt{{\varpi_0^2 \over r^2 + \varpi_0^2}} \varpi^2\right),\nonumber \\
B^z&=& {B_0 \over \varpi\sqrt{\gamma}}
{\pa \over \pa \varpi}
\left(\sqrt{{\varpi_0^2 \over r^2 + \varpi_0^2}} \varpi^2\right),\label{eqB}
\eeqn
and 
\beqn
B^\varpi&=&-{B_0 \over \varpi\sqrt{\gamma}}
{\pa \over \pa z}
\left({\varpi_0^2 \over r^2 + \varpi_0^2} \varpi^2\right),\nonumber \\
B^z&=& {B_0 \over \varpi\sqrt{\gamma}}
{\pa \over \pa \varpi}
\left({\varpi_0^2 \over r^2 + \varpi_0^2} \varpi^2\right),\label{eqB2}
\eeqn
where $r=\sqrt{\varpi^2 + z^2}$. With these settings, the magnetic-field strength on the horizon can be set to be initially identical with that with Eq.~(\ref{eqBz}), but the field strength in the outer region becomes weaker. Specifically, the magnetic field strength along the $z$ axis for the distant region is $\propto z^0$, $\propto z^{-1}$, and $\propto z^{-2}$ with Eqs.~(\ref{eqBz}), (\ref{eqB}), and (\ref{eqB2}), respectively. We will illustrate that the evolution of the magnetic-field strength on the horizon depends strongly on the initial field configurations. In particular, in the initial condition of Eq.~(\ref{eqB2}) with $\varpi_0=10^3$\,km, the magnetic-field strength on the horizon does not increase significantly with time due to the matter accretion in the central region, and hence, even for an initially high magnetic-field strength, a jet is not quickly launched. For Eq.~(\ref{eqB2}), the magnetic pressure along the $z$-axis is proportional to $z^{-4}$ for the distant region, which is steeper than that for the gas pressure. This is also disadvantageous for launching a jet along the $z$-axis. For the choice of Eq.~(\ref{eqB2}), we also perform simulations varying the value of $\varpi_0$ to confirm that higher values of $\varpi_0$ are advantageous for the jet launch.  

We specify the models by the maximum magnetic-field strength, $B_\mathrm{max}$. We choose it as $B_\mathrm{max}=3\times 10^{11}$, $2 \times 10^{11}$, $1 \times 10^{11}$, $3\times 10^{10}$, and $1\times 10^{10}$\,G for the magnetic field of Eq.~(\ref{eqBz}), and each model is referred to as models B11.5, B11.3, B11.0, B10.5, and B10.0. For Eq.~(\ref{eqB}), we choose $B_\mathrm{max}=1\times 10^{11}$ and $3\times 10^{10}$\,G, and refer to each model as Br11.0 and Br10.5. For Eq.~(\ref{eqB2}), we choose $B_\mathrm{max}=3\times 10^{12}$, $1\times 10^{12}$, and $1 \times 10^{11}$\,G, and refer to each model as Bq12.5, Bq12.0, and Bq11.0 for $\varpi_0=10^3$\,km. For Eq.~(\ref{eqB2}) with $B_\mathrm{max}=10^{11}$\,G, we also prepare the models with $\varpi_{0,3}=\varpi_0/10^3\,{\rm km}=5$ and $10$, which are referred to as Bq11.0b and Bq11.0c. Table~\ref{tab:model} summarizes the models and their parameters.  

Since the initial electromagnetic pressure is weaker than the gas pressure and ram pressure of the infalling matter for these choices, the effect of the magnetic field is always negligible in the early stage of the simulations; in other words, the total electromagnetic energy is much smaller than the internal and kinetic energies. 
The choice of $B_\mathrm{max} \leq 3 \times 10^{12}$\,G is likely to be reasonable, because the maximum magnetic-field strength for neutron stars is typically $10^{11}$--$10^{13}$\,G~\cite{Lorimer:2008se} and the black hole is likely to be formed through a shorter-term protoneutron star stage, although we have to keep in mind that it is not very clear whether an aligned poloidal magnetic field is established during the evolution of the black hole by the matter accretion. We also note that in the presence of a strong magnetic field, the explosion may take place in the protoneutron-star stage~\cite{Burrows2007,Obergaulinger2020,Obergaulinger2021,Obergaulinger2022may} or in an early evolution stage of a new-born black hole. The present choice of the relatively weakly magnetic fields stems partly from excluding this possibility. 

For $B_\mathrm{max} \agt 1 \times 10^{11}$\,G with Eqs.~(\ref{eqBz}) and (\ref{eqB}) or with Eq.~(\ref{eqB2}) and $\varpi_{0,3}=10$, a jet is generated by the Blandford-Znajek mechanism before a torus (geometrically thick disk) is developed around the black hole. For this case, the initially-given magnetic field is amplified by the winding associated with the black-hole spin and by the compression of the magnetic field associated with the infalling matter motion. Thus, this is considered to be the models that a fossil magnetic field, which is strong enough, induces the jet. This scenario is possible only in the presence of the strong fossil poloidal magnetic field in the progenitor star. 
For some of other models (B10.5, Bq12.5, and Bq11.0b), a jet is launched after the formation of a disk and a torus. For this case, the evolution of the torus partly plays a role for enhancing the strength of the magnetic fields that penetrate the horizon. These models indicate the importance of the co-evolution of the torus and black-hole magnetosphere, which is the key to an eventual jet launch. 

For the even smaller initial field strength or with the initial condition of Eq.~(\ref{eqB2}) with $B_\mathrm{max} \leq 10^{12}$\,G and $\varpi_{0,3}=1$, we do not find the launch of a jet/outflow in the simulation time although it may be driven after long-term torus evolution in reality (see the discussion in Sec.~\ref{sec3A2}). Since our simulation is carried out assuming axisymmetry and thus it cannot fully follow the magnetorotational-instability (MRI) turbulence~\cite{Balbus:1998ja} due to the anti-dynamo theorem~\cite{1978mfge.book.....M}, the enhancement of the magnetic-field strength on the horizon is limited. 
It is natural to suppose, in reality, that a turbulence should be developed after the disk/torus formation, magnetic-field strength is quickly amplified in the disk/torus, and eventually a strong poloidal magnetic field that penetrates the black hole and can be the source of the Blandford-Znajek mechanism is developed (see, e.g., Refs.~\cite{Christie2019dec,Hayashi:2021oxy,Gottlieb:2023est} for related issues). This scenario may be more realistic, but we cannot study it in the present setting. 

\section{Numerical Results}\label{sec3}

\begin{figure*}[t]
\includegraphics[width=0.32\textwidth]{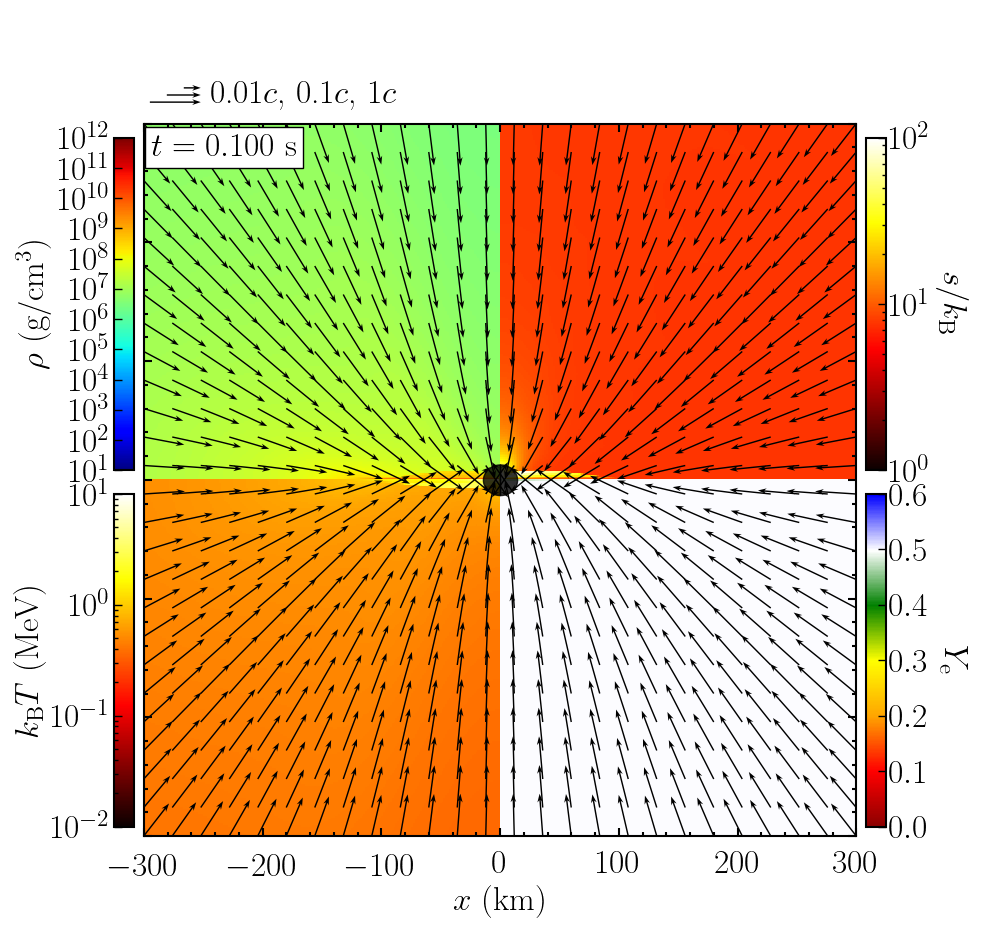}
\includegraphics[width=0.32\textwidth]{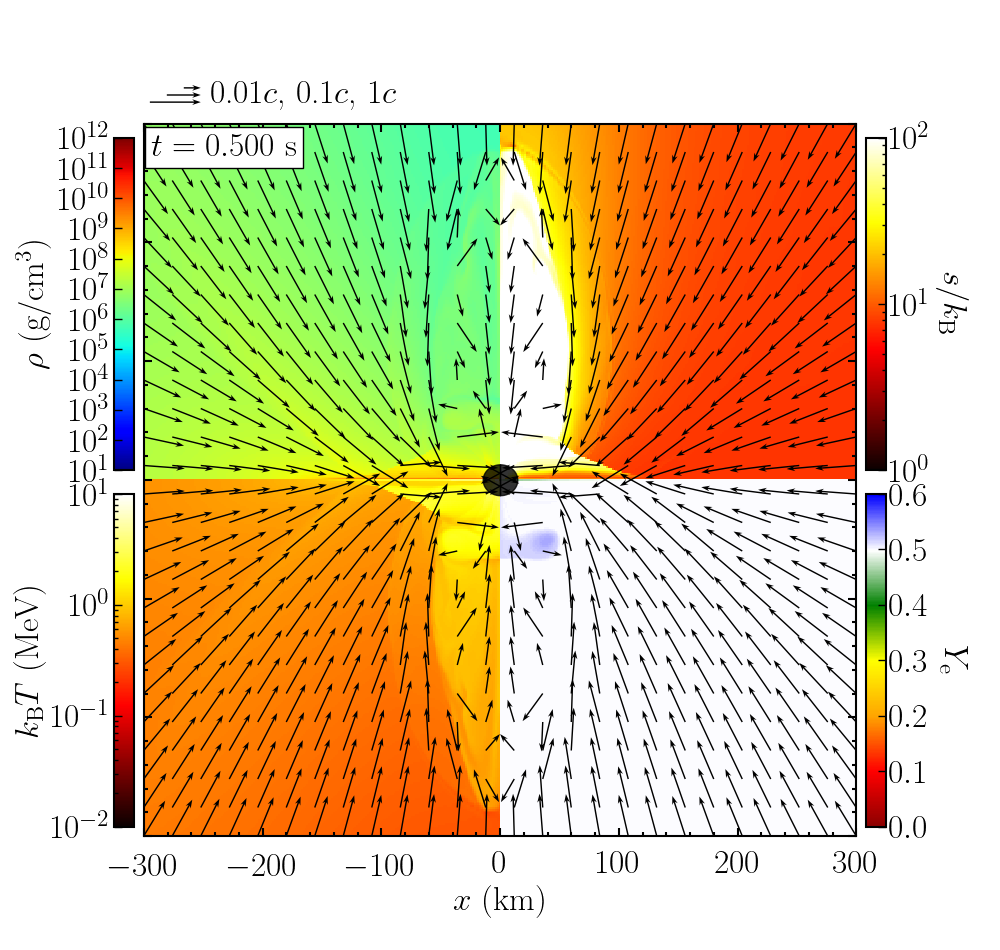}
\includegraphics[width=0.32\textwidth]{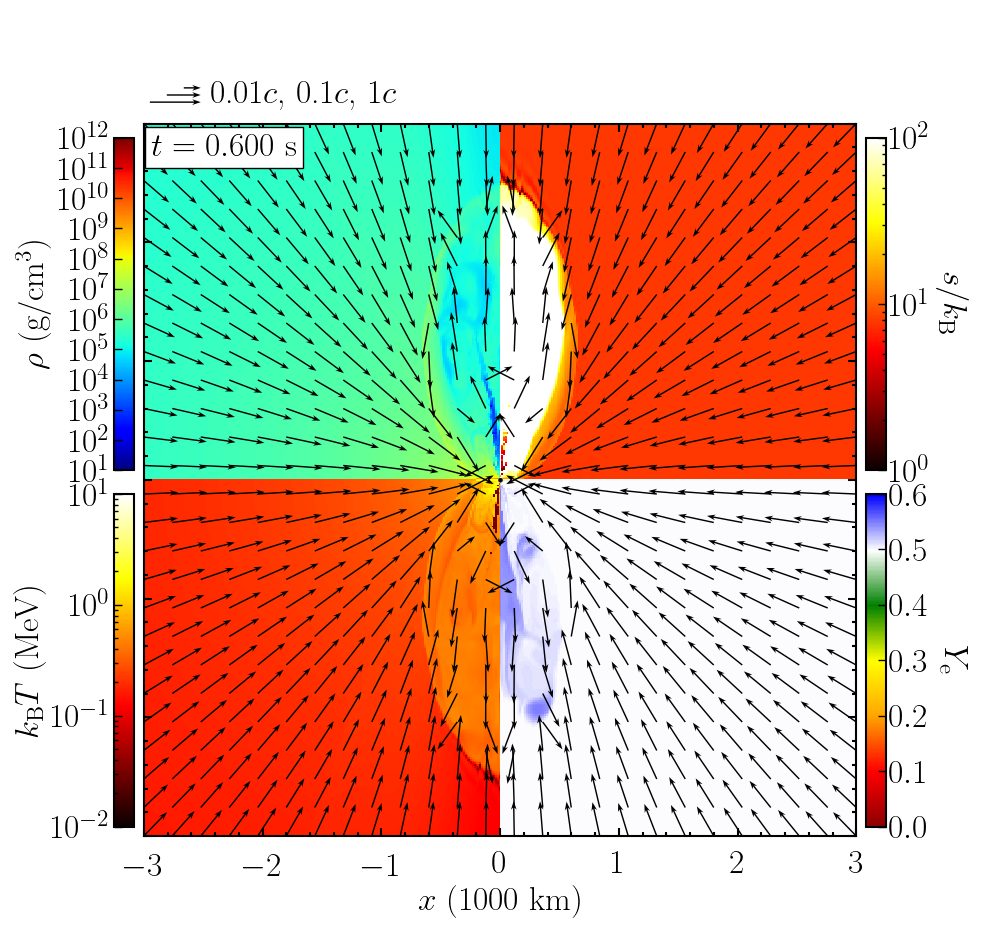}\\
\includegraphics[width=0.32\textwidth]{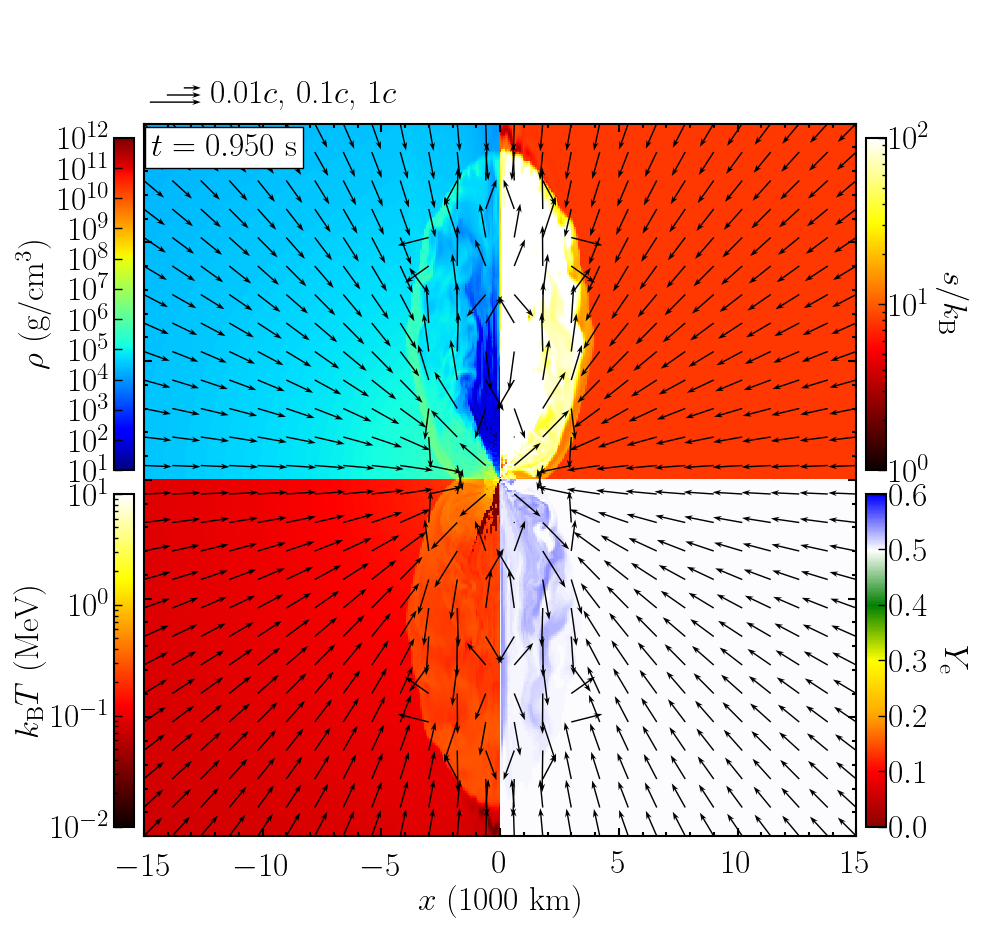}
\includegraphics[width=0.32\textwidth]{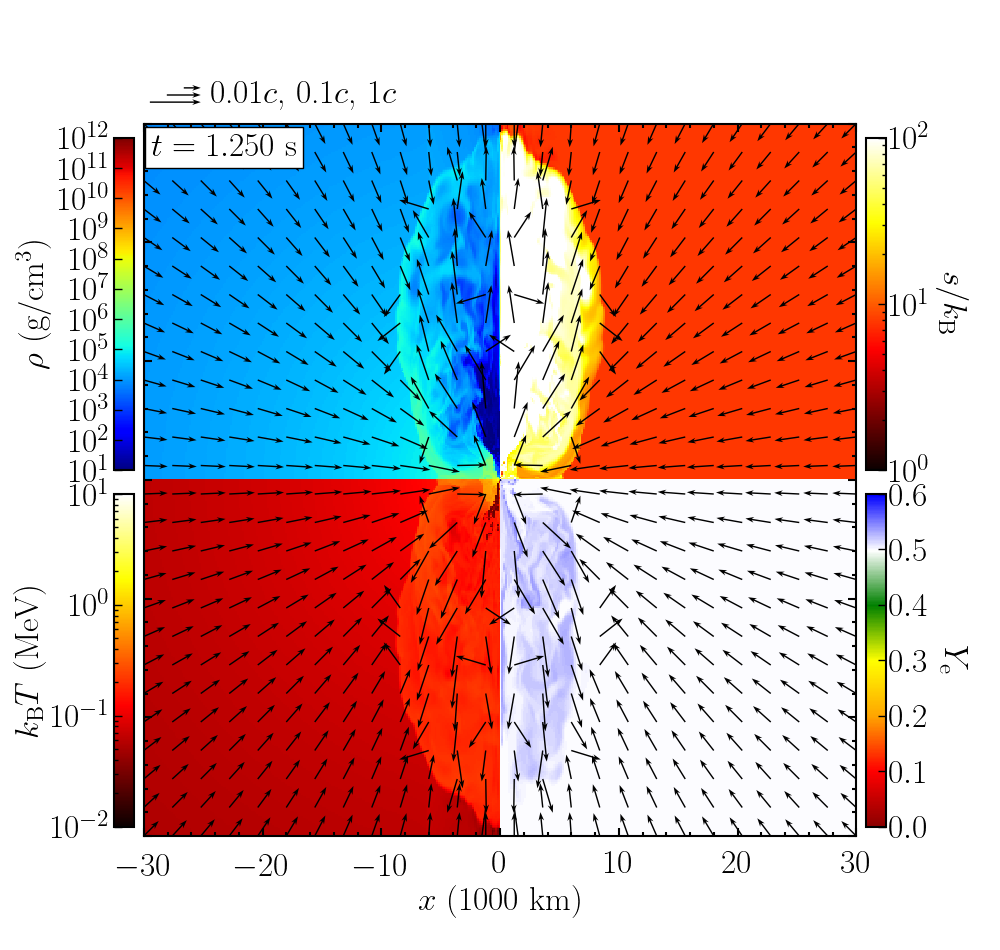}
\includegraphics[width=0.32\textwidth]{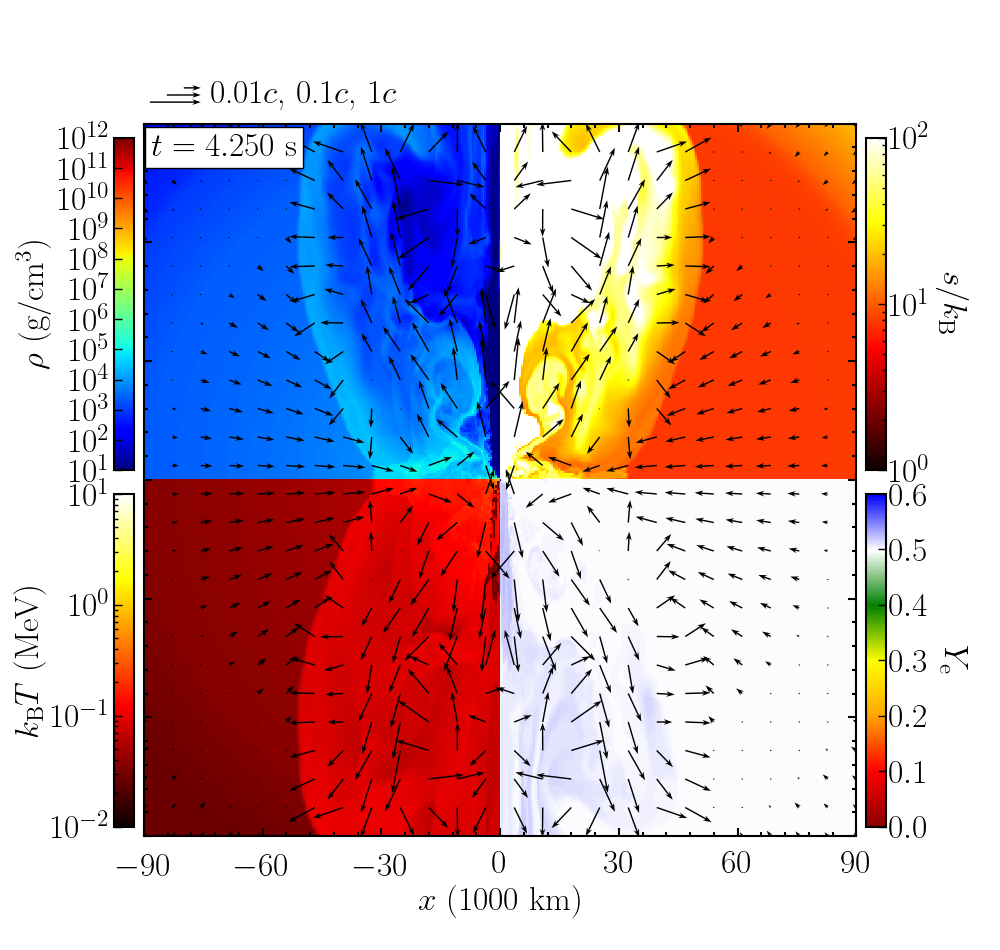}
\caption{Snapshots of the rest-mass density (top-left), entropy per baryon (top-right), temperature (bottom-left), and electron fraction (bottom-right) on the $\varpi$-$z$ plane are shown at selected time slices for model B11.5. Note that for each panel (except for the first two panels), the regions displayed are different. The black filled circles in the first two panels denote the region inside the black hole. An animation for this model is found at \url{https://www2.yukawa.kyoto-u.ac.jp/~sho.fujibayashi/share/B11.5-multiscale.mp4}
}
\label{figA}
\end{figure*}

\begin{figure*}[t]
\includegraphics[width=0.32\textwidth]{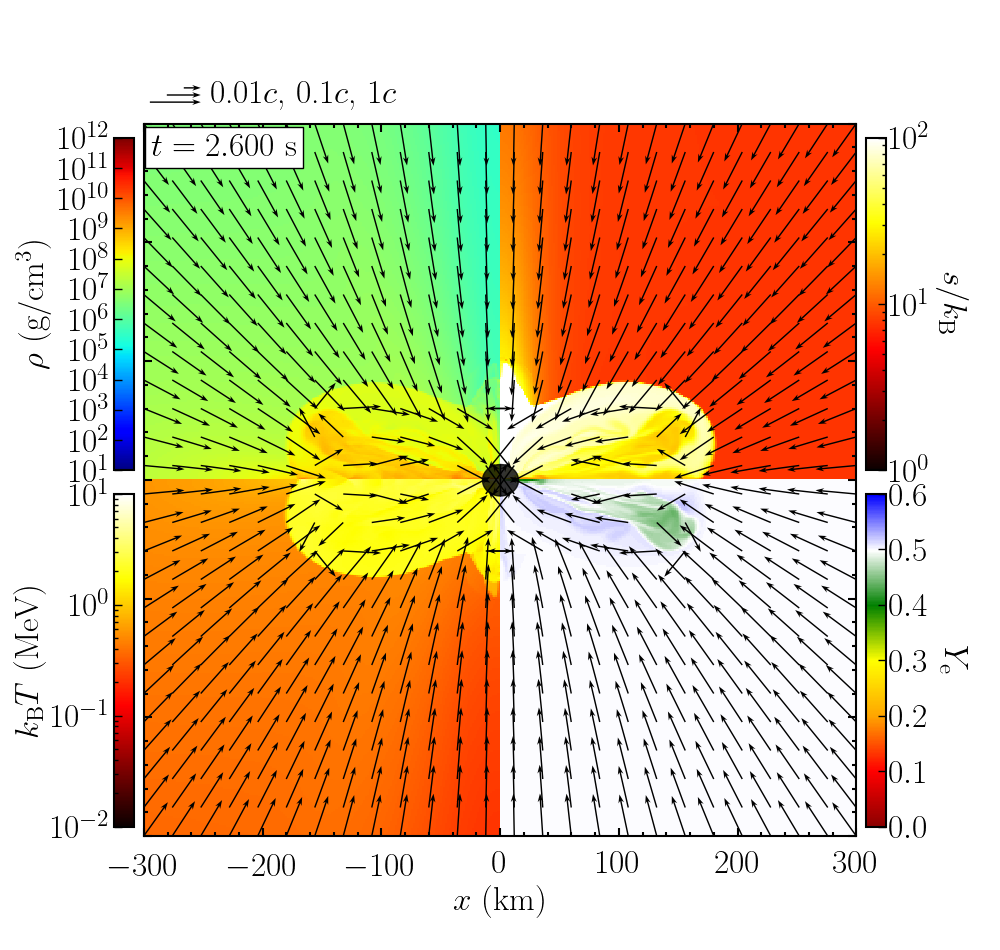}
\includegraphics[width=0.32\textwidth]{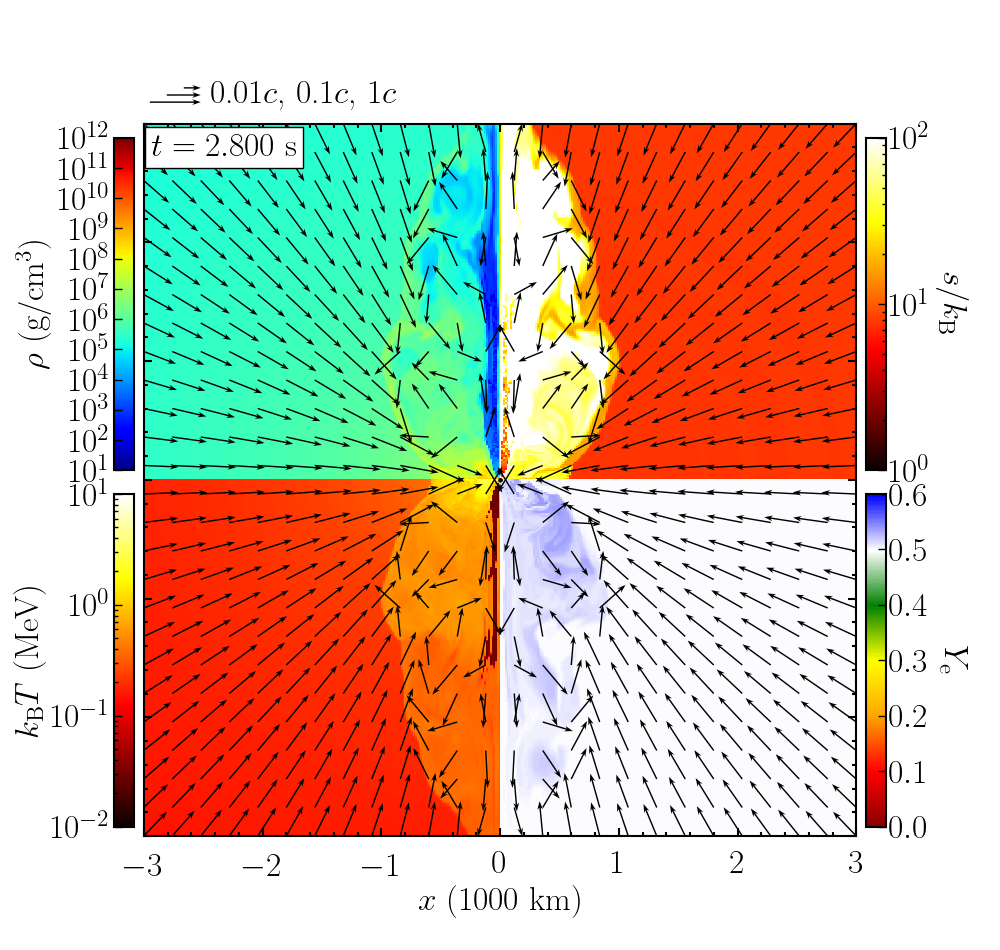}
\includegraphics[width=0.32\textwidth]{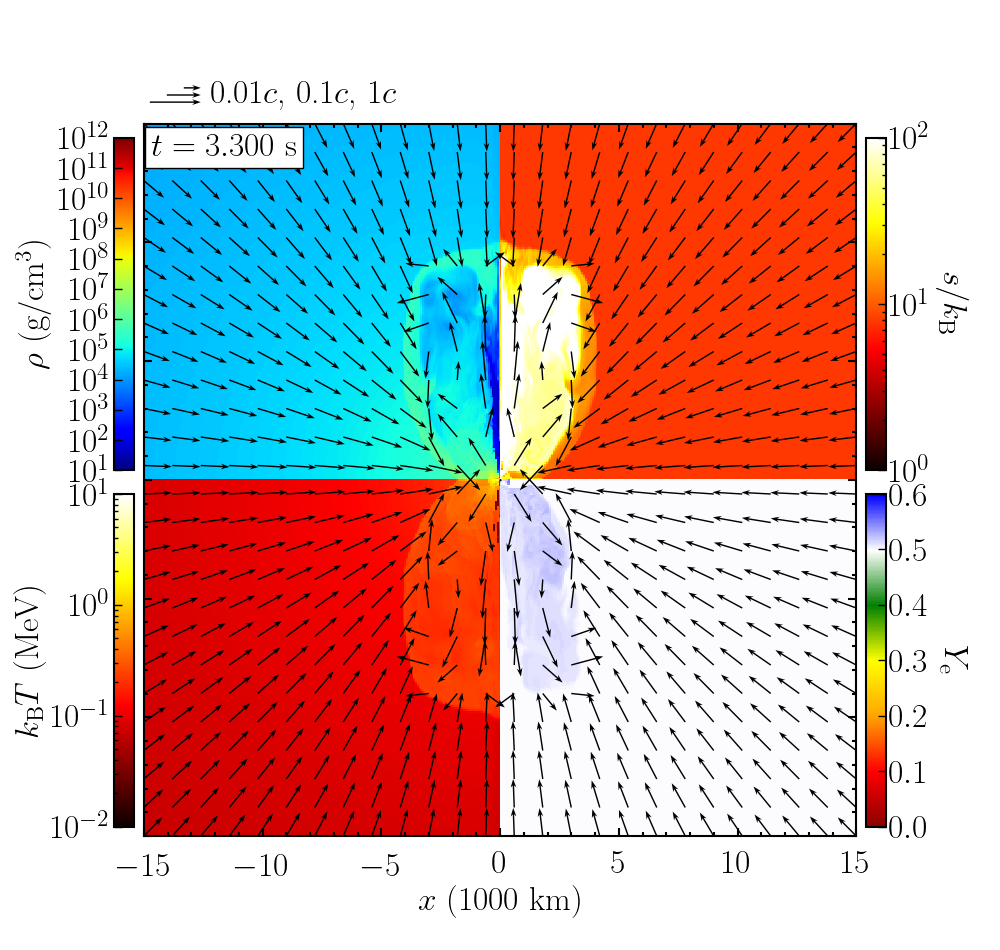}\\
\includegraphics[width=0.32\textwidth]{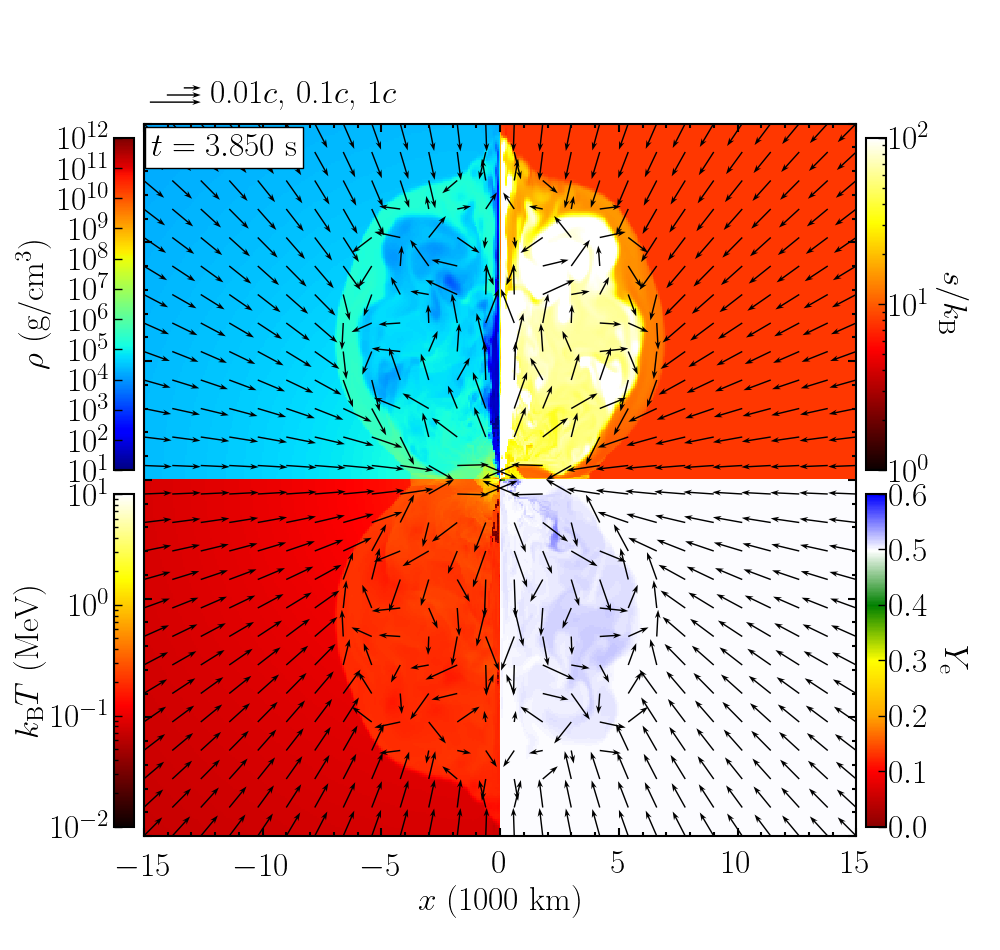}
\includegraphics[width=0.32\textwidth]{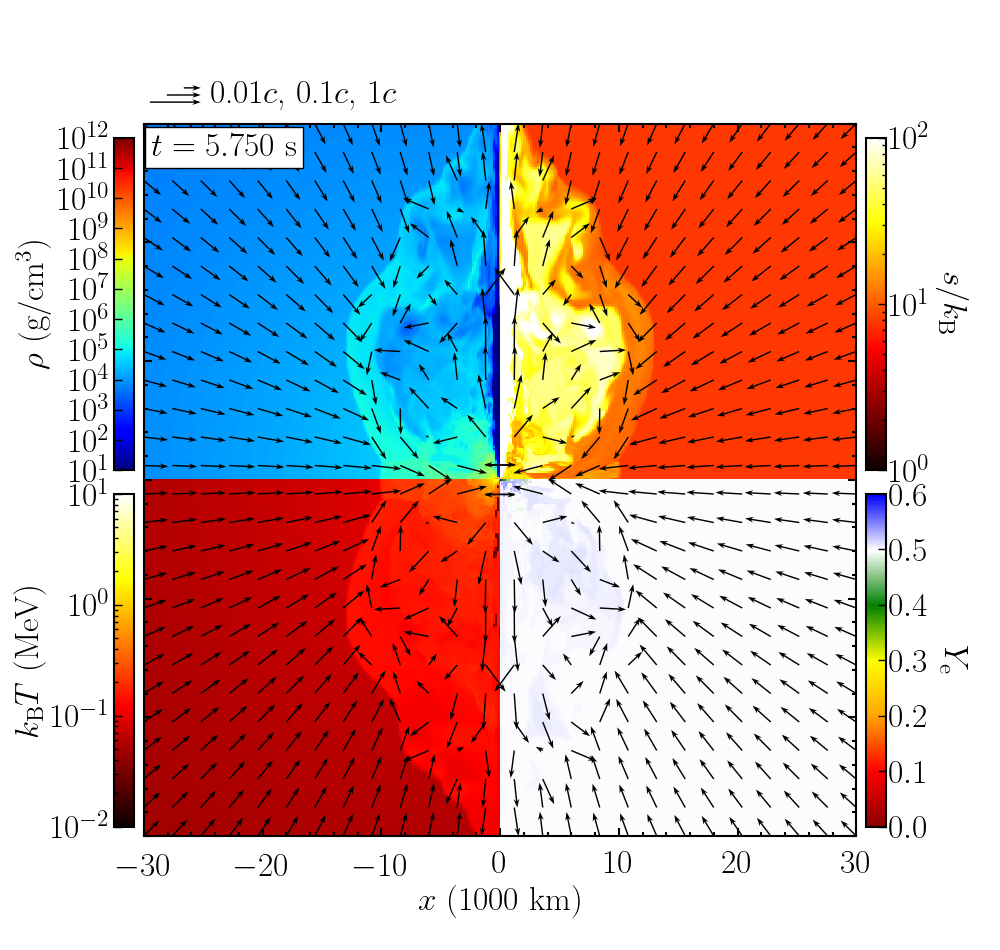}
\includegraphics[width=0.32\textwidth]{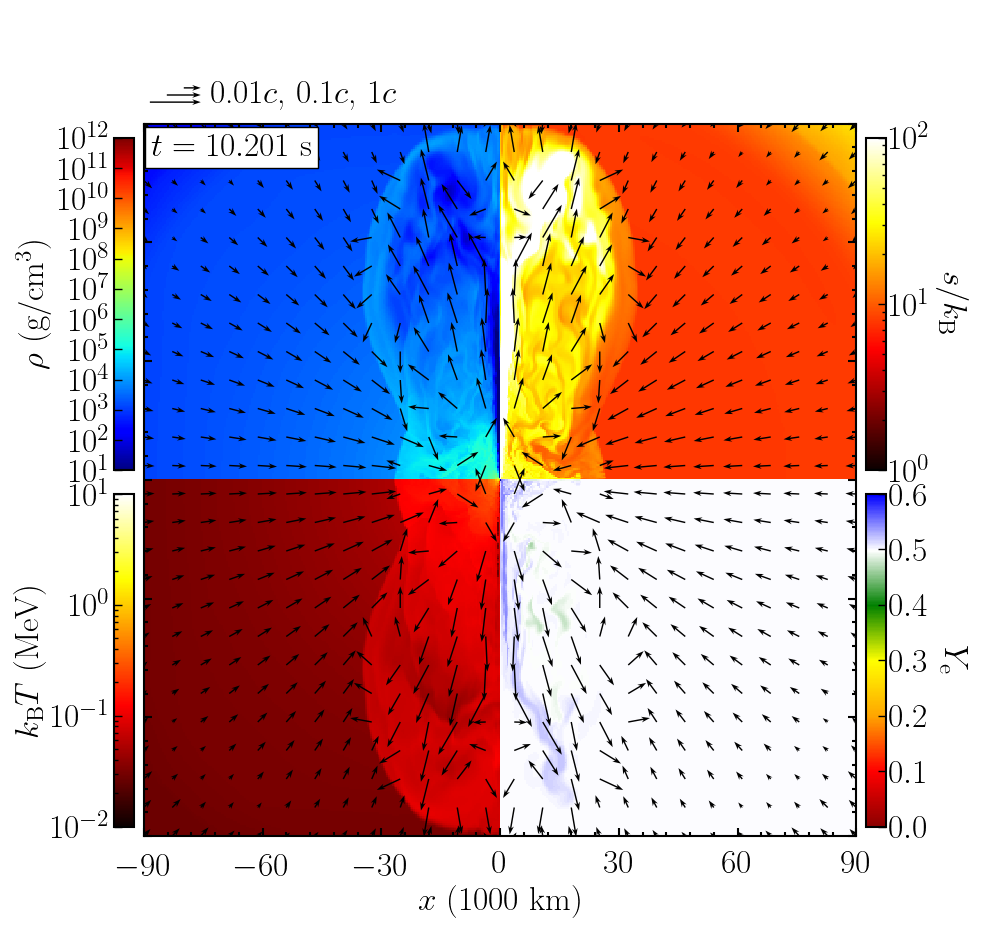}
\caption{The same as Fig.~\ref{figA} but for model B10.5. 
An animation is found at \url{https://www2.yukawa.kyoto-u.ac.jp/~sho.fujibayashi/share/B10.5-multiscale.mp4}
}
\label{figB}
\end{figure*}

\begin{figure*}[t]
\includegraphics[width=0.355\textwidth]{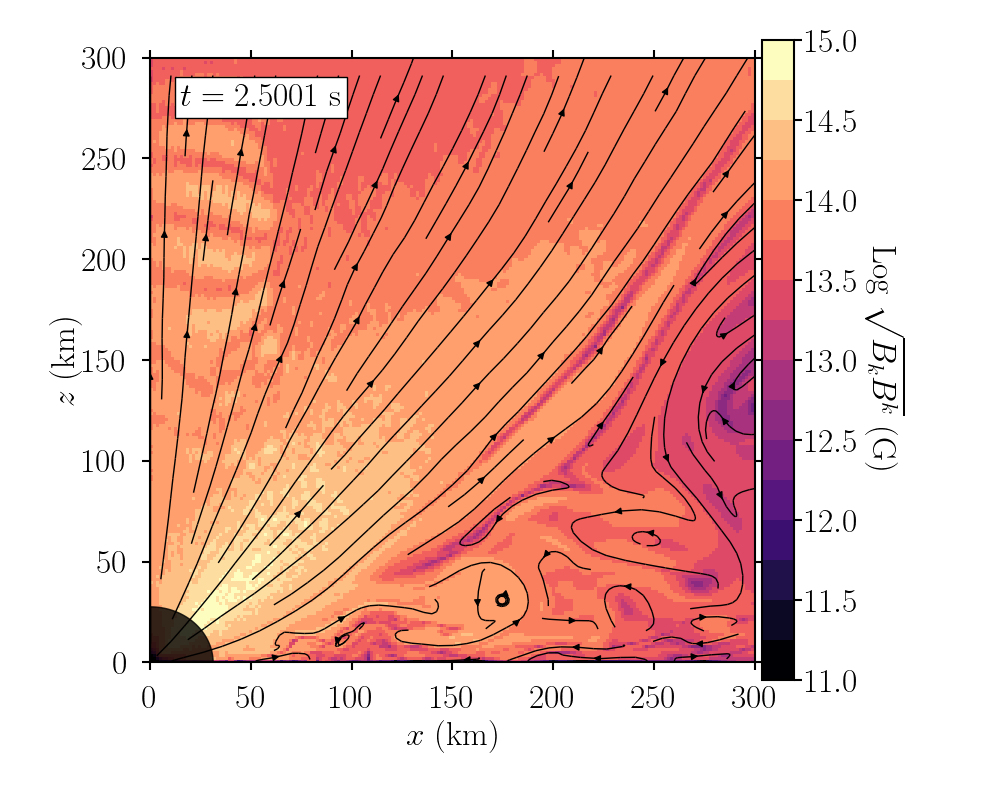}
\hspace{-8mm}
\includegraphics[width=0.355\textwidth]{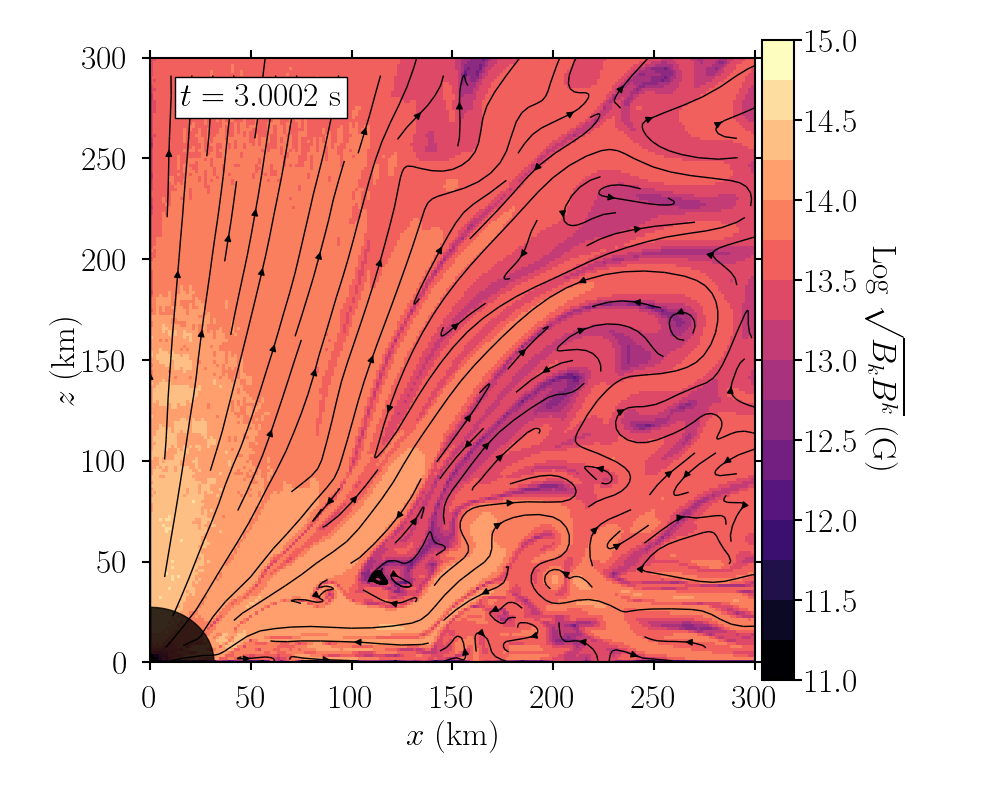}
\hspace{-8mm}
\includegraphics[width=0.355\textwidth]{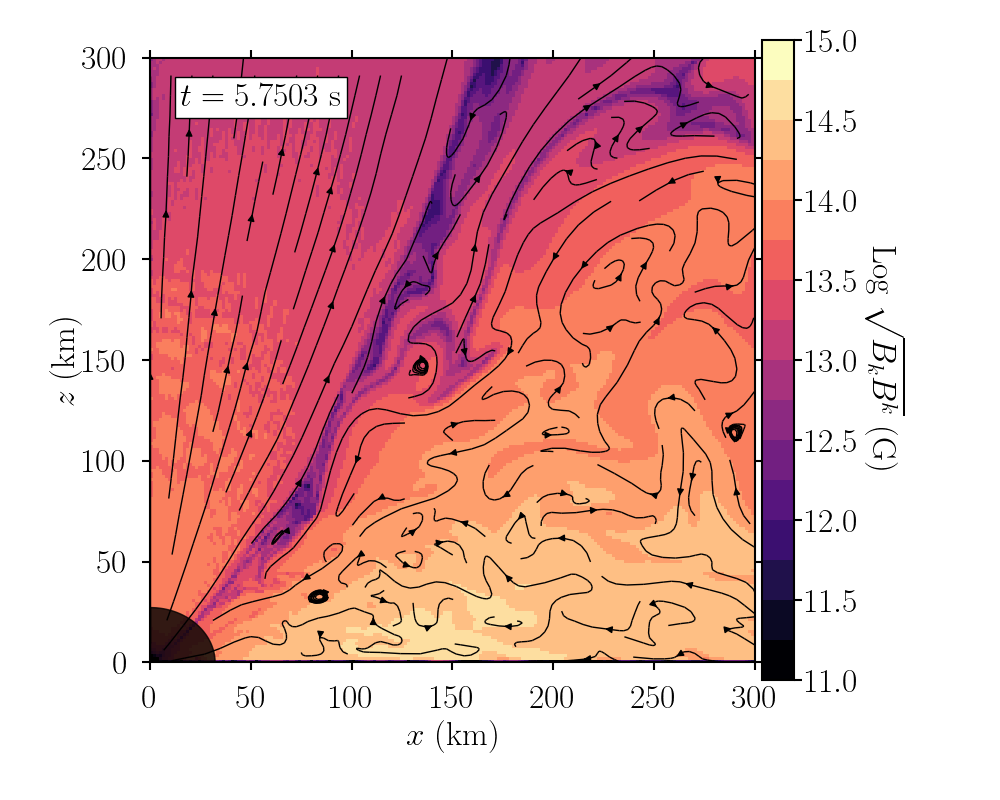}
\caption{The magnetic-field lines and field strength in an inner region of $300$\,km$\times 300$\,km are shown for the stage at which the outgoing jet is established for models B11.5 (left), B11.0 (middle), and B10.5 (right) on the $\varpi$-$z$ plane. 
}
\label{figC}
\end{figure*}

\begin{figure*}[t]
\includegraphics[width=0.32\textwidth]{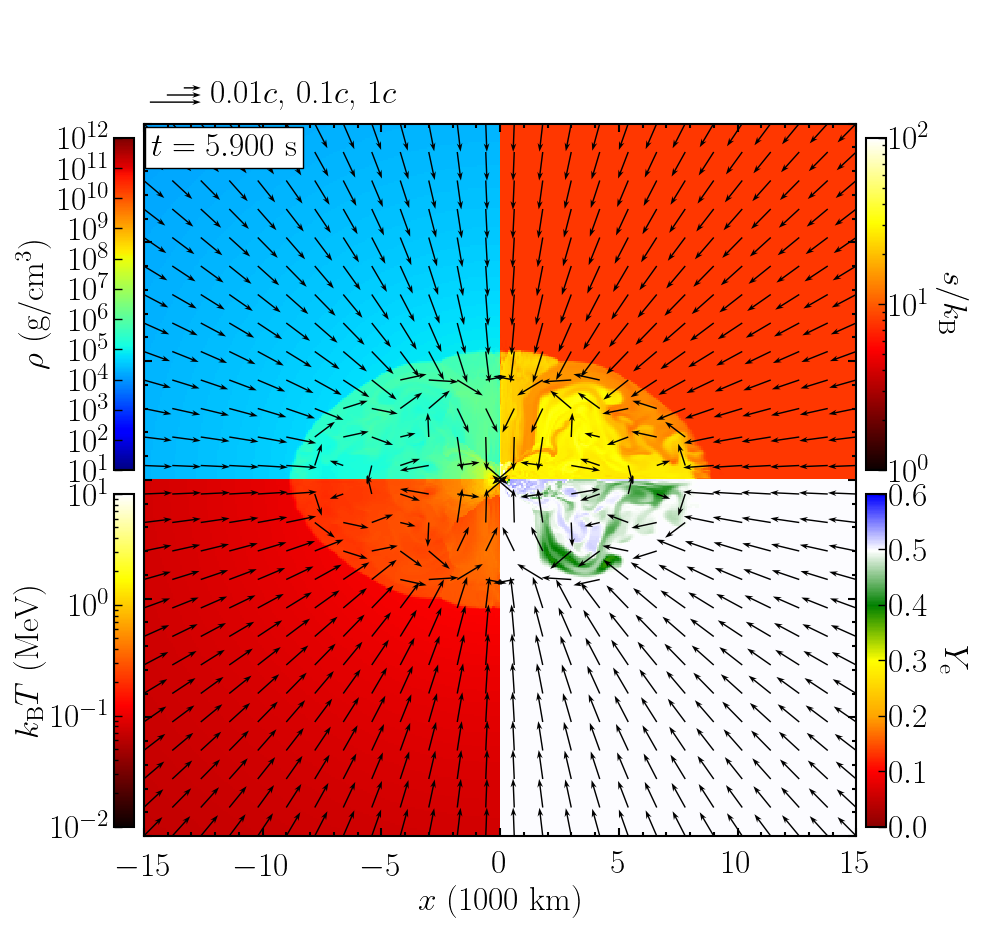}
\includegraphics[width=0.32\textwidth]{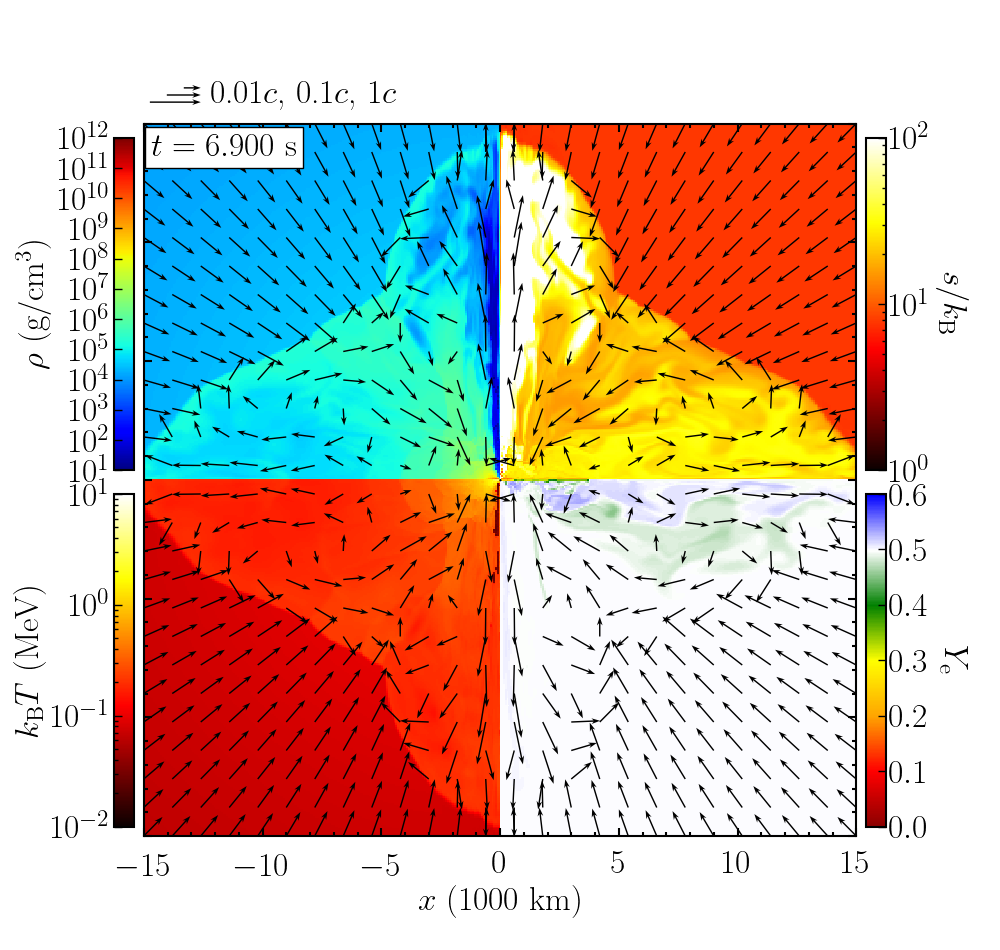}
\includegraphics[width=0.32\textwidth]{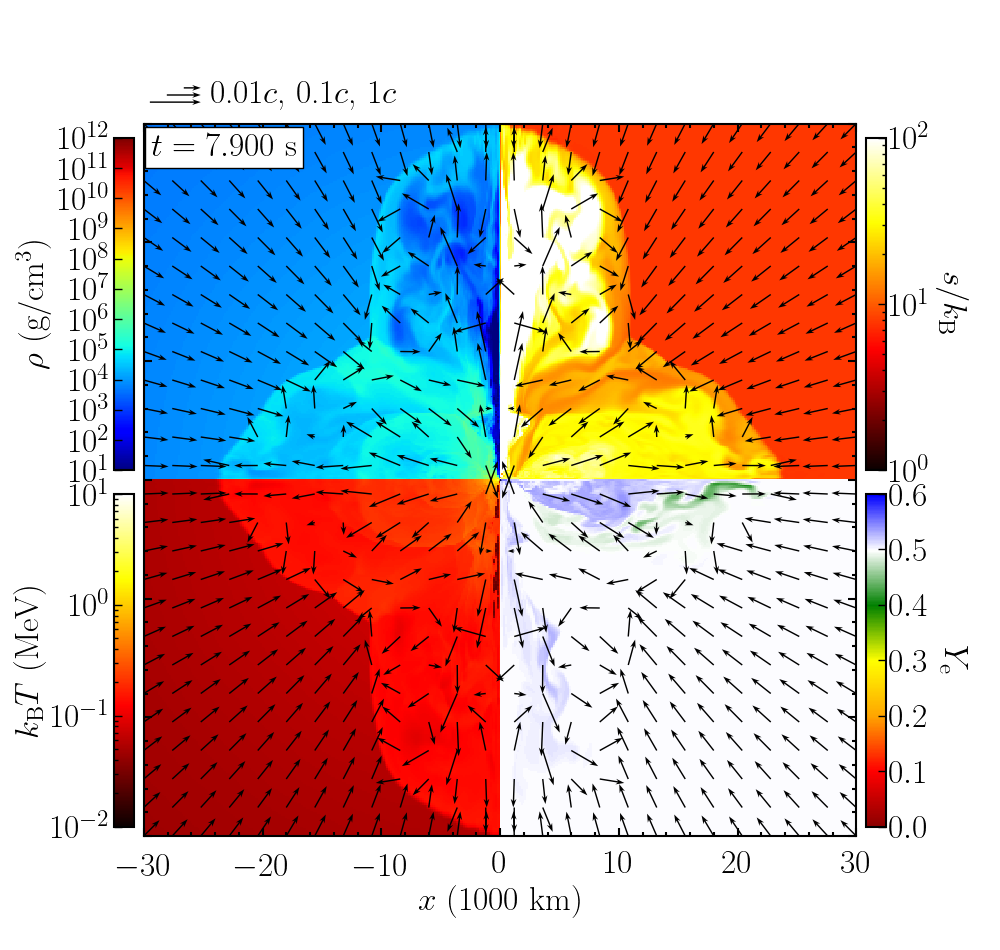}\\
\includegraphics[width=0.32\textwidth]{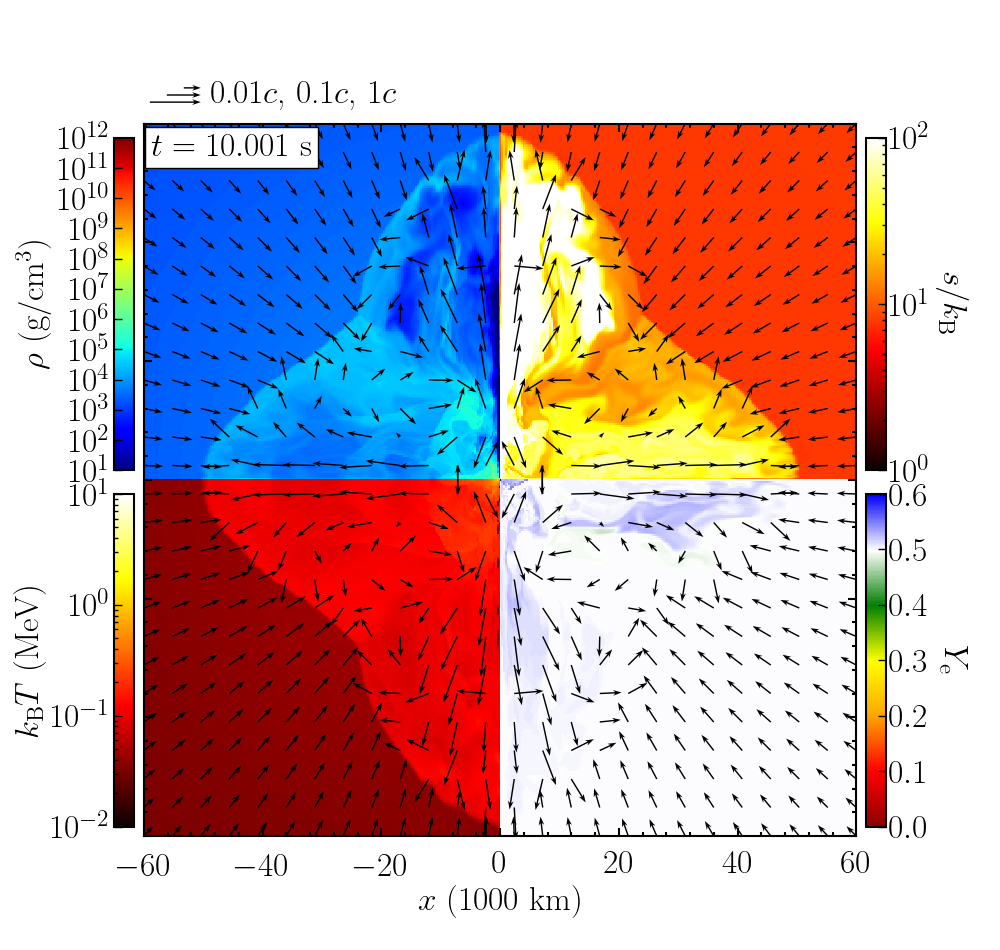}
\includegraphics[width=0.32\textwidth]{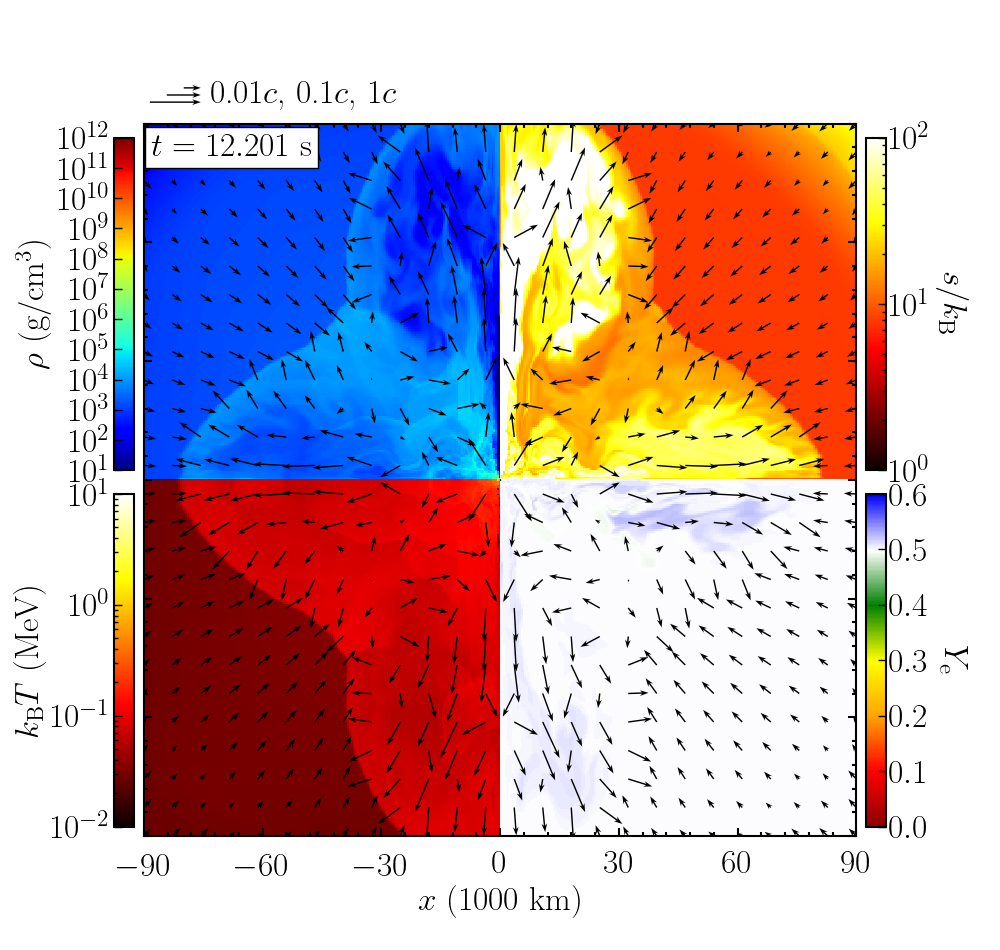}
\includegraphics[width=0.335\textwidth]{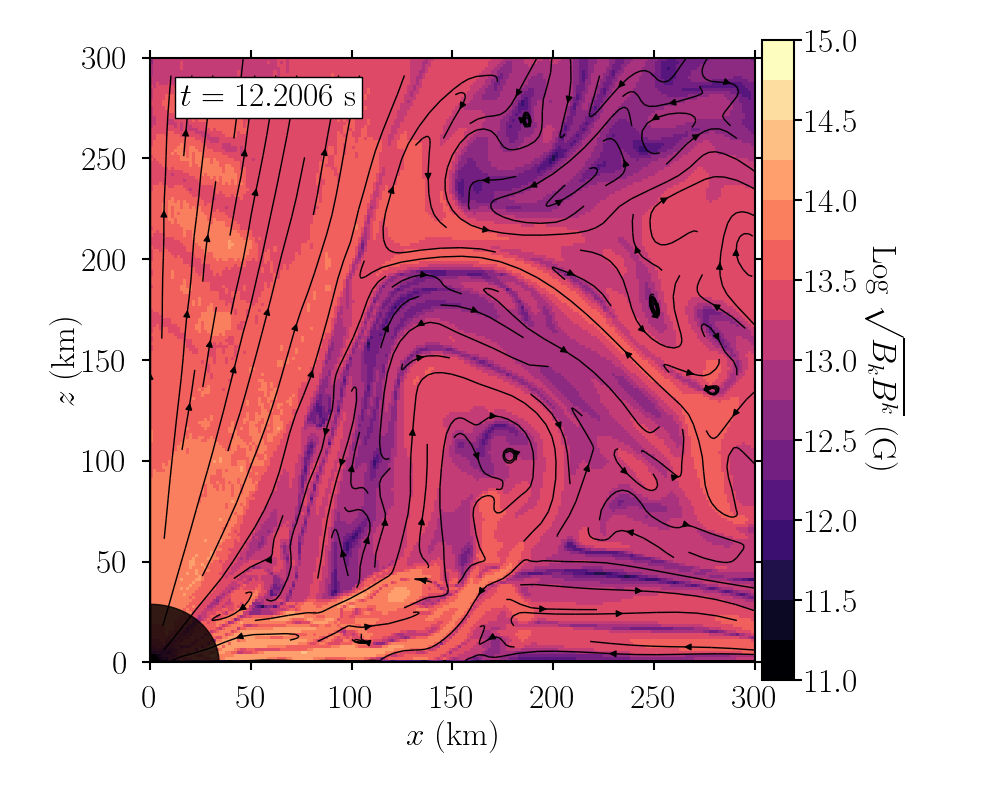}
\caption{The first 5 panels are the same as Fig.~\ref{figA} and 
the panel in the bottom right is the same as Fig.~\ref{figC} but for model Bq11.0b. An animation is found at \url{https://www2.yukawa.kyoto-u.ac.jp/~sho.fujibayashi/share/Bq11.0b-multiscale.mp4}
}
\label{figE}
\end{figure*}

\begin{figure*}[t]
\includegraphics[width=0.32\textwidth]{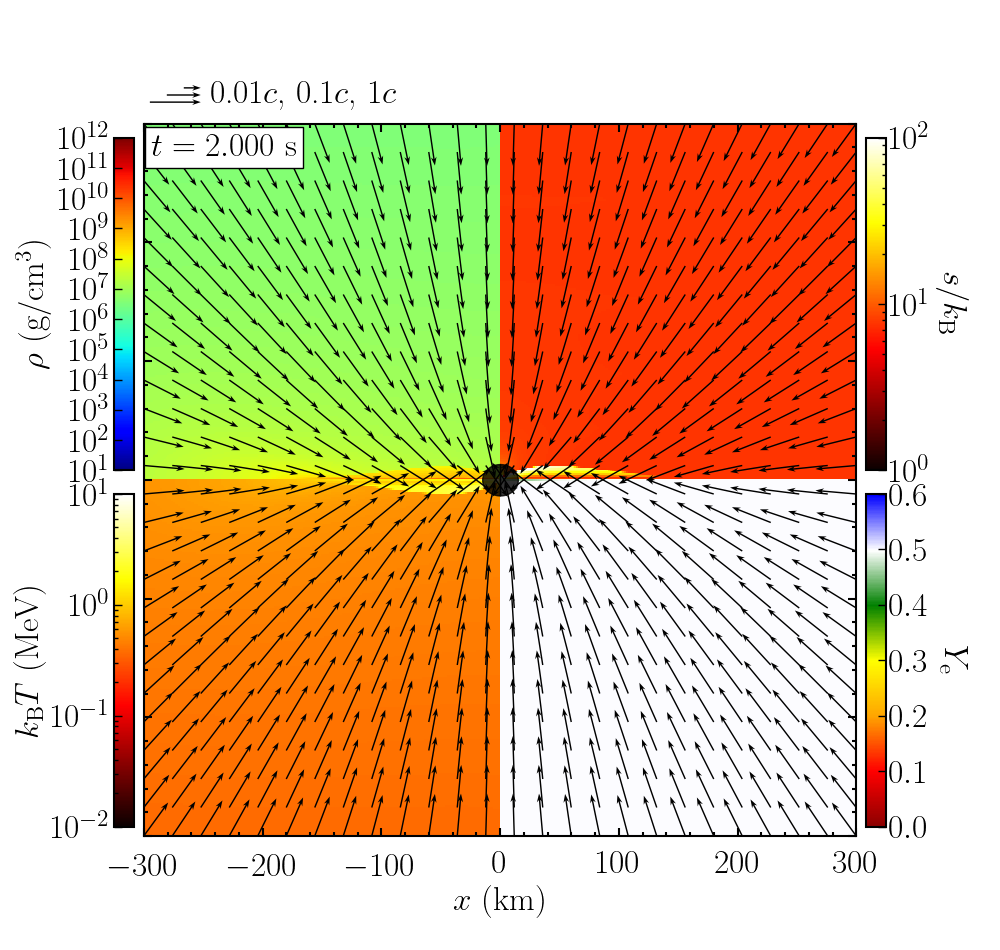}
\includegraphics[width=0.32\textwidth]{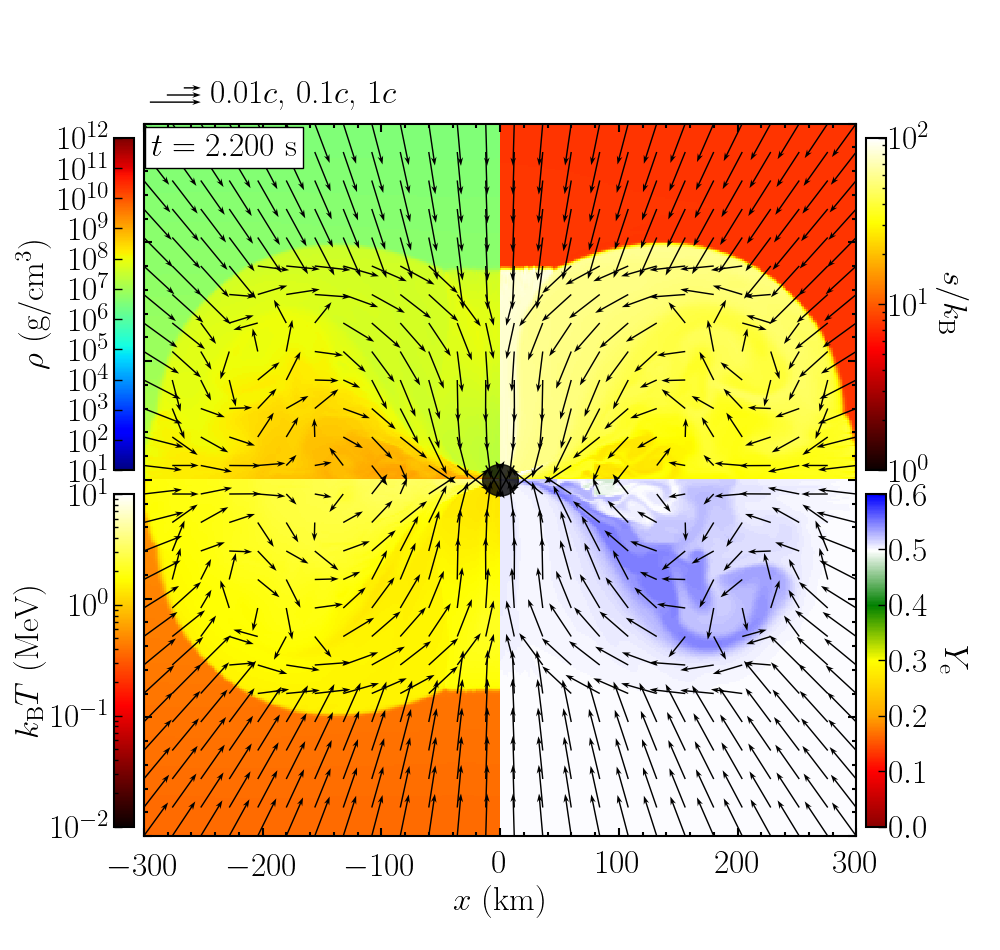}
\includegraphics[width=0.32\textwidth]{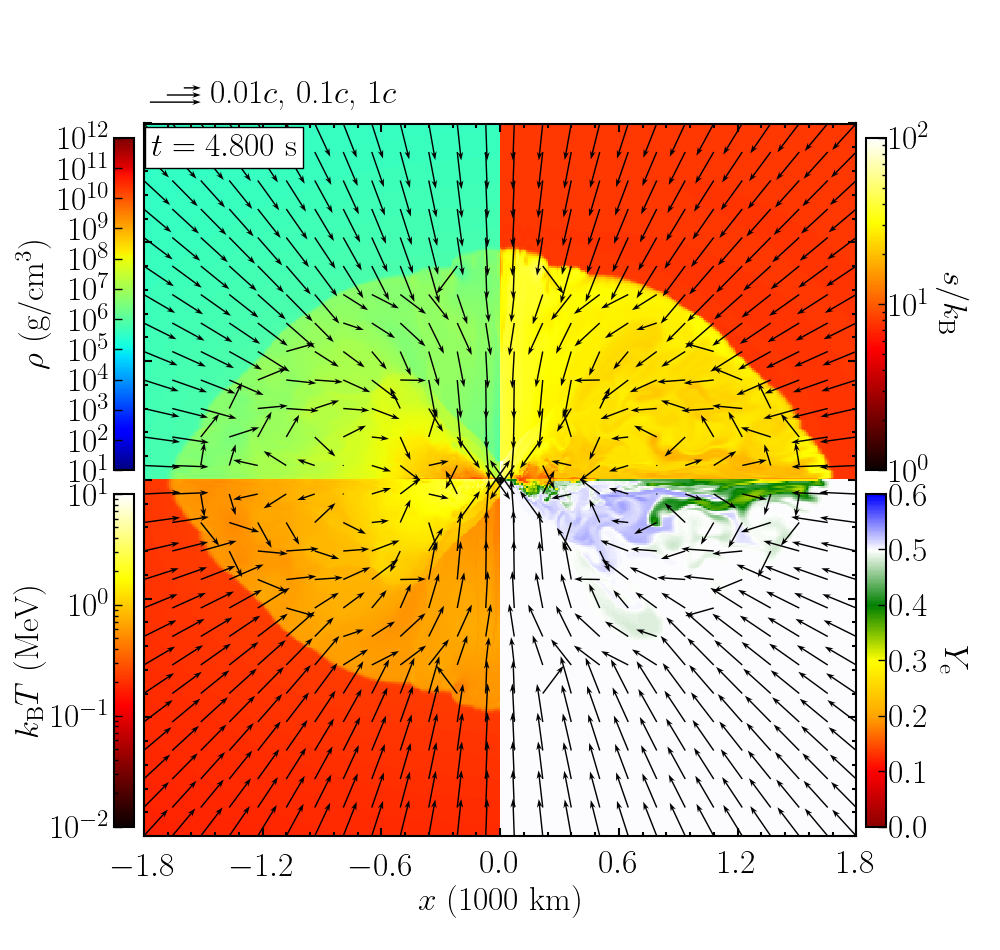}\\
\includegraphics[width=0.32\textwidth]{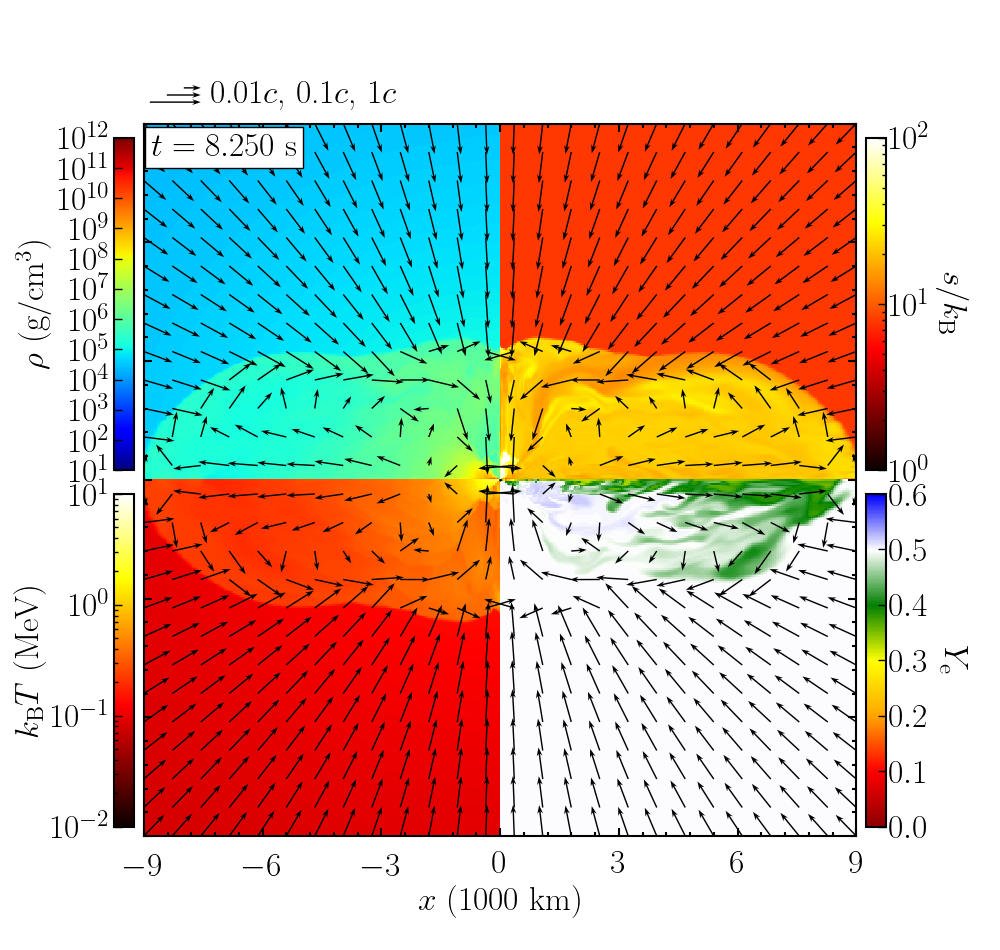}
\includegraphics[width=0.32\textwidth]{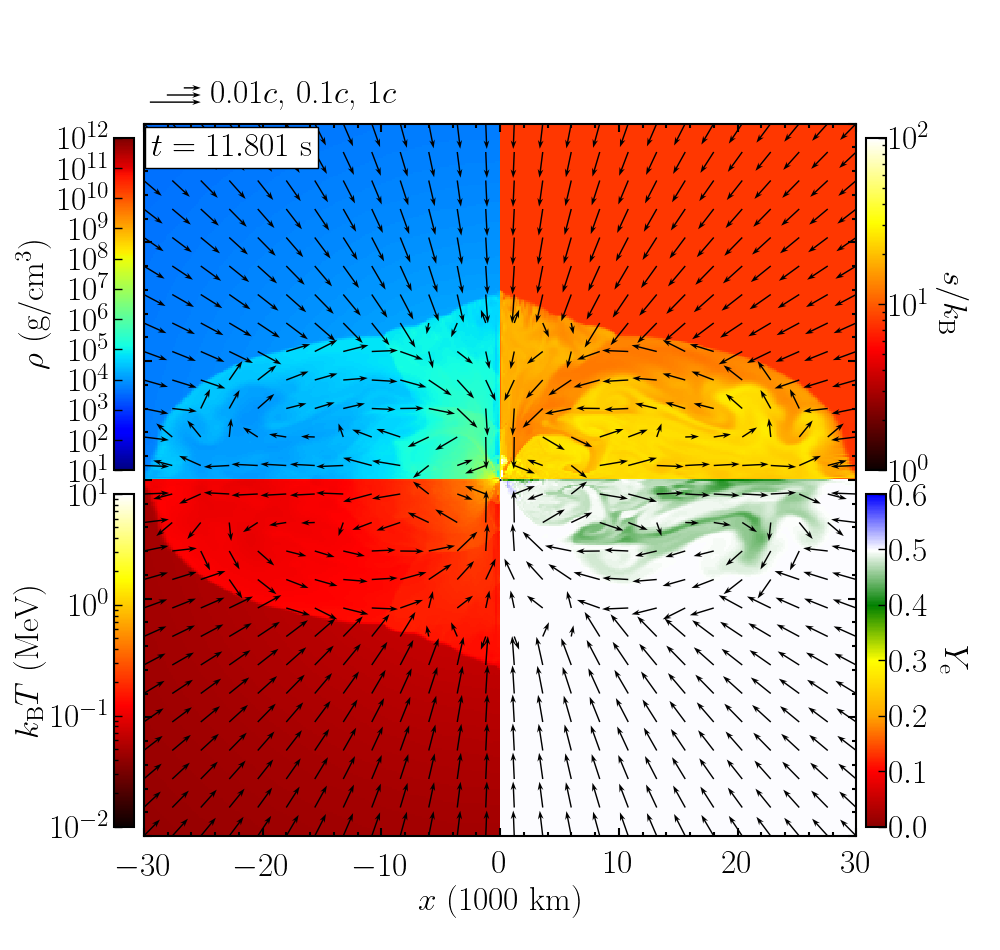}
\includegraphics[width=0.335\textwidth]{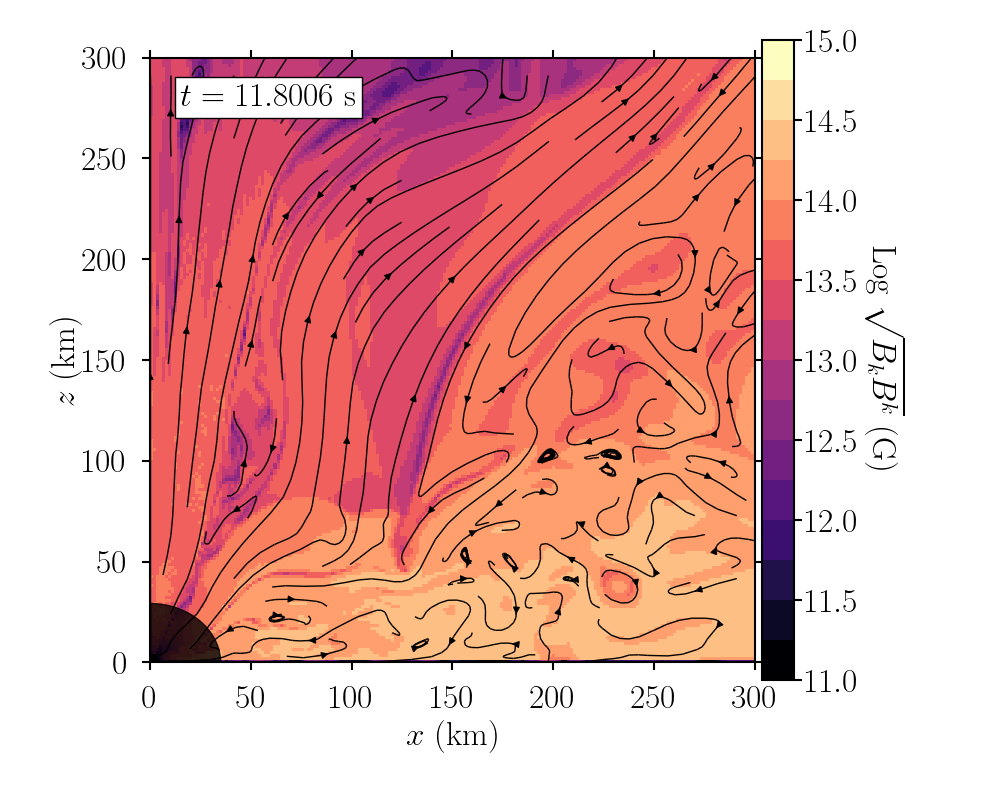}
\caption{The same as Fig.~\ref{figE} but for model Br10.5, for which the jet launch was not found in the simulation time of 12\,s. An animation is found at \url{https://www2.yukawa.kyoto-u.ac.jp/~sho.fujibayashi/share/Br10.5-multiscale.mp4}
}
\label{figD}
\end{figure*}

As in our recent paper~\cite{Fujibayashi2023BH}, the simulation is performed on a two-dimensional domain of $\varpi$ and $z$ (see also Refs.~\cite{Fujibayashi2020a,Fujibayashi2020b}). For the $\varpi$ and $z$ directions, the following non-uniform grid is employed for the present numerical simulations: For $x \alt 7GM_\mathrm{BH,0}/4c^2$ ($x=\varpi$ or $z$), a uniform grid is used, while outside this region, the grid spacing $\Delta x_i$ is increased uniformly as $\Delta x_{i+1}=1.01\Delta x_i$, where the subscript $i$ denotes the $i$-th grid. 
The black-hole horizon (apparent horizon) is always located in the uniform grid zone, and the outer boundaries along the $\varpi$ and $z$ axes are located at $\approx 10^5$\,km. The grid resolution of the uniform grid zone is $\Delta x =360\,\mathrm{m} \approx 0.016GM_\mathrm{BH,0}/c^2$, which is chosen to derive a reliable result for the black-hole spin evolution (see Appendix B of Ref.~\cite{Fujibayashi2023BH}). For two models (B11.5 and B11.3) we perform higher resolution runs with $\Delta x =300\,\mathrm{m}\approx 0.0135GM_\mathrm{BH,0}/c^2$ to confirm that the rate of the spin-down of black holes are computed in a fair accuracy irrespective of the grid resolution. We refer to these models as B11.5hi and B11.3hi, respectively. 

\subsection{Jet launch or not}\label{sec3A}

For our present setting, a mildly relativistic jet is found except for (i) the models for which the initial magnetic-field strength is too weak (B10.0 and Br10.5) and the field amplification is not large enough in the simulation time ($\sim 10$\,s), and (ii) the models with the initial field configuration of Eq.~(\ref{eqB2}) with $B_\mathrm{max} \leq 1 \times 10^{12}$\,G and $\varpi_0=10^3$\,km. The mechanism for launching the jet depends on the initial magnetic-field strength and configuration (see Ref.~\cite{2008ApJ...678.1180B} for a related topic). Thus, we will describe it separately. 

\subsubsection{Strong initial field cases}\label{sec3A1}

For the initial conditions with strong magnetic fields aligned well with the black-hole spin axis, a jet is launched in a short timescale after the magnetic-field amplification near the black hole by the winding associated with the black-hole spin and by the compression due to the matter infall onto the black hole.

Figure~\ref{figA} displays the snapshots for the rest-mass density, temperature, entropy per baryon, and electron fraction at 6 selected time slices for model B11.5 for which the initial magnetic-field strength is high enough to launch a jet in a short timescale ($\sim 0.4$\,s). We note that the displayed range is wider for the later-stage snapshots in this figure. 

For this model, the magnetic field is quickly amplified by the winding associated with the black-hole spin far before the formation of a disk around the black hole (note that the pancake structure for the rest-mass density at the second panel of Fig.~\ref{figA} does not imply the formation of the orbiting disk, because it is compact enough to be  subsequently swallowed by the black hole). Since the angular velocity of the black hole is approximately
\beqn
\Omega_\mathrm{BH} &\approx& 2.8 \times 10^3\left({\chi \over 0.7}\right)
\left({M_\mathrm{BH}\over 15M_\odot}\right)^{-1}
\left({\hat r_+ \over 1.7}\right) \,{\rm rad/s},
\nonumber \\
&&~~~~~~~~~~~~~~~ 
\eeqn
and the magnetic-field strength can increase approximately in proportional to  $\Omega_\mathrm{BH}t$ near the black hole, the maximum field strength can increase by $10^3$ times in $\sim 0.4$\,s. Indeed, the maximum magnetic-field strength in the polar region of the black hole exceeds $10^{14}$\,G at $t\approx 0.4$\,s, leading to high magnetic pressure of $\approx 4 \times 10^{26}(B/10^{14}\,{\rm G})^2\,{\rm dyn/cm^2}$. On the other hand, the ram pressure of the infalling matter is $\rho v_\mathrm{infall}^2=4 \times 10^{26}\,{\rm dyn/cm^2}$ for $\rho=10^6\,{\rm g/cm^3}$ and $v_\mathrm{infall}=2c/3 \approx 2 \times 10^5\,{\rm km/s}$ (see Eq.~(\ref{eq:ram1})). 
Thus, when the density of the infalling matter decreases below $\sim 10^6\,{\rm g/cm^3}$, the magnetic pressure overcomes the ram pressure in this case. Indeed, for model B11.5, the rest-mass density along the $z$-axis is a few times $10^6\,{\rm g/cm^3}$ at the launch of the jet (cf. the second panel of Fig.~\ref{figA}). 

Once the jet is launched, subsequently, it quickly goes outward because the ram pressure decreases with the radius while the magnetic-field strength does not steeply decrease for this model. The magnetic-field strength near the black-hole horizon also becomes strong enough to halt the mass accretion from the equatorial direction as well as from the polar direction. Thus, a magnetically arrested disk (MAD)~\cite{2003PASJ...55L..69N,Igumenshchev:2003rt,Tchekhovskoy:2011zx} structure is established after the jet launch. Indeed, the dimensionless MAD parameter defined by
\beq
\phi_\mathrm{AH}:={\Phi_\mathrm{AH} \over \sqrt{4\pi G^2 c^{-3} \dot M_\mathrm{BH,*} M_\mathrm{BH}^2}},
\eeq
is high $\phi_\mathrm{BH} \agt 20$ up to $\agt 10^3$ after the jet launch (cf.~Fig.~\ref{fig2}). Here, $\Phi_\mathrm{AH}$ is the magnetic flux that penetrates the black-hole horizon (practically apparent horizon) and $\dot M_\mathrm{BH,*}$ denotes the rest-mass infall rate across the horizon. In later stages, $\dot M_\mathrm{BH,*}$ becomes quite low $\alt 10^{-4}M_\odot$/s for models B11.5, B11.3, and Bq11.0c. 


We also find similar jet generation mechanisms for model B11.3, for which $|\Phi_\mathrm{AH}|$ is only slightly smaller than that for model B11.5 (cf.~Fig.~\ref{fig2}). Since the magnetic-field strength near the horizon for given time is higher for models with higher initial field strength, the jet launch is earlier for higher values of $B_\mathrm{max}$. In other words, the jet launch is delayed until the formation of a disk at $t \sim 2$\,s, if $B_\mathrm{max}$ is smaller than a threshold value. 

Models B11.0 and Br11.0 have values of $B_\mathrm{max}$ which are close to such a threshold value, and hence, the jet launch times ($t \sim 2$\,s) are appreciably later than those for models B11.5 and B11.3. However, a jet is launched before the disk formation for these models. 
It is worthy to emphasize again that for higher values of $B_\mathrm{max}$, the magnetic-field strength on the horizon is higher after the jet launch (cf.~Fig.~\ref{figC}). This stems from the fact that the ram pressure at the jet launch is higher for the earlier jet-launch case (i.e., for larger values of $B_\mathrm{max}$). This results in  higher Poynting luminosity by the Blandford-Znajek mechanism during the jet propagation for larger values of $B_\mathrm{max}$ (see below). 


For models B11.3, B11.0, and Br11.0, we followed the evolution of the collapsing envelope for a long timescale and found that the star entirely explodes together with the jet propagation (see, e.g.,  Refs.~\cite{2006ApJ...647.1192U,2008MNRAS.385L..28B,2019ApJ...871L..25P,Eisenberg2022nov} for related issues). For these models, poloidal magnetic-field lines that penetrate the spinning black hole are present not only along the polar region but also for the other regions. In such a situation, the magneto-centrifugal force associated with the black-hole spin plays an important role in transporting the angular momentum from the inner to the outer region, which can be an engine of the stellar explosion. Also strong toroidal magnetic fields enhanced by the winding associated with the black-hole spin can be the source of the Tayler instability~\cite{1973MNRAS.161..365T,1978RSPTA.289..459A,1985MNRAS.216..139P}. The Tayler instability can induce a convective motion for redistributing the entropy and the angular momentum of the fluid elements~\cite{Kiuchi:2008ss}, and hence, it may also contribute to the stellar explosion. The Tayler instability appears to play a more important role for the models with an initially large cylindrical component of the magnetic fields, e.g., for models Br11.0, Br10.5, Bq12.5, and Bq11.0b (see below). 

In the presence of efficient neutrino cooling, the jet propagation may be decelerated by the reduction of the thermal pressure. However, for the early jet-launch models considered in this subsection, the maximum neutrino luminosity is of order $10^{50}$\,erg/s, which is smaller than the Poynting luminosity associated with the Blandford-Znajek mechanism (cf.~Sec.~\ref{secIIIC}). Hence, it is unlikely that the neutrino cooling gives a seriously negative effect for the jet launch. It should be also mentioned that the jets are driven by the magnetohydrodynamical effect, and hence, the thermal pressure does not play a primary role.

\subsubsection{Weak initial field cases}\label{sec3A2}

For the models with initially weak magnetic fields such as models B10.5, B10.0,  Br10.5, Bq12.5, Bq12.0, and Bq11.0b, a jet is generated after the formation of a disk/torus or no jet formation is found in the simulation time. Also, the evolution process is qualitatively different from that for the initially strong-field cases discussed in the previous subsection. 

Figure~\ref{figB} displays the same plots as Fig.~\ref{figA} but for model B10.5. For this case, a jet is not driven before the formation of a disk around the black hole, and for an early stage, the formation of a disk and a torus proceeds (see the first panel). Because the disk/torus evolves only quasi-steadily and orbits the black hole with less angular velocity than $\Omega_\mathrm{BH}$, the magnetic stress is enhanced due to the winding of the magnetic-field lines connecting between a black hole and orbiting matter and also the angular momentum is transformed from the black hole to the matter\footnote{For slowly spinning black holes with $\chi \alt 0.36$, the angular velocity of the matter orbiting the black holes can be larger than $\Omega_\mathrm{BH}$, and hence, the angular momentum may not be transported outward. In this case, the orbiting matter may contribute to spin-up of the black holes.}. At the formation of the torus (see the first panel of Fig.~\ref{figB}), in addition, an oblique shock is formed around its surface and enhances the matter flow toward the polar region. This also enhances the magnetic-flux inflow toward the black hole, and consequently, the magnetic-field strength near the polar region of the black hole is increased. When the magnetic pressure exceeds the ram pressure at $t \agt 2.6$\,s near the horizon, a jet is driven from the vicinity of the black hole toward the polar region (cf. the second and third panels of Fig.~\ref{figB}). 

However, for this case, the magnetic-field strength on the horizon achieved until the jet launch is not as high as those for models B11.5, B11.3, B11.0, Br11.0, and Bq11.0c (cf.~Fig.~\ref{figC}), and hence, the jet is once decelerated on the way of the propagation (cf.~the third and fourth panels of Fig.~\ref{figB}). During this stage, the opening angle of the jet is widen to the equatorial region and the magnetic flux on the horizon decreases (see Fig.~\ref{fig2}). Nevertheless, the winding of the magnetic-field lines by the black-hole spin continuously enhances the magnetic pressure near the polar region and, at the same time, the ram pressure of the infalling matter decreases with time. This eventually causes the revival of the jet (cf. the fifth and sixth panels of Fig.~\ref{figB}) although the propagation speed is much lower than those for models B11.5, B11.3, B11.0, Br110.0, and Bq11.0c. 

The situation for models Bq12.5 and Bq11.0b is similar to that of model B10.5 although the disk/torus evolution stage is longer (see Fig.~\ref{figE} for model Bq11.0b). For these models, due to the angular momentum transport associated with magneto-centrifugal effects by the black-hole spin (around the equatorial plane) and orbital motion, the torus expands gradually with time, in particular toward the equatorial direction. Also due to the matter infall, the black-hole mass and spin increase with time. 
For models B10.5, Bq12.5, and Bq11.0b, the MAD parameter is $\sim 5$--10 in the late stage of the jet propagation (see Fig.~\ref{fig2}) because the magnetic flux on the horizon is by one order of magnitude smaller than those for models B11.5 and B11.3. The evolution processes for models B10.5, Bq12.5, and Bq11.0b indicate that in the presence of a poloidal magnetic field that penetrates a spinning black hole, a jet may be always generated after long-term winding of the magnetic-field lines even if the initial magnetic field strength is not very strong.



Figure~\ref{figC} displays the magnetic-field lines and field strength in the vicinity of the black hole on the $\varpi$-$z$ plane for the stages at which a jet was already launched for models B11.5, B11.0, and B10.5 (see also the bottom-right panel of Fig.~\ref{figE}). This clearly shows that for higher initial magnetic-field models (i.e., for earlier jet-launch models), the magnetic fields around the black hole are stronger reflecting the ram pressure at the jet launch. For model B10.5, the magnetic-field strength is highest around the equatorial plane at the selected time slice because a torus is present there and the magnetic field is amplified by the winding and partly by the MRI (note that due to the anti-dynamo nature in the axisymmetric simulation~\cite{1978mfge.book.....M}, the MRI dynamo cannot be developed in this simulation). 
For model B11.0, an orbiting disk is not formed around the black hole before the jet launch, but mass accretion proceeds from the equatorial region, gradually increasing the black-hole mass (cf. Fig.~\ref{fig5}). 

\begin{figure*}[t]
\includegraphics[width=0.49\textwidth]{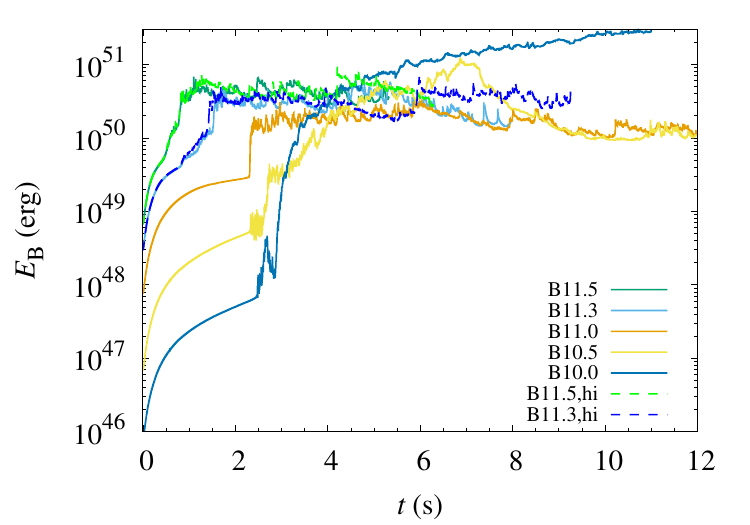}
\includegraphics[width=0.49\textwidth]{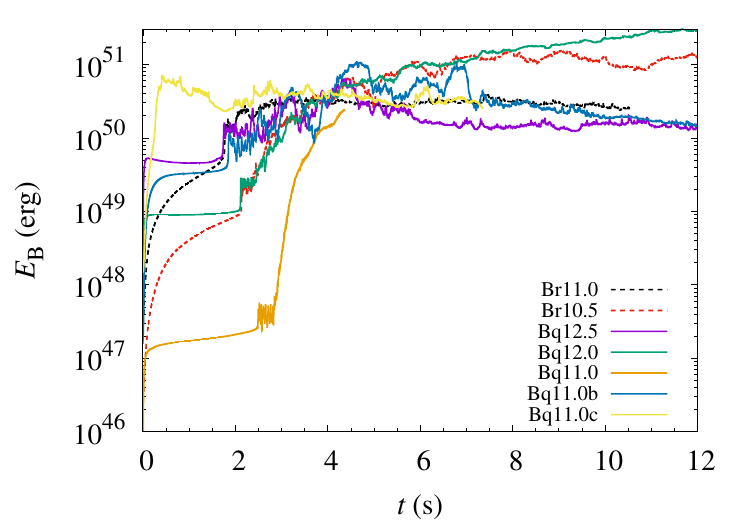}
\caption{Evolution of the electromagnetic energy for the models with Eq.~(\ref{eqBz}) (left) and with Eqs.~(\ref{eqB}) and~(\ref{eqB2}) (right). We stopped the simulation for model Bq11.0 at $t\approx 4.3$\,s because the evolution path looks similar to that for model Bq12.0 after the disk formation. 
}
\label{fig1}
\end{figure*}

For models B10.0, Br10.5, and Bq12.0, neither a jet nor an outflow is  launched in the simulation time of $\sim 10$\,s. For these models, the magnetic-field strength on the black-hole horizon is not enhanced enough to launch a jet during the torus formation, and the torus is simply evolved around the black hole (see Fig.~\ref{figD} for model Br10.5). In particular for the initial condition with Eq.~(\ref{eqB2}) with $\varpi_0 \leq 5 \times 10^3$\,km, the magnetic-field strength on the black hole does not significantly increase with time in the early stage before the torus formation even for very high values of $B_\mathrm{max}$ (cf. Figs.~\ref{fig1} and \ref{fig2}). Only for a high value of $\varpi_0$, e.g., $10^4$\,km, a jet is quickly launched even for Eq.~(\ref{eqB2}) with $B_\mathrm{max}\agt 10^{11}$\,G (this is likely to be also the case for very high values of $B_\mathrm{max}$ with $\varpi_0=10^3$\,km). This indicates that the magnetic-field strength on the horizon before the formation of a disk/torus depends strongly on the field profile in the progenitor star. 
In the present axisymmetric simulation, the dynamo mechanism does not work, and hence, the poloidal magnetic fields are not amplified sufficiently in the torus and on the black hole in a short timescale. As a consequence, the strong poloidal magnetic field that penetrates the black hole and launches a jet, is not developed quickly in the absence of initially strong fields (compare the last panel of Fig.~\ref{figD} with Fig.~\ref{figC}). In these cases, the poloidal magnetic field is not well aligned near the rotation axis. 

However, in reality (i.e., in non-axisymmetric simulations in which dynamo and resultant turbulence can be modelled), a strong poloidal field could be developed after the formation of a disk/torus within a certain timescale and we may expect that a jet is launched eventually (see, e.g., Refs.~\cite{Christie2019dec,Hayashi:2021oxy,Gottlieb:2023est} for recent relevant works). An accretion disk/torus for which the equipartition is established in the turbulent state would have the relation (e.g., Ref.~\cite{Hayashi:2021oxy})
\beq
{B_\mathrm{disk}^2 \over 8\pi} \sim f_\mathrm{eq} \rho_\mathrm{disk} c_s^2,
\eeq
where $B_\mathrm{disk}$ is the typical magnetic-field strength inside the disk/torus, $c_s$ is the typical sound speed, and $f_\mathrm{eq}$ is approximately constant with 0.02--0.05 for disks/tori in equipartition. Hence, for the typical values at the inner region of the disk/torus, we expect the magnetic-field strength
\beqn
B_\mathrm{disk} &\sim& 1 \times 10^{14} \left({f_\mathrm{eq} \over 0.04}\right)^{1/2}
\left({\rho_\mathrm{disk} \over 10^{10}\,{\rm g/cm^3}}\right)^{1/2}
\nonumber \\
&&~~~~~~~~~~\times \left({c_s \over 10^9\,{\rm cm/s}}\right)\,{\rm G}. \label{guessB}
\eeqn
Thus, by the accretion of the turbulent matter onto the black hole with a coherent magnetic-field polarity, a poloidal magnetic field that penetrates the black hole with $B\sim B_\mathrm{disk} \sim 10^{14}$\,G can be formed as illustrated in recent simulation works~\cite{Christie2019dec,Hayashi:2021oxy,Gottlieb:2023est}. This can be strong enough to launch a jet if the density of the infalling matter is low enough, satisfying $B^2/8\pi > \rho_\mathrm{infall} v_\mathrm{infall}^2$, near the black hole, i.e., 
\beq
\rho_\mathrm{infall} < f_\mathrm{eq}\rho_\mathrm{disk}\left({c_s \over v_\mathrm{infall}}\right)^2.
\eeq
Indeed this condition is satisfied in the late stage, e.g., for models Br10.5 and B10.0 if $f_\mathrm{eq}=O(0.01)$, i.e., $\rho_\mathrm{infall} \alt 10^{-4}\rho_\mathrm{disk}$ assuming $c_s/v_\mathrm{infall}\sim 0.1$. Therefore, for the initially weak field cases, a jet may be launched after a turbulent state is established in the disk/torus. 

Models Bq12.5 and Bq11.0b show not only a jet launch but also an explosion of the entire star (cf. Fig.~\ref{figE}) . As in the cases of the torus-formed models such as B10.0, Br10.5, and Bq12.0, initially a disk and subsequently a torus are developed in the early stage of the evolution for these models. Then, the magneto-centrifugal force associated with the black-hole spin around the equatorial plane appears to play an important role for developing a gradually expanding torus because the magnetic-field strength is high around the equatorial plane. Subsequently, the toroidal magnetic field is amplified by the winding associated with the black-hole spin and inside the torus. A convective motion resulting from the Tayler instability~\cite{1973MNRAS.161..365T,Kiuchi:2008ss} is also seen. As a result of these effects, the torus starts exploding approximately simultaneously with the jet launch. This explosive motion is accelerated with the decrease of the ram pressure of the infalling matter. This result suggests that, although the explosion was not observed for the models such as B10.0, Br10.5, and Bq12.0, in the longer-term evolution, these models may lead to the explosion eventually by the long-term winding of the magnetic field lines as well.

For all the massive disk/torus-formation models considered in this subsection, neutrino luminosity is enhanced to $\sim 2 \times 10^{52}$\,erg/s, and in the presence of the stellar explosion, it subsequently decreases with time. The energy source for this neutrino emission is the shock heating on the shock surface around the torus. This neutrino cooling may not play an important role in the jet launch that can be driven primarily by the magnetohydrodynamical effect. However, it can decelerate the stellar explosion because the neutrino luminosity is much higher than the Poynting luminosity by the Blandford-Znajek mechanism which is smaller than $10^{51}$\,erg/s for the disk/torus-formation models (see Fig.~\ref{fig3}). This fact indicates that, for the stellar explosion found in the current study, not the thermal pressure but the magnetohydrodynamical effect associated with the extraction of the rotational kinetic energy of the black hole plays a major role.


\subsection{Magnetic-field energy and magnetic flux on the horizon}\label{secIIIB}

\begin{figure*}[t]
\includegraphics[width=0.49\textwidth]{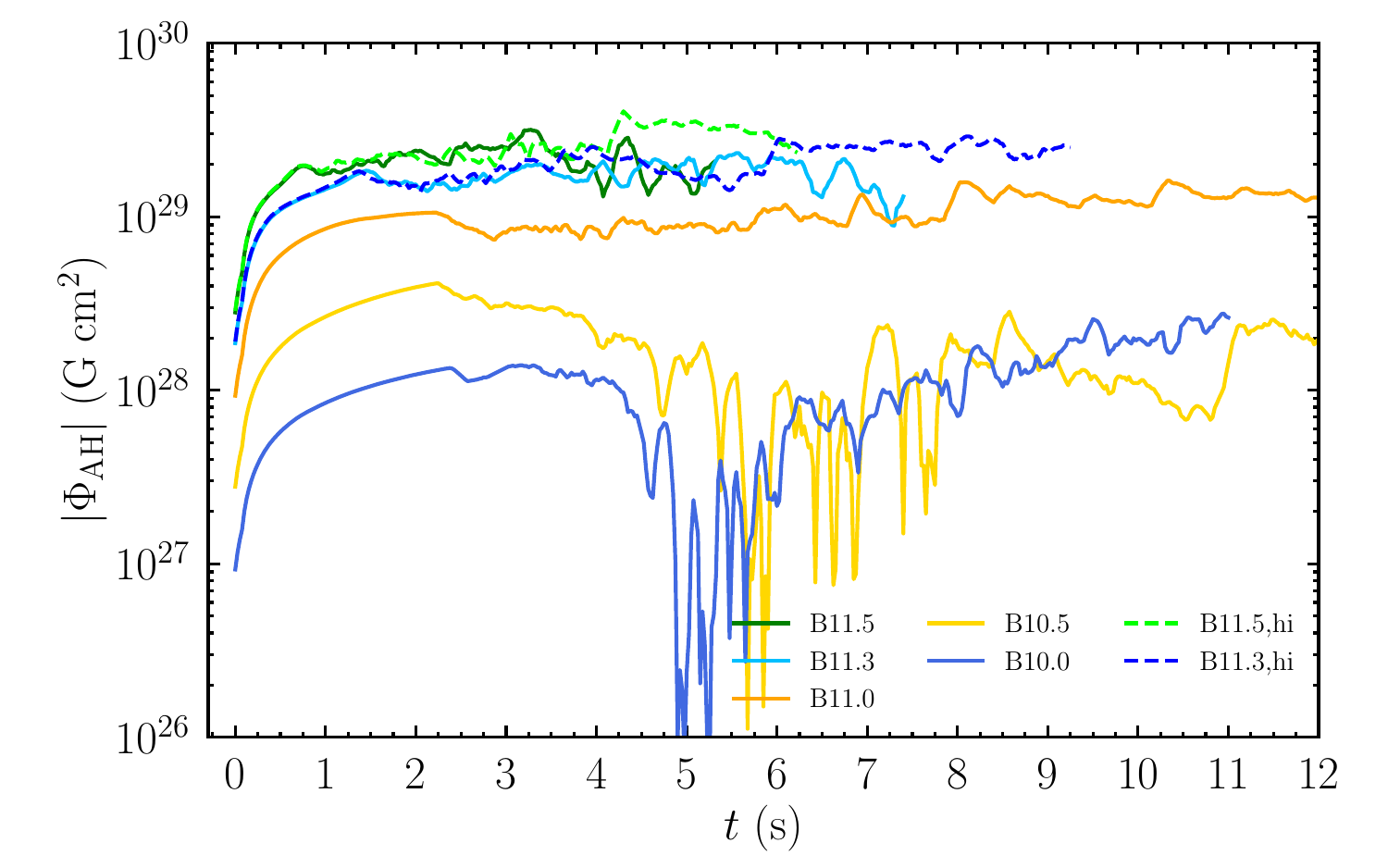}
\includegraphics[width=0.49\textwidth]{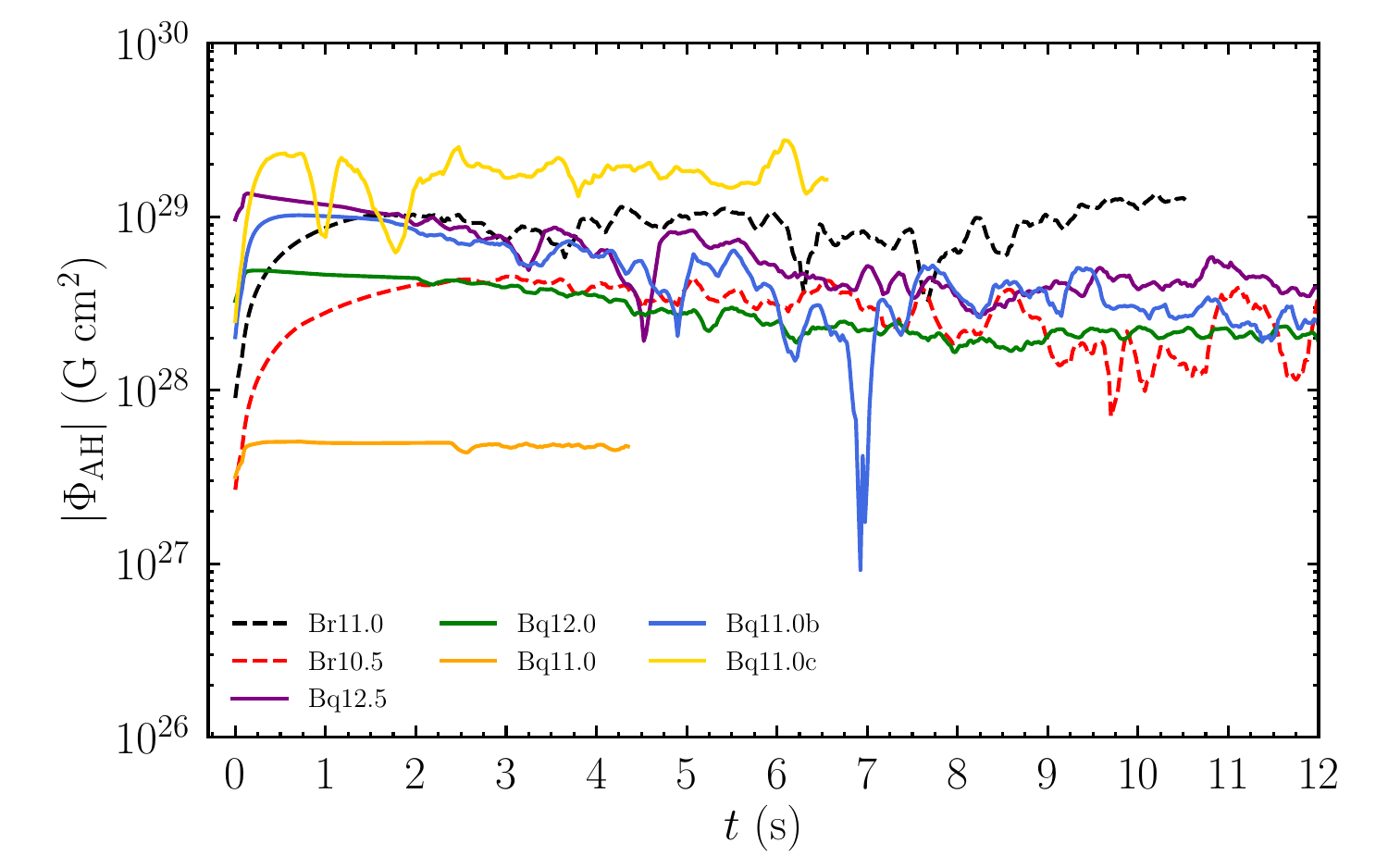}\\
\includegraphics[width=0.49\textwidth]{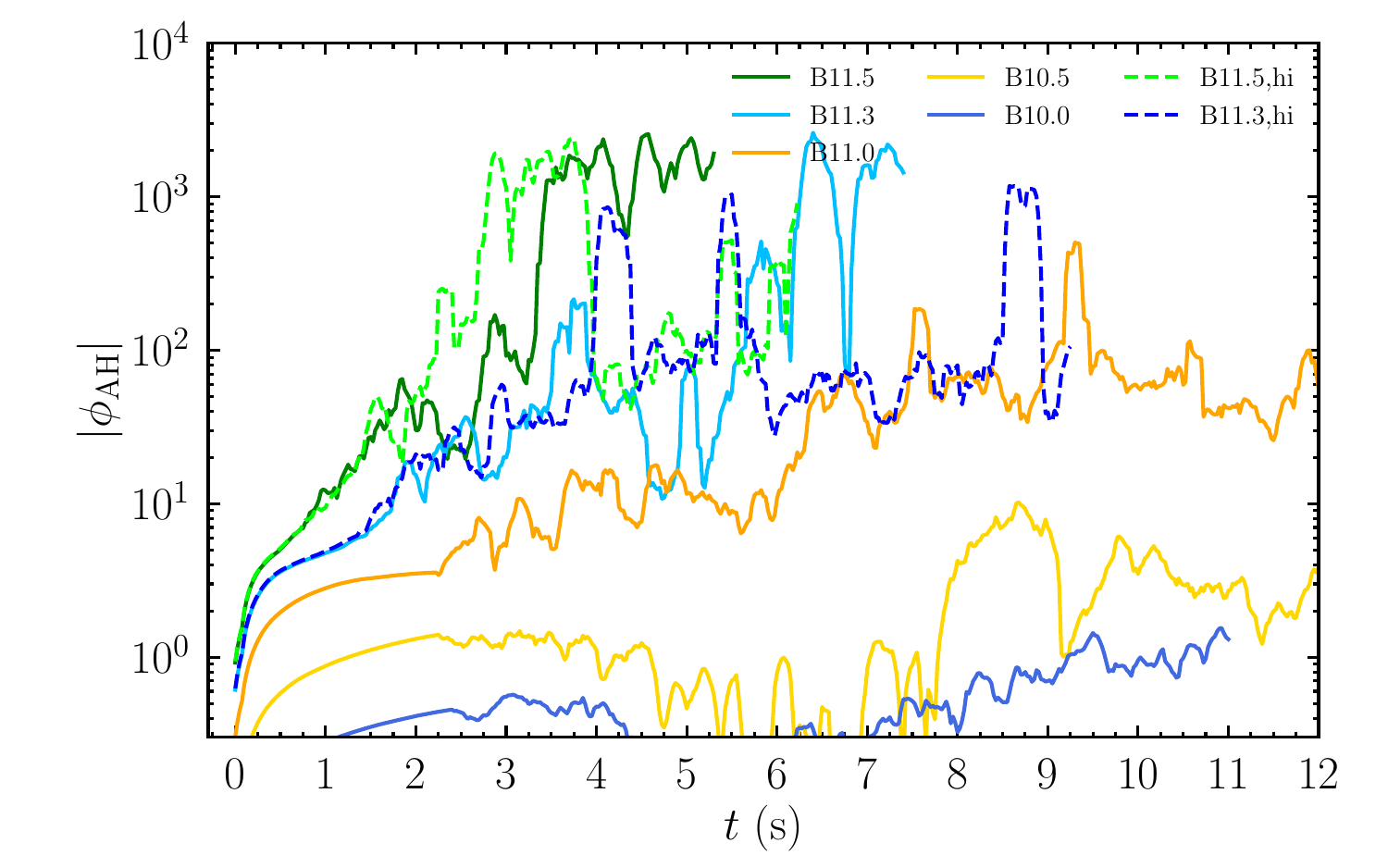}
\includegraphics[width=0.49\textwidth]{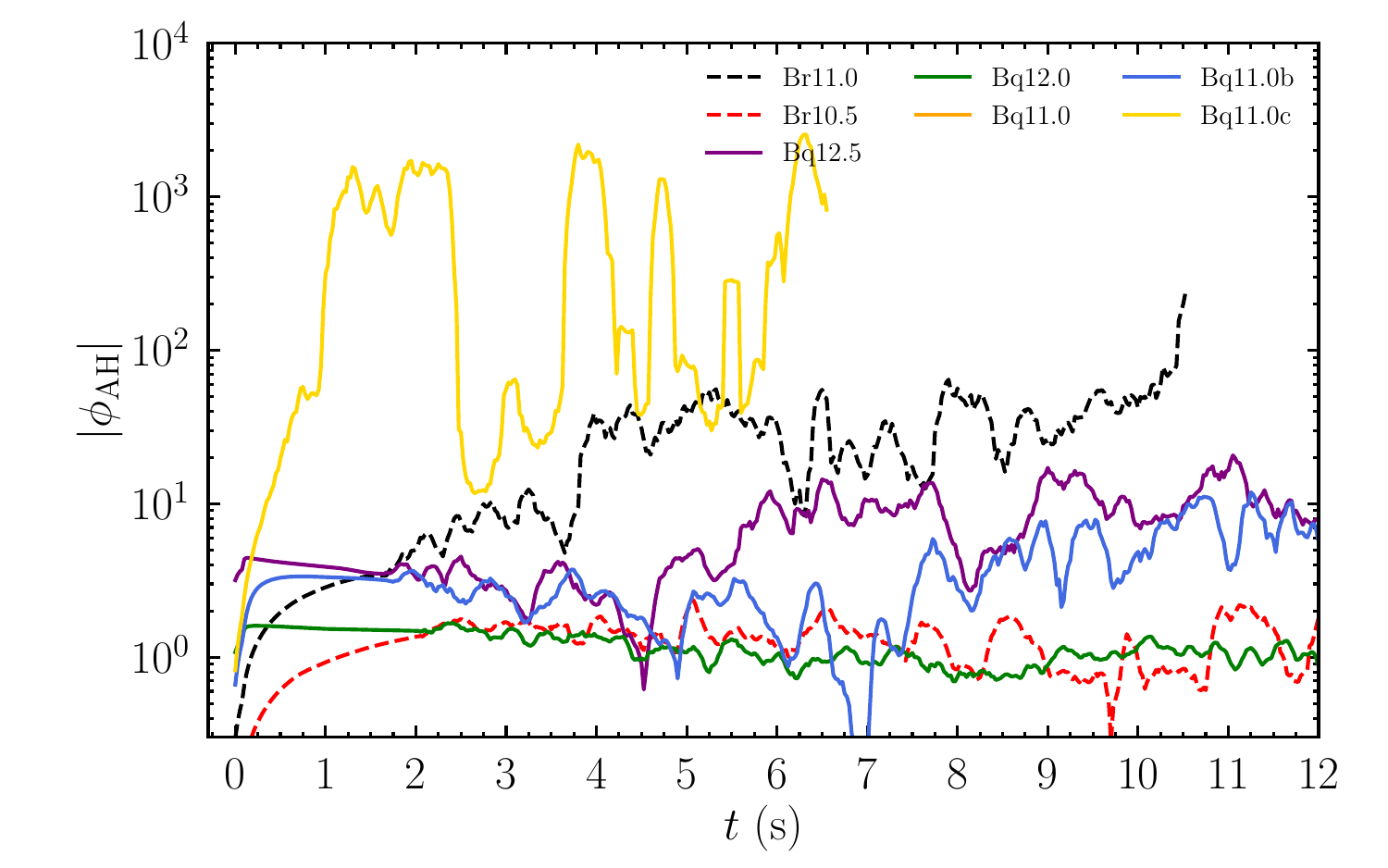}
\caption{Evolution of the magnetic flux on apparent horizons (top) and resultant MAD parameter (bottom) as functions of time. The left and right panels show the results for the models with Eq.~(\ref{eqBz}) (left) and with Eqs.~(\ref{eqB}) and~(\ref{eqB2}) (right), respectively. To see the trend clearly, moving averages are taken with the time interval 0.2~s. Note that for model Bq11.0 $|\phi_\mathrm{AH}|$ is smaller than 0.3. 
}
\label{fig2}
\end{figure*}

Figure~\ref{fig1} shows the evolution of the electromagnetic energy $E_\mathrm{B}$. Here the electromagnetic energy is defined in the same way as in Ref.~\cite{Shibata2021b}. For all the models with the initial magnetic-field configuration by Eq.~(\ref{eqBz}) or (\ref{eqB}), $E_\mathrm{B}$ monotonically increases with the compression due to the matter infall and winding in an early stage. For initially strong and well-aligned  magnetic-field models (B11.5, B11.3, B11.0, Br11.0, and Bq11.0c), a jet is launched along the $z$-axis before the formation of a disk/torus by this magnetic-field amplification. The jet subsequently makes a cocoon around it and a convective motion is developed. Associated with this motion, the magnetic fields are wound and compressed, and hence, the magnetic energy is quickly enhanced until a saturation is reached. Here, at the saturation the electromagnetic energy becomes comparable to the rotational kinetic energy of the matter, $10^{50}$--$10^{51}$\,erg. The saturated values of $E_\mathrm{B}$ are slightly larger for higher initial field strengths reflecting the ram pressure at the jet launch.

For initially weaker magnetic-field models (B10.5, Br10.5, and B10.0) with Eq.~(\ref{eqBz}) or (\ref{eqB}) and models with Eq.~(\ref{eqB2}) and $\varpi_0 \leq 5 \times 10^3$\,km, a significant amplification of the electromagnetic energy found at $t \agt 2$\,s takes place due to the formation of a disk and a torus, in which the magnetic fields are amplified by the compression and winding. When the torus (geometrically thick disk) is formed, the matter velocity vector converges to the spin-axis direction, and hence, the magnetic flux is also converged, leading to the enhancement of the magnetic-field strength. Also, the magnetic field in the torus has a substantial fraction of the $\varpi$ component, which plays an important role in the magnetic-field amplification by the winding. In addition, the MRI may partly play a role for the magnetic-field amplification after the $z$-component of the magnetic field becomes high enough to resolve the fastest growing mode of the MRI in the limited grid resolution. As already mentioned, a torus is developed from a geometrically thin disk and the oblique shock on the shock surface around the torus enhances the matter and magnetic-flux accretions onto the black hole. After the magnetic flux that penetrates the black hole becomes high enough, a slowly-expanding jet can be eventually launched from the vicinity of the black hole, although this is found only for models B10.5, Bq12.5, and Bq11.0b. After the jet launch, a cocoon and associated convective motion are developed until a saturation at which the electromagnetic energy relaxes to $\sim 10^{50}$\,erg, which is comparable to the rotational kinetic energy of the matter as well. 

For the initial magnetic field of Eq.~(\ref{eqB2}), a steep increase of the electromagnetic energy is not found in an early stage. For this case, the magnetic-field strength rather decreases with time in the vicinity of the black hole because the magnetic flux decreases with the matter infall due to the magnetic-field configuration initially given (see also Fig.~\ref{fig2}). Only for model Bq11.0c, for which $\varpi_0$ is large ($10^4$\,km), a steep increase of the electromagnetic energy takes place, leading to an early jet launch. This result illustrates that the timing and mechanism of the jet launch depend strongly on the initial magnetic-field configuration. 

For the initially weak magnetic-field models with the initial configurations of Eq.~(\ref{eqBz}) or (\ref{eqB}) and for most of the models with Eq.~(\ref{eqB2}) (except for model Bq11.0c), the electromagnetic energy in the torus contributes substantially to the total one. This is developed mainly by the magnetic winding. Because of the anti-dynamo nature in the axisymmetric simulation, the poloidal magnetic field in the torus does not increase with time significantly.


\begin{figure*}[t]
\includegraphics[width=0.49\textwidth]{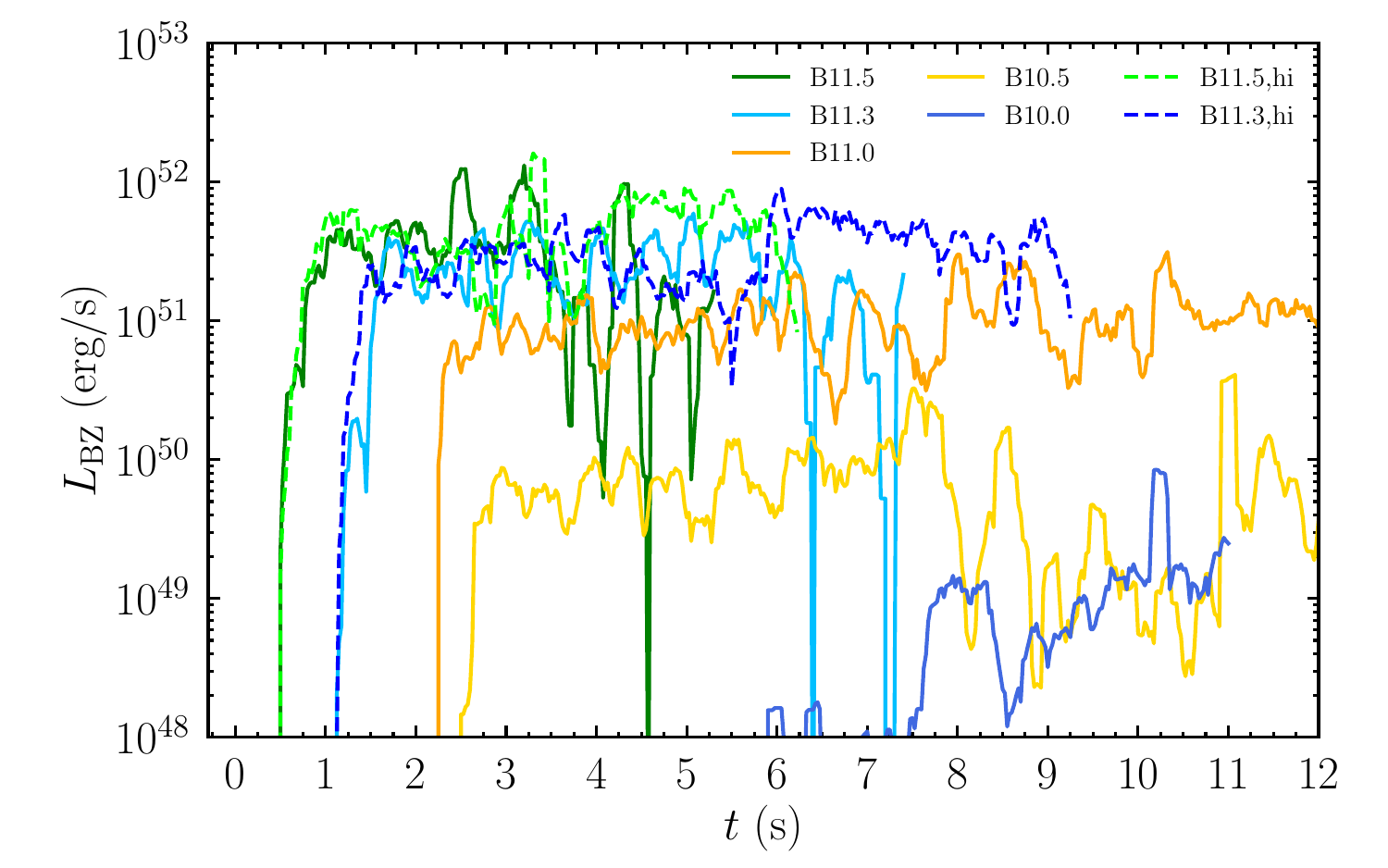}
\includegraphics[width=0.49\textwidth]{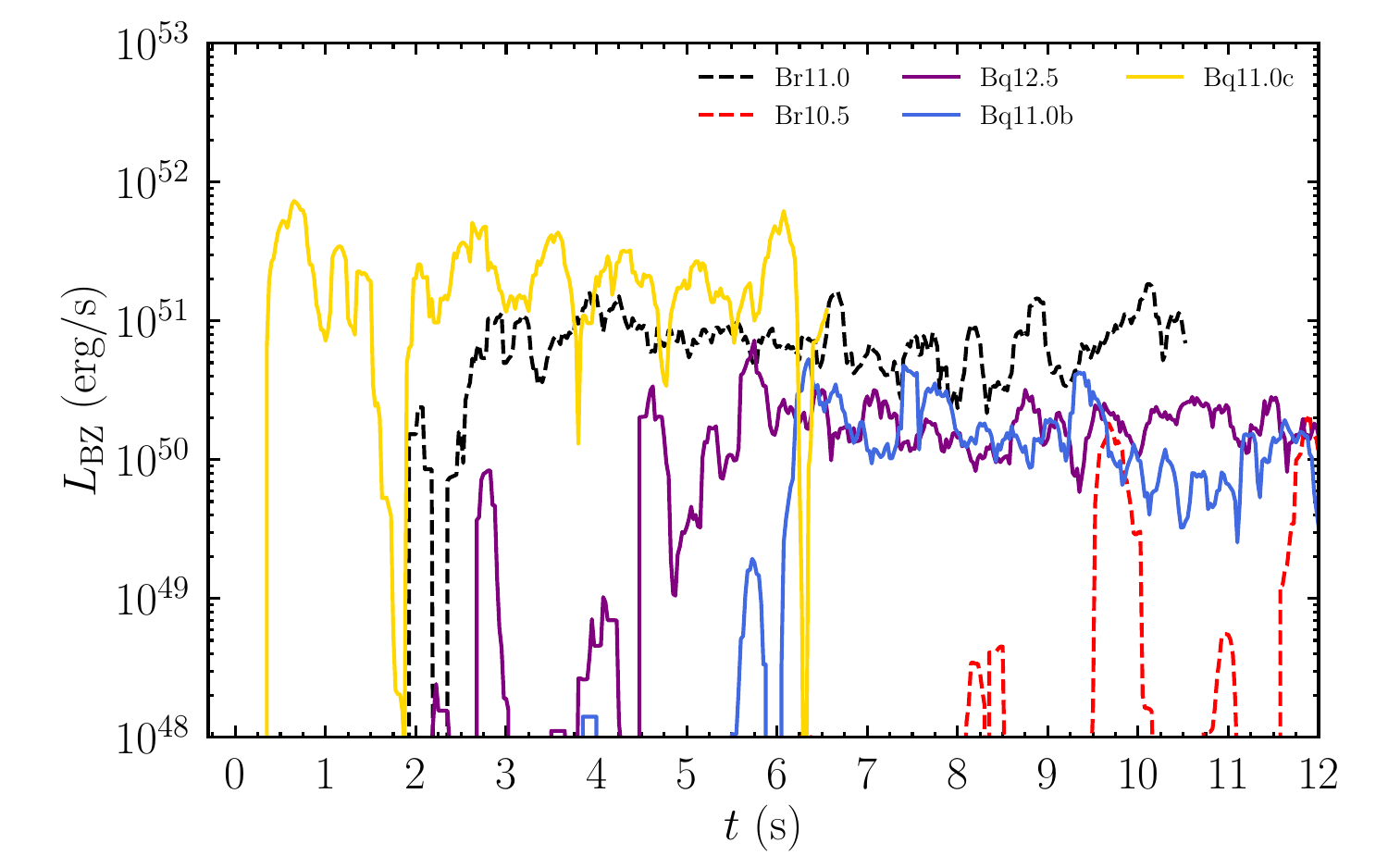}
\caption{Poynting luminosity measured on apparent horizons as a function of time for selected models. The left and right panels show the results of the models with the initial magnetic field given by Eq.~(\ref{eqBz}) (left) and by Eqs.~(\ref{eqB}) and~(\ref{eqB2}) (right), respectively. We note that for models Bq12.0 and Bq11.0, $L_\mathrm{BZ}$ is smaller than $10^{48}$\,erg/s. 
}
\label{fig3}
\end{figure*}

Figure~\ref{fig2} displays the evolution of the magnetic flux, $\Phi_\mathrm{AH}$, on apparent horizons and resultant MAD parameter, $\phi_\mathrm{AH}$, as functions of time. It is found that for the models with the initial magnetic-field profiles given by Eqs.~(\ref{eqBz}) and (\ref{eqB}) the magnetic flux on the horizon steeply increases soon after the onset of the simulations and eventually approaches a relaxed value. Only if the magnetic flux exceeds $\approx 1\times 10^{29}\,\mathrm{G\,cm^2}$ a jet is launched before the disk formation in the present stellar model.
For most of the models with the initial field configuration of Eq.~(\ref{eqB2}) (except for model Bq11.0c), the magnetic flux relaxes to low values. In particular, for model Bq12.5, the magnetic flux appreciably decreases for $t \alt 3$\,s. This is the reason why we do not find the quick jet launch for this model in spite of the large initial field strength on the horizon. 

The condition of the jet launch can be also discussed in terms of the MAD parameter~\cite{Tchekhovskoy:2011zx}. In the present study, jets are launched for the models only with $|\phi_\mathrm{AH}|\agt 5$. The MAD parameter is high for the models with the initial field configuration of Eq.~(\ref{eqBz}), and for models B11.5, B11.3, and Bq11.0c, it can be extremely high $\agt 10^3$, reflecting that the mass accretion onto the black hole is significantly halted. For models B11.0 and Br11.0 in which the initial magnetic-field strength is weaker, the MAD parameter is typically 10--100 and jets are steadily generated from a relatively early stage. For model B10.5, by contrast, the jet (or outflow along the rotation axis) is launched in the early stage but it is stalled in the middle of the outward propagation. This may be interpreted as an insufficient MAD parameter $\sim 1$ in such a stage. In the later stage of this model, the MAD parameter increases to $\sim 10$, in which the outward propagation of the jet/outflow is observed. The situation is similar to those of models Bq12.5 and Bq11.0b, in which a jet is launched after the MAD parameter increases beyond $\sim 5$. 

As we find from Fig.~\ref{fig2}, the magnetic flux on the horizon and MAD parameter are good indicators to determine whether a jet can be launched or not. This clearly shows that the magnetic flux on the horizon is one of the crucial quantities. To obtain the high value of the magnetic flux on the horizon in the collapsar model, we may need a suitable initial magnetic-field profile. However, this does not give us the comprehensive scenario for generating jets from black holes because the magnetic-field strength and configuration should have a wide variety in the progenitor stars. The other possibility to universally generate the strong poloidal magnetic field that penetrates black holes is the mechanism throughout the enhancement of the magnetic-field strength in the torus due to the MRI turbulence and accretion of the strong magnetic flux onto the horizon. In the axisymmetric study, we cannot explore this possibility due to the anti-dynamo nature, and thus, obviously we need a simulation in which the dynamo effect is taken into account (a  high-resolution three-dimensional simulation or a phenomenological simulation; e.g., Ref.~\cite{Shibata2021b}) to confirm this possibility.

\begin{table*}[t]
    \centering
\caption{Average increase rates of the ejecta mass, $\dot M_\mathrm{ej}$, and explosion energy, $\dot E_\mathrm{exp}$, the Poynting luminosity, $L_\mathrm{BZ}$, and the ratio, $L_\mathrm{BZ}/\dot E_\mathrm{exp}$, for the models in which a jet is launched. The quantities are averaged over the last 5 seconds of each simulation. The last two columns list the approximate spin-down timescale and expected total electromagnetic energy carried by the Blandford-Znajek mechanism for models in which the spin-down is found. $\dagger$ specifies the models for which a jet is launched after the formation of a disk/torus and the spin-down is not found. 
    \label{tab:exp}
    }
    \begin{tabular}{lccccccc}
    \hline\hline
        Model & ~$\langle \dot{M}_\mathrm{ej}\rangle$ ($M_\odot$/s)~ & $~\langle \dot{E}_\mathrm{exp}\rangle$ ($10^{51}$~erg/s)~ & ~$\langle L_\mathrm{BZ}\rangle$ ($10^{51}$~erg/s)~ & ~$\langle L_\mathrm{BZ}\rangle/\langle \dot{E}_\mathrm{exp}\rangle$~ & ~~~$\tau$\,(s)~~~ & $\langle L_\mathrm{BZ}\rangle\,\tau\,(10^{53}$\,erg)\\
        \hline
        B11.5,hi & 0.90   & 5.2 & 5.0 & 1.0 & 30 & 1.5\\
        B11.3,hi & 0.57   & 4.2 & 3.5 & 0.8 & 60 & 2.1 \\
        B11.5 & 0.86      & 4.2 & 3.2 & 0.8 & 50 & 1.6\\
        B11.3 & 0.88      & 4.0 & 2.4 & 0.6 & 70 & 1.8 \\
        B11.0 & 0.60      & 1.3 & 1.1 & 0.9 & 175 & 2.0\\
        B10.5$\dagger$ & 0.34     & 1.1 & 0.07& 0.1 & -- & --\\
        Br11.0 & 0.68    & 1.4 & 0.7 & 0.5 & 250 & 1.7\\
        Bq12.5$\dagger$ & 0.49    & 1.5 & 0.20 & 0.1 & --& --\\
        Bq11.0b$\dagger$& 0.62   & 0.41 & 0.10 & 0.2 &--& --\\
        Bq11.0c& 0.76     & 2.4 & 1.9 & 0.8 & 75 & 1.4\\
        \hline
    \end{tabular}\\
\end{table*}

\begin{figure*}[t]
\includegraphics[width=0.49\textwidth]{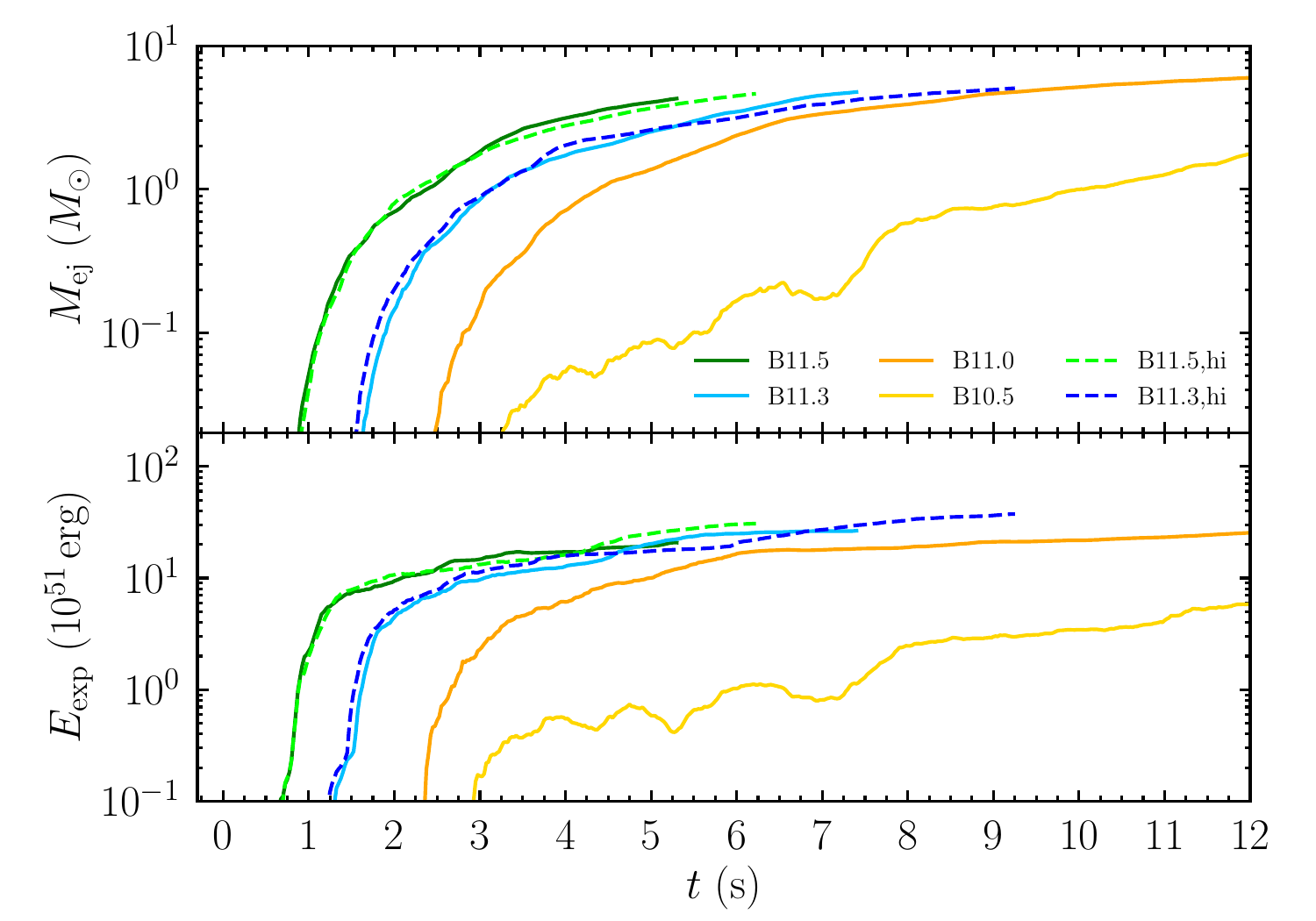}
\includegraphics[width=0.49\textwidth]{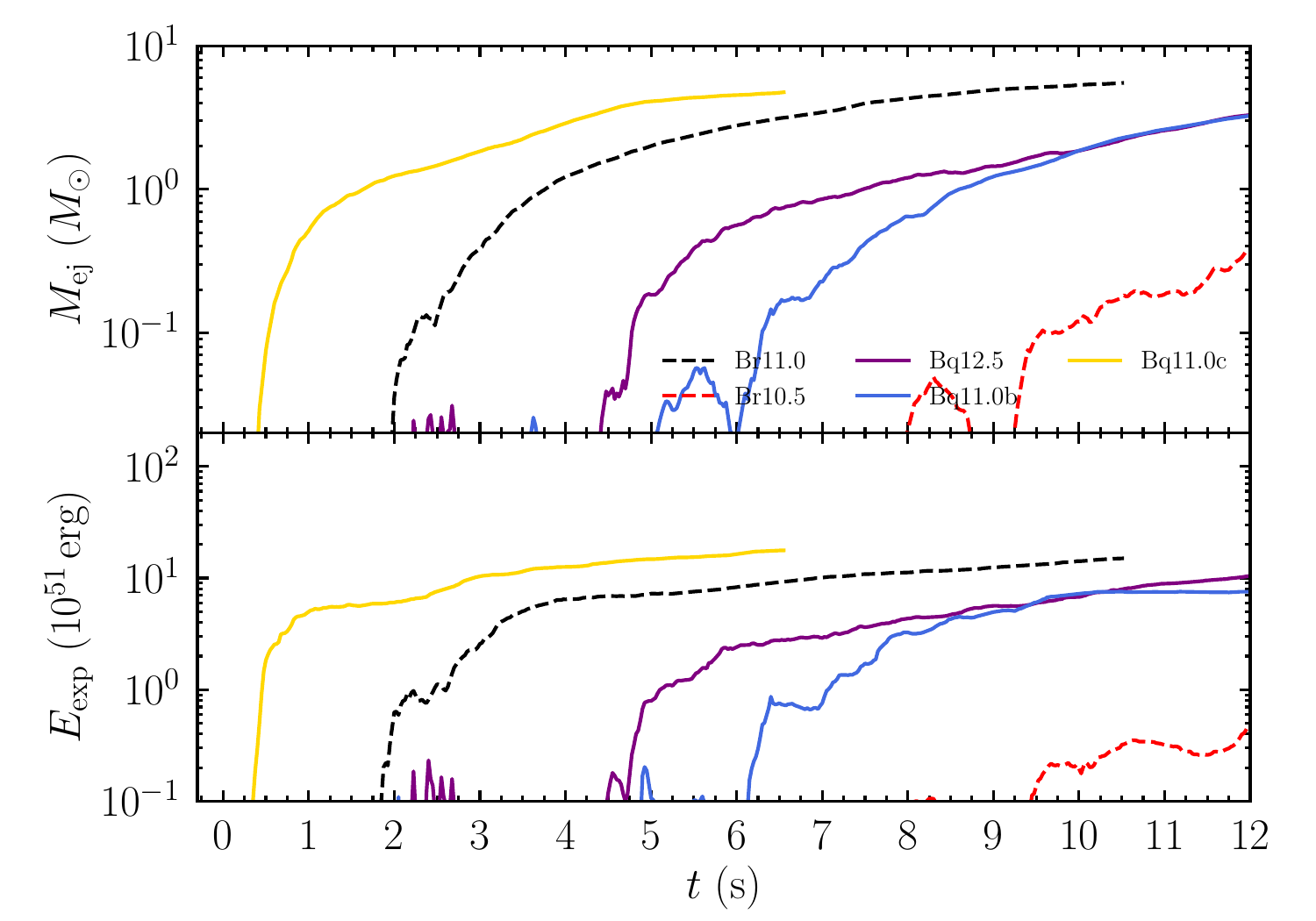}\\
\includegraphics[width=0.49\textwidth]{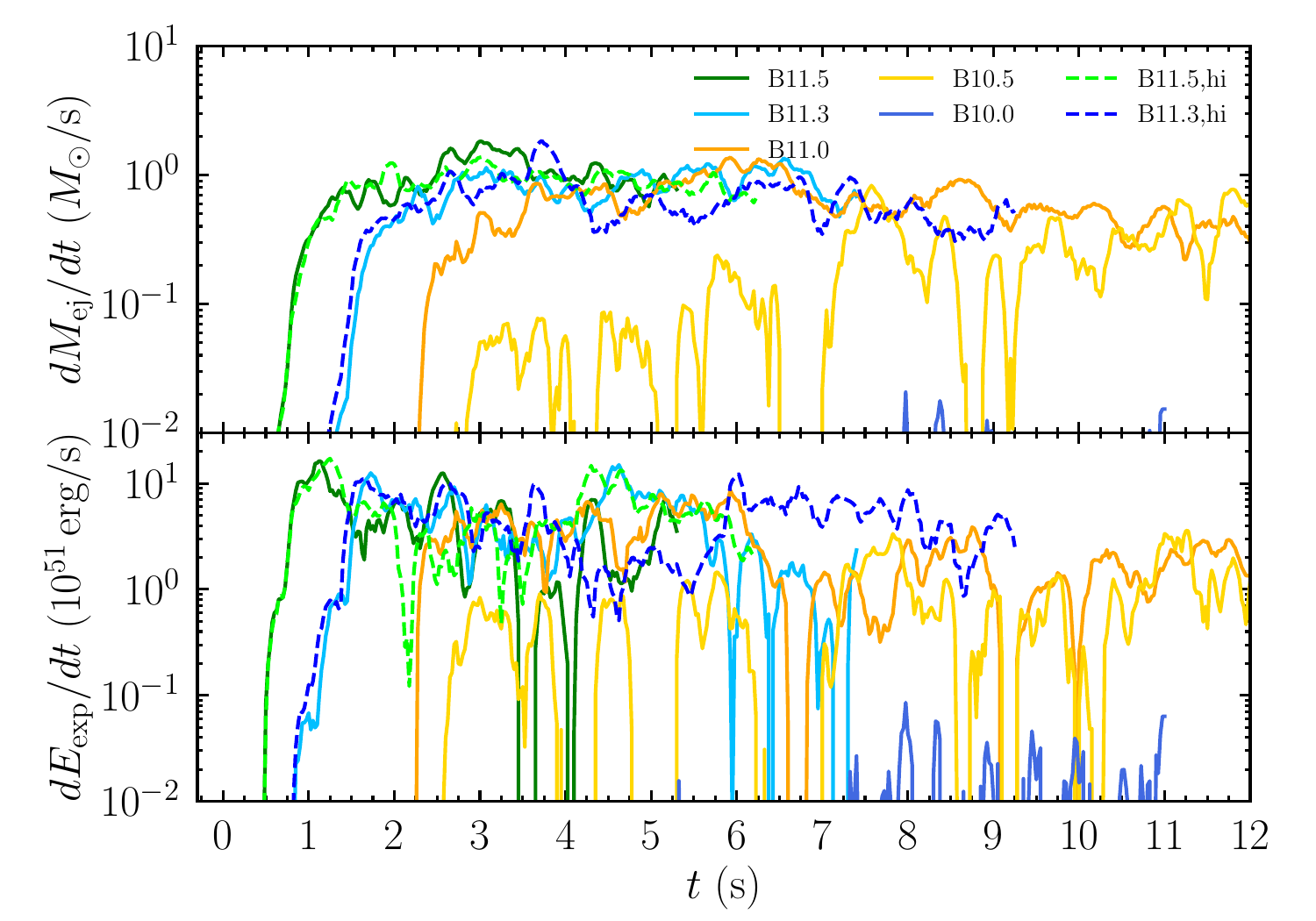}
\includegraphics[width=0.49\textwidth]{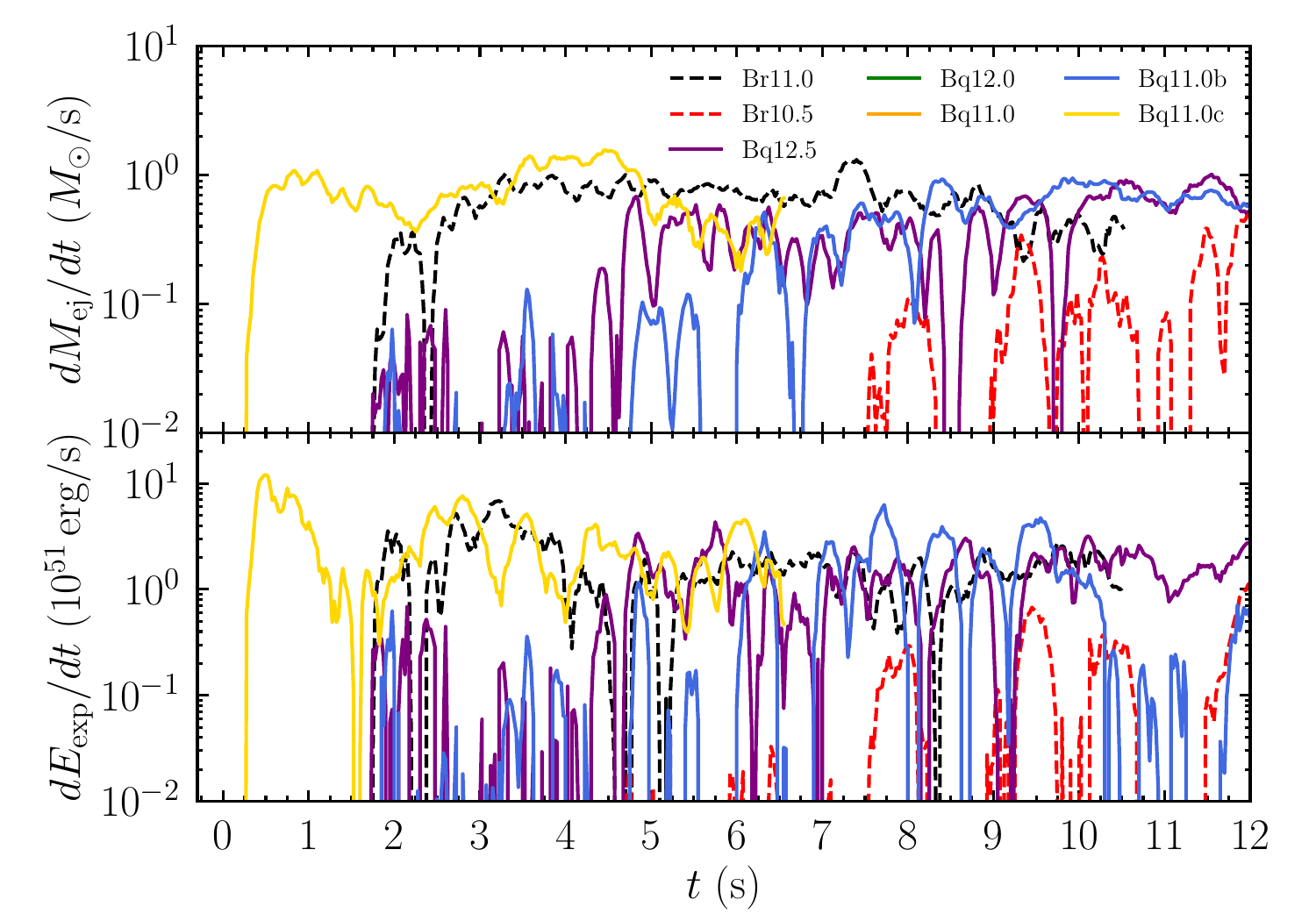}
\caption{$M_\mathrm{ej}$, $E_\mathrm{exp}$ (top panels), $\dot M_\mathrm{ej}$, and $\dot E_\mathrm{exp}$ (bottom panels) for the models with Eq.~(\ref{eqBz}) (left) and with Eqs.~(\ref{eqB}) and~(\ref{eqB2}) (right), respectively. We note that for models Bq12.0 and Bq11.0, the jet launch and stellar explosion are not found, and hence, $dM_\mathrm{ej}/dt$ and $dE_\mathrm{exp}/dt$ are smaller than $10^{-2}M_\odot$/s and $10^{48}$\,erg/s, respectively. 
}
\label{fig4}
\end{figure*}

For higher resolution runs of models B11.5 and B11.3, the magnetic-field energy and field strength on the horizon are larger than those for the corresponding lower resolution runs in later stages. Our interpretation for this is that for the lower grid resolution, the numerical dissipation and diffusion of the magnetic field are stronger. This results in lower Poynting luminosity and slower spin-down of the black hole with the lower resolution runs (see below). 

\subsection{Poynting luminosity, ejecta mass, and outflow energy}\label{secIIIC}

Figure~\ref{fig3} shows the Poynting luminosity, $L_\mathrm{BZ}$, extracted from the spinning black hole by the Blandford-Znajek mechanism. The surface integral of the Poynting flux is performed on apparent horizons. Only the portion that the outgoing energy flux including the matter energy flux is positive (i.e., net energy is extracted from the black hole) contributes to the surface integral (see Appendix~\ref{appB} for the definition of $L_\mathrm{BZ}$). $L_\mathrm{BZ}$ is naturally higher for higher values of $|\Phi_\mathrm{AH}|$ (see Fig.~\ref{fig2}) for the models in which a jet is launched. 


Figure~\ref{fig3} illustrates that, broadly speaking, the Poynting flux is steadily generated for the early-jet-launch models with the luminosity of $\agt 10^{51}$\,erg/s. This reflects that the poloidal magnetic field that penetrates the black hole is in a quasi-steady state during the jet generation. It is also found that the Poynting luminosity is higher for the higher initial field strength, which results in the higher strength of the magnetic field that penetrates the black hole during the jet generation: For models B11.5, B11.3, and Bq11.0c, for which a jet is launched in early stages of the evolution, the Poynting luminosity can be much higher than $10^{51}\,\mathrm{erg/s}$ after the jet launch, and for models B11.0 and Br11.0, $L_\mathrm{BZ} \sim 10^{51}\,\mathrm{erg/s}$. 

However, the total energy carried away by the Poynting flux seems to depend only weakly on the initial field strength, because the spin-down timescale of the black hole is shorter for the higher initial field strength (see Sec.~\ref{SecIIId}). Irrespective of the initial condition, the predicted total energy emitted by the Poynting flux, i.e., $\Delta E=$(the average Poynting luminosity)$\times$(spin-down timescale of Sec.~\ref{SecIIId}), is  1.4--$2.1\times 10^{53}$\,erg, if the Poynting luminosity is assumed to be approximately constant in the spin-down timescale (see Table~\ref{tab:exp}). Taking into account the uncertainty for the value of $f$ and for the validity of the force-free approximation, the order of the magnitude for this is consistent with Eq.~(\ref{eq7}), indicating that the estimate carried out in Sec.~\ref{sec:intro} is good. 

The estimated values of $\Delta E$ are about 20--30\% of the result from Eq.~(\ref{eq7}). The primary reason for this is that the poloidal magnetic field coherently penetrates only a portion of the black-hole horizon (see, e.g, Fig.~\ref{figC}) so that the Poynting luminosity should be smaller than that of Eq.~(\ref{eq1}). Near the equatorial plane the magnetic-field lines are not very coherently aligned and moreover the force-free condition is not well satisfied because of the presence of infalling matter. Even with such magnetic-field lines, the angular momentum of the black hole can be extracted because the angular velocity of the black hole is larger than that of the matter around the black hole, while the energy extraction may be less efficient in the presence of dense matter (see, e.g., Fig.~9 of Ref.~\cite{McKinney:2004ka} and Ref.~\cite{Huang:2019wqv} for a discussion on the matter effect). 
It should be also pointed out that $L_\mathrm{BZ}$ is appreciably smaller than $L_\mathrm{BZ}^\mathrm{full}$ (see Appendix~\ref{appB}): On a large portion of the surface of apparent horizon, the total energy flux (matter plus electromagnetic energy flux) outgoing from the horizon is negative. Thus, the effect of the matter infall plays a significantly negative role for the extraction of the rotational kinetic energy of the black hole in the collapsar scenario. 

However, the total amount of the outgoing energy is still larger than $10^{53}$\,erg, which is much larger than the typical energy of gamma-ray bursts (including the afterglow and associated supernova). This suggests that the energy injection from the black hole has to be stopped before the entire spin-down of the black hole (see a discussion in Sec.~\ref{sec4}).

Figure~\ref{fig4} shows the evolution of the ejecta mass, $M_\mathrm{ej}$, and outflow energy (including the explosion energy of the star), $E_\mathrm{exp}$, as well as the increase rates of the ejecta mass and outflow energy for selected models. The ejecta mass and outflow energy are calculated using the similar formulae as in Ref.~\cite{Fujibayashi2022dec} with the extraction radius of $1\times 10^5$\,km (see Appendix~\ref{appB} for the formulae). In the present context, the contribution from the jet is appreciable in the ejecta mass and outflow energy. Since these quantities increase monotonically with time until the end of the simulations in the present study (i.e., they do not relax to constants), we also plot the time derivative of them, i.e., $\dot M_\mathrm{ej}$ and $\dot E_\mathrm{exp}$, in Fig.~\ref{fig4}. Table~\ref{tab:exp} also shows average values of $\dot M_\mathrm{ej}$ and $\dot E_\mathrm{exp}$ as well as of $L_\mathrm{BZ}$.  

It is found that $\dot E_\mathrm{exp}$ is of the same order of magnitude as $L_\mathrm{BZ}$ for models B11.5, B11.3, B11.0, Br11.0, and Bq11.0c for which a jet is launched before the disk formation. For these models, thus, the Blandford-Znajek mechanism is the major central engine for the jet launch. For models B11.3, B11.0 and Br11.0, we confirmed that the entire star explodes, indicating that the Blandford-Znajek mechanism can also be the engine of the stellar explosion (but see discussions in Sec.~\ref{sec4}). It is also found that the mass ejection rate is quite high for these models as $\sim M_\odot$/s. Since the total mass outside the black hole in the present models is $M_\mathrm{env} \sim 10M_\odot$, and hence, $M_\mathrm{env}/\dot M_\mathrm{ej} \sim 10$\,s is much shorter than the spin-down timescale by the Blandford-Znajek mechanism (see Sec.~\ref{SecIIId} for the spin-down timescale). Thus, in the late stage of the evolution of the system, the Poynting flux will be used to accelerate the ejected matter if the Blandford-Znajek mechanism works until the complete spin-down of the black hole. 

For models B10.5, Bq12.5, and Bq11.0b for which a jet is launched after the formation of the disk and torus, $L_\mathrm{BZ} \sim 10^{50}$\,erg/s, which is by one or two orders of magnitude lower than those for the early-jet-launch models such as B11.5 and B11.3. For these models, the magnetic-field strength is weaker around the polar region on the horizon (see Fig.~\ref{fig2}), and thus, this result is quite reasonable. However, $\dot E_\mathrm{exp}$ for these models is not very small, and thus, the ratio of $L_\mathrm{BZ}/\dot E_\mathrm{exp}$ is much lower than those for the early-jet-launch models. The reason for this is that for these models the magnetohydrodynamical effects such as magneto-centrifugal effect and Tayler instability play an important role not only in the jet launch but also in the stellar explosion for a substantial fraction of the stellar envelope. Specifically, the angular momentum transport from the black hole to matter around the equatorial region through the winding of the magnetic-field lines associated with the black-hole spin (like a propeller effect by a rotating neutron star~\cite{2011ApJ...736..108P}) appears to play an important role for extracting the angular momentum (and rotational kinetic energy) of the black hole (see, e.g., Ref.~\cite{1990ApJ...363..206T} for the related issue). 

As already mentioned, the dynamo effect is not taken into account in the present axisymmetric modelling. In reality, the dynamo and resulting turbulence that effectively generate the viscous effect will contribute to the activity of the torus, likely leading to more efficient mass ejection and energetic  explosion~\cite{Fujibayashi2023BH}. In addition, an enhanced magnetic-field strength on the horizon would increase the Poynting luminosity. Thus, for models B10.5, Bq12.5, and Bq11.0b, the values of $L_\mathrm{BZ}$ and $E_\mathrm{exp}$ may be even higher in reality. Since $E_\mathrm{exp}$ was already much higher than the typical supernova energy ($\sim 10^{51}$\,erg) at $t=10$\,s, these can be models for powerful supernovae with the explosion energy higher than $\agt 10^{52}$\,erg (but the overproduced energy by the extraction of huge rotational kinetic energy of spinning black holes can be an issue as well; see a discussion in Sec.~\ref{sec4}).  

\begin{figure*}[t]
\includegraphics[width=0.48\textwidth]{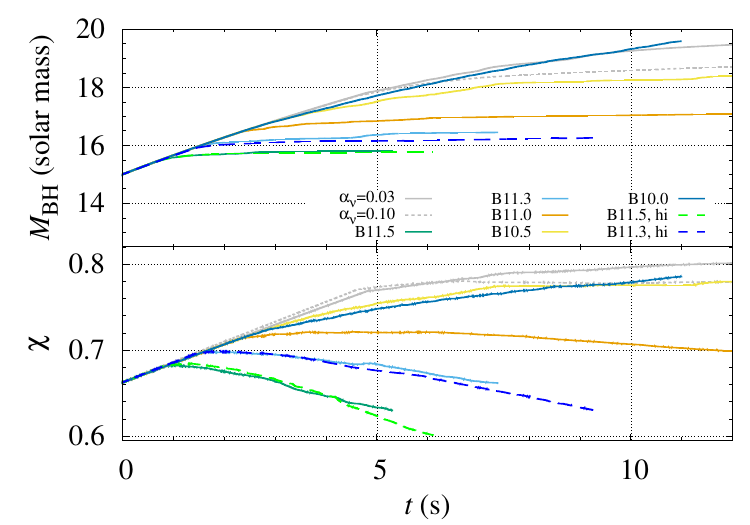}
\includegraphics[width=0.48\textwidth]{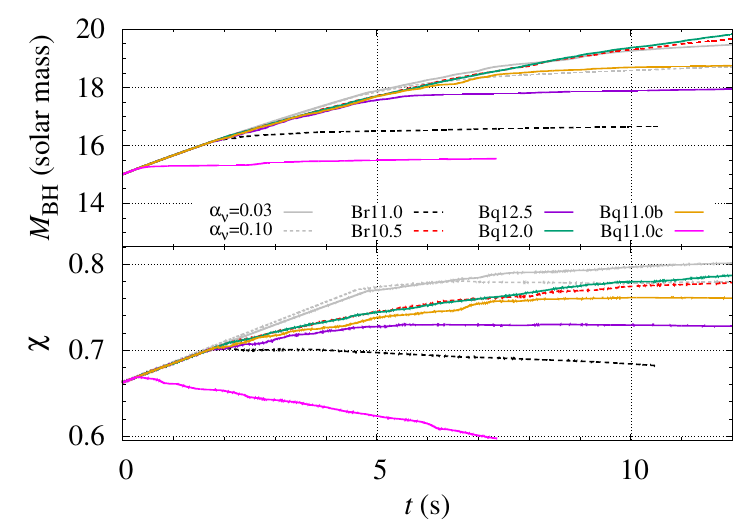}\\
\includegraphics[width=0.65\textwidth]{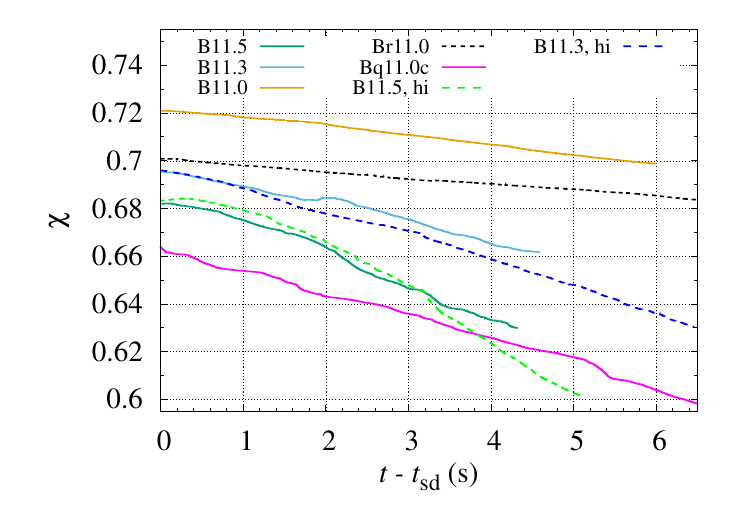}
\caption{Top left: Evolution of the mass (upper panel) and dimensionless spin (lower panel) of spinning black holes for all the models with the initial conditions of Eq.~(\ref{eqBz}) as well as a viscous-hydrodynamics model with the alpha parameter of 0.03 (solid curve) and 0.10 (dotted curve) of Ref.~\cite{Fujibayashi2023BH}. 
Top right: The same as top left panel but for the selected models with the initial configuration of Eqs.~(\ref{eqB}) and~(\ref{eqB2}). The curves for models Br10.5 and Bq12.0 are accidentally very similar. 
Bottom: Zoom-up of the spin evolution for the selected models as a function of $t-t_\mathrm{sd}$ where $t_\mathrm{sd}$ denotes the approximate time at which the spin starts decreasing. 
}
\label{fig5}
\end{figure*}

\subsection{Spin evolution of black holes}\label{SecIIId}

Figure~\ref{fig5} shows the evolution of the mass and dimensionless spin of black holes  for the models with the initial magnetic-field configuration of Eq.~(\ref{eqBz}) (top left) and with those of Eqs.~(\ref{eqB}) and~(\ref{eqB2}) (top right). For comparison, we plot results in viscous-hydrodynamics simulations with the alpha parameters of $0.03$  and $0.10$ of Ref.~\cite{Fujibayashi2023BH}. 
For the better view of the spin-down, we also plot the evolution of the dimensionless spin focusing only on the models for which the spin eventually decreases with time (bottom panel). 

For all the models, both the mass and dimensionless spin initially increase with time due to the matter accretion onto the black hole. For models B11.5 and B11.3 for which the magnetic-field lines are well aligned with the black-hole spin axis and its strength is very high initially, a MAD state is quickly established as a result of the amplification of the magnetic field that penetrates the black hole and launches a jet (cf. Fig.~\ref{fig2}). 
The evolution process of the black hole for model Bq11.0c is similar to these models. For this model, the cylindrical component of the magnetic field is present from the beginning,  and thus, the strong magneto-centrifugal force also plays a role in halting the matter accretion from the equatorial direction. 
After the jet launch for these models, the mass accretion onto the black hole essentially ceases except for an intermittent accretion from the equatorial direction, leading to a state of $dM_\mathrm{BH}/dt < 0.1M_\odot/$s. For this stage, the black hole is evolved primarily by the Blandford-Znajek mechanism, and the dimensionless spin decreases with time (see the bottom panel of Fig.~\ref{fig5} for the zoom-up view). We note that for model B11.3 (standard resolution run), an intermittent spin-up stage is seen at $t \sim 4.7$--5.0\,s at which the MAD state was disrupted for a while. This is due to an accidental large-mass accretion from the equatorial region. However, for $t \agt 5$\,s, the MAD state is recovered and the dimensionless spin steadily decreases again. 

The timescale of the spin-down is evaluated using Eq.~(\ref{eq10}). For the present numerical results, $\Delta J=J_0-J_\mathrm{BH}~(\Delta J>0)$ is a small fraction of $J_0$. Thus, $|\Delta  J_\mathrm{BH}^2| \approx 2J_0\Delta J$, and hence,  $\tau$ is determined approximately by  
\beq
\tau=\Delta t {J_0 \over \Delta J}\approx \Delta t {\chi_0 \over \Delta \chi},
\eeq
where $\Delta t$ denotes a time duration in the spin-down stage and $\Delta \chi=\chi_0-\chi$. 

For models B11.5, B11.3, and Bq11.0c for which the spin-down sets in soon after the jet launch, we find $\tau < 100$\,s from Fig.~\ref{fig5} (see Table~\ref{tab:exp} for the results). These results illustrate that a short-term spin-down, i.e., the case that the spin-down timescale is comparable or shorter than the typical time duration of long gamma-ray bursts $\alt 100$\,s, is possible if a MAD state is developed in an early stage of gravitational collapse. For lower resolution runs of B11.5 and B11.3, the spin-down timescale is longer than for the corresponding higher resolution runs. This is due to a numerical dissipation and diffusion of the magnetic fields.  

For models B11.0 and Br11.0, for which the initial field strength is high enough to launch a jet before the disk formation, the mass accretion onto the black hole is also suppressed due to the strong magnetic pressure near the horizon after the jet launch. However, the mass accretion still proceeds for a while, and associated with it, the dimensionless spin increases with time overcoming the spin-down by the Blandford-Znajek mechanism in an early stage. Only for late stages the spin-down by the Blandford-Znajek mechanism overcomes the spin-up by the mass accretion. For models B11.0 and Br11.0, the MAD state appears to be developed only for $t \agt 6$\,s and 4\,s, respectively, after which $\dot M_\mathrm{BH}$ is less than $0.1M_\odot$/s, although a strong jet is launched earlier, and the spin-down rate is low: $\Delta \chi \sim 0.01$ for $\Delta t \sim 3$--$4$\,s, and the estimated spin-down timescale is longer than $100$\,s, much longer than those for models B11.5, B11.3, and Bq11.0c. This is due to a weaker magnetic-field strength achieved around the black hole after the jet launch. 

For the initially weak magnetic-field models (B10.5, B10.0, and Br10.5) as well as for most of the models with the initial magnetic-field configuration of Eq.~(\ref{eqB2}), for which a MAD state is not achieved or only weakly achieved in the simulation time, the black-hole mass increases with the mass accretion from the equatorial direction and the dimensionless spin does not decrease even after the jet launch (for models B10.5, Bq12.5, and Bq11.0b). Even for these models, after the mass accretion onto the black hole ceases, the black-hole spin may eventually decrease with time due to the Blandford-Znajek mechanism. However, for these models the magnetic-field strength on the horizon is much weaker than those for models B11.5, B11.3, B11.0, Br11.0, and Bq11.0c for which the jet is launched after the sufficient enhancement of the magnetic-field strength before the disk formation. In reality the magnetic-field strength on the horizon can be increased if the MRI turbulence in the disk after its formation is fully resolved. As we estimated at Eq.~(\ref{guessB}), however, the magnetic-field strength would not be as high as those for models B11.5, B11.3, B11.0, Br11.0, and Bq11.0c for such cases. Hence, the spin-down time scale would be $\gg 100$\,s. 

As we discussed in Sec.~\ref{sec:intro}, suppose that a fossil magnetic field in the progenitor star is not extremely strong, it is natural to consider that the strong poloidal magnetic field that penetrates the black hole and is responsible to the Blandford-Znajek mechanism should be developed after the disk formation, in which the magnetic-field amplification takes place and from which the magnetic flux is provided to the black hole. Our present numerical results indicate that the rapid spin-down is achieved only for the case that the magnetic-field strength is high even in the absence of the disk. This suggests that the spin-down effect of the black hole might be minor in the typical collapsar scenario during the typical duration of long gamma-ray bursts of 10--100\,s. 

For models Br10.5, B10.0, Bq12.5, Bq12.0, and Bq11.0b, the evolution process of the black hole is similar to that for viscous hydrodynamics models with different viscous efficiency. For these models, the dimensionless spin achieved after the evolution is slightly smaller than that for the viscous hydrodynamics models, indicating that the outward angular momentum transport in the disk/torus becomes more efficient by the magnetohydrodynamical effect, e.g., by the magneto-centrifugal effect associated with the black-hole spin, than by the viscous effect. This is in particular the case for model Bq12.5. Nevertheless, the evolution path of the black hole in the magnetohydrodynamics models is similar to those in the viscous hydrodynamics models. This suggests that in the absence of jets by the Blandford-Znajek mechanism, the evolution of the system is similar irrespective of the physical mechanisms of the angular momentum transport. On the other hand, in the presence of a strong jet, the growth of the black hole by the mass accretion could be suppressed, and hence, amount of the matter outside the black hole, which could be the ejecta and energy source of electromagnetic signals, may be larger. 

\section{Summary and discussion}\label{sec4}

\begin{figure*}[t]
\includegraphics[width=0.355\textwidth]{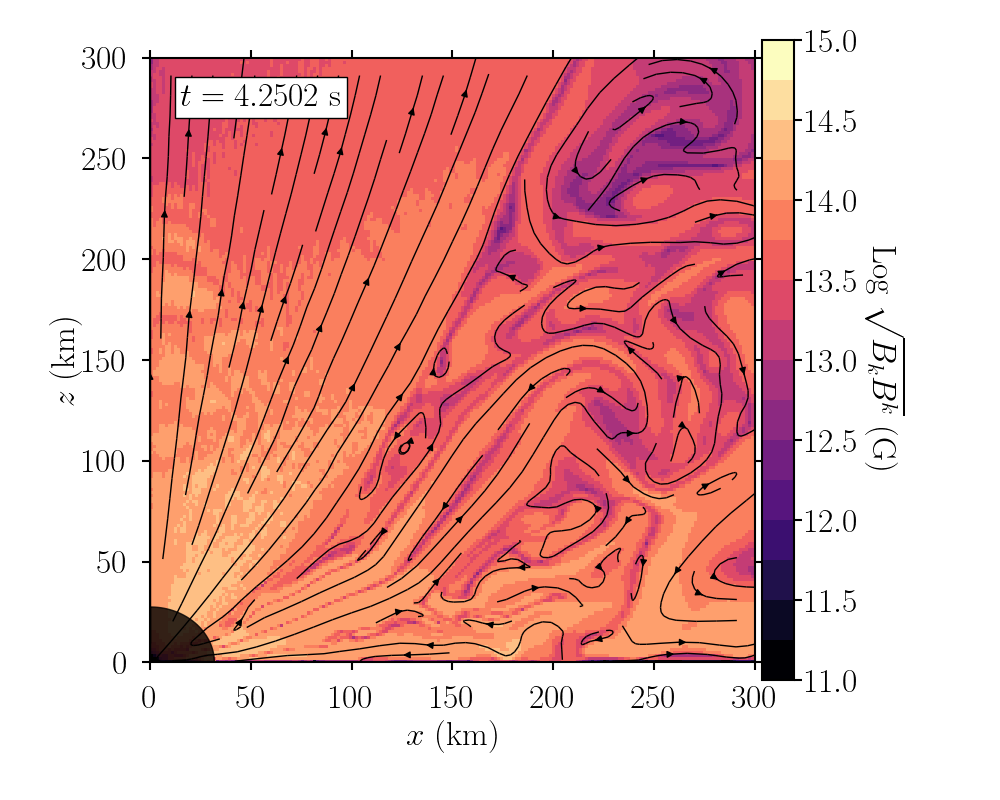}
\hspace{-8mm}
\includegraphics[width=0.355\textwidth]{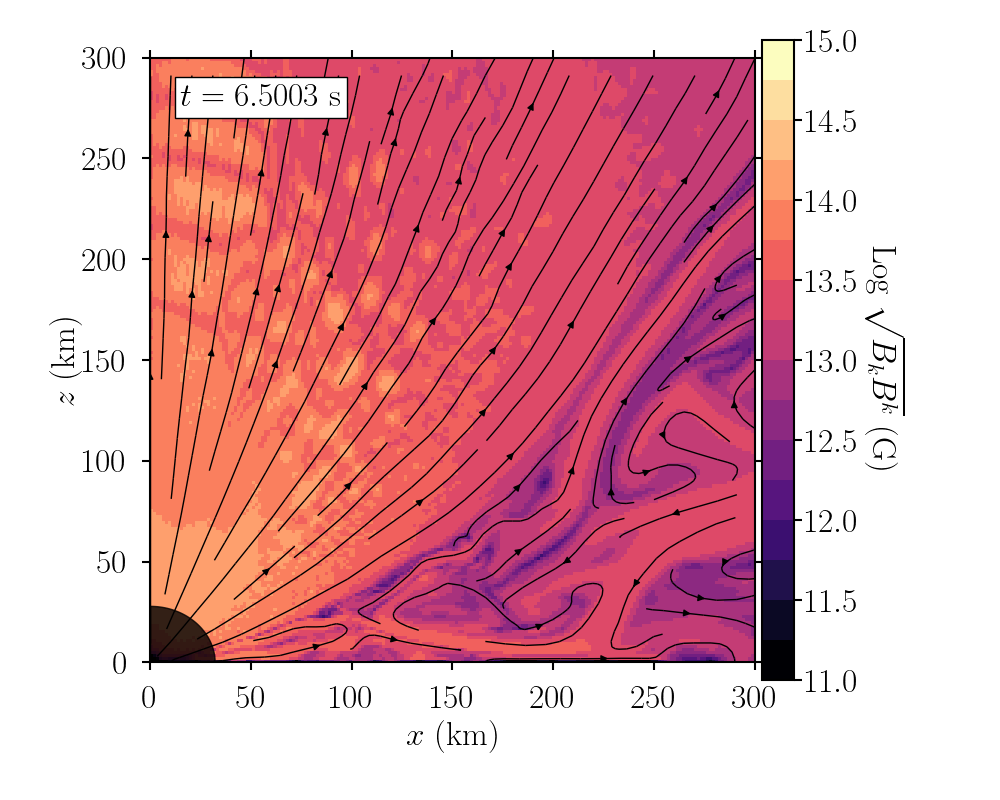}
\hspace{-8mm}
\includegraphics[width=0.355\textwidth]{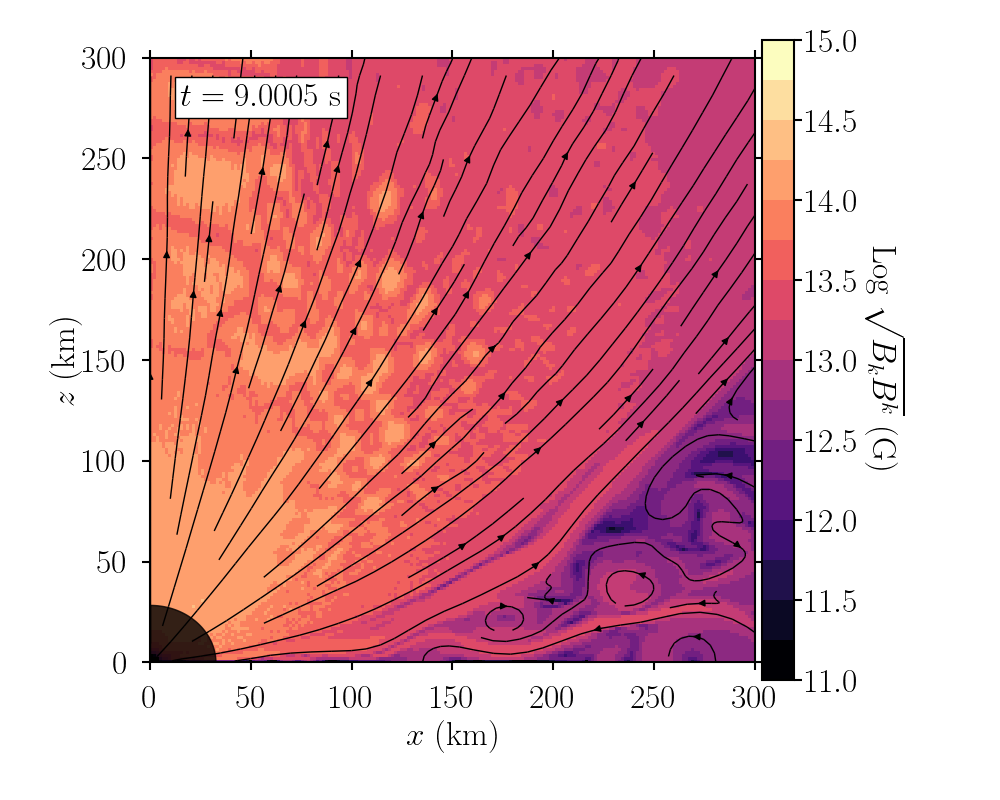}

\caption{The same as Fig.~\ref{figC} but for the evolution of the magnetic-field profile for model B11.0. 
}
\label{fig6}
\end{figure*}

We performed neutrino-radiation magnetohydrodynamics simulations in full general relativity in the context of the collapsar scenario. The simulations were started from a system of a moderately-rapidly spinning black hole and infalling matter, which are prepared based on a stellar evolution model~\cite{Aguilera-Dena2020oct}. Poloidal magnetic fields with a variety of the field strengths and configurations are superimposed initially. Axial symmetry is assumed to achieve long-term evolution with $\agt 10$\,s duration. 

We found that the evolution process of the system depends strongly on the magnetic-field strength and configuration initially given. For the models with initially strong magnetic fields of $B_\mathrm{max} \geq 10^{11}$\,G and with the field aligned well with the black-hole spin, a jet is launched due to the Blandford-Znajek mechanism in a short timescale after the magnetic-field amplification by the winding associated primarily with the black-hole spin. For this case, the jet is launched before the formation of a  disk around the black hole and a MAD state is eventually established after the jet launch due to the strong magnetic field achieved, which halts the matter accretion onto the black hole. The black hole subsequently evolves primarily by the Blandford-Znajek mechanism with subdominant matter accretion, and its dimensionless spin decreases with time. The timescale of the spin-down depends on the field strength initially given, because the magnetic-field strength at the jet launch, which is determined by the ram pressure of the infalling matter, is higher for the stronger initial field strength. For a sufficiently strong field strength, the timescale of the spin-down is shorter than 100\,s in the present models, i.e., shorter than or comparable to the typical duration of long gamma-ray bursts. However, for the models initially with a lower magnetic-field strength, the timescale is longer than 100\,s. 

The expected total energy emitted by the Blandford-Znajek mechanism depends weakly on the initial profiles of the magnetic field, because for models with shorter spin-down timescales the Poynting luminosity is higher. The expected total energy is about 20--30\% of the rotational kinetic energy of black holes that can be liberated by the Blandford-Znajek mechanism (see Eq.~(\ref{eq7}) with $f_{1/2}=1$). The most likely reason for this reduction is that although the outgoing Poynging flux is generated on the horizon, the matter infall onto the horizon (primarily from the equatorial region) prevents the outward emission of electromagnetic waves. 

For several models (B10.5, Bq12.5, and Bq11.0b), a jet is launched after the formation of a disk/torus. For these models, the strength of the magnetic field that penetrates the black hole increases by the winding associated with the black-hole spin and by the accumulation of the magnetic flux from the torus in a long timescale of $\sim 10$\,s. Because the matter accretion from the equatorial region continues even after the jet launch, the spin-down of the black hole is not found for these models. For models Bq12.5 and Bq11.0b, an entire stellar explosion after the expansion of the torus is found together with the jet launch. This results from the winding of the magnetic-field lines around the equatorial region associated with the black-hole spin, which enhances the toroidal magnetic-field strength and the magneto-centrifugal force to the torus. Because the toroidal magnetic fields become very high, the Tayler instability appears to play an important role in inducing a convective motion in the torus and infalling matter, which appears to contribute to the stellar explosion. 

For models with initially weak magnetic-field strengths or with the field configuration of Eq.~(\ref{eqB2}), $B_\mathrm{max} \leq 10^{12}$\,G, and $\varpi_0=10^3$\,km, jets are not launched in the simulation time. For these models, a disk/torus is formed and its size increases gradually with time due to the matter  infall and magneto-centrifugal effect associated with the black-hole spin. The mass and dimensionless spin for the black hole also increase simply with time. For a very long-term run, a jet launch and stellar explosion may occur for these models due to the continuous injection of the energy and angular momentum from the black hole. However, this is likely to be the results associated with axial symmetry imposed in this work. In non-axisymmetric simulations, MRI and associated turbulence would be developed, enhancing angular momentum transport, mass ejection, and mass accretion onto the black hole. Associated with the mass accretion onto the black hole, a strong poloidal magnetic field that penetrates the black hole is likely to be developed as previous simulation works demonstrated  (e.g., Refs.~\cite{Christie2019dec,Hayashi:2021oxy,Gottlieb:2023est}). If this is the case, a jet may be driven after the evolution of the disk/torus in a relatively early stage. However, in this scenario, the ram pressure at the jet launch should be weaker than those in the earlier jet-launch models (models B11.5, B11.3, B11.0, Br11.0, and Bq11.0c), and hence, the spin-down timescale will be $\gg 100$\,s (or spin-down may not be found as in models B10.5, Bq12.5, and Bq11.0b). 


As discussed in Sec.~\ref{sec:intro}, the gamma-ray burst energy had to be much larger than the observed values if a substantial fraction of the black-hole spin were extracted by the Blandford-Znajek mechanism and the corresponding rotational kinetic energy is distributed to the matter surrounding the black hole. Taking into account the non-observation of such extremely energetic gamma-ray bursts, afterglows, and supernovae, the initially strong magnetic-field models with the short spin-down timescales are not suitable for the models of long gamma-ray bursts. This suggests that the magnetic field, which penetrates black hole and is the source of the Blandford-Znajek mechanism, is likely to be generated after the disk/torus formation, its evolution, and subsequent amplification of the magnetic field in it in long gamma-ray bursts. In this scenario, the degree of the black-hole spin-down during the generation of gamma-ray bursts should not be appreciable. 

However we still have an issue. Although the spin-down timescale of the black hole is likely to be longer than the typical duration of long gamma-ray bursts, the spin-down should proceed for a long timescale as long as the poloidal magnetic field that penetrates the black hole is present. If a substantial fraction of the rotational kinetic energy of the black hole is transported to the matter surrounding the black hole, an extremely bright electromagnetic signal, which has not been observed, had to be emitted. To avoid this possibility, magnetic fields have to be dissipated within the spin-down timescale. One possible mechanism to reduce the poloidal magnetic-field strength on the horizon is the reconnection around the equatorial plane~\cite{Kisaka:2015sya}. During the jet generation inside the funnel region, the magnetic pressure balances with the gas pressure of the infalling matter or torus. In the late stage of the evolution of the system, the density and pressure of these matter fields decrease. Then, the opening angle of the magnetic-field lines around the rotational axis should increase. This can take place after most of the progenitor-star matter falls onto the central region and the torus matter are ejected outward or accreted onto the black hole by a (effectively) viscous process. The infall process proceeds approximately in the dynamical timescale of the system as
\beq
t_\mathrm{ff}=\sqrt{{R_*^3 \over GM_*}},
\eeq
where $R_*$ and $M_*$ denote the stellar radius and mass of the progenitor star at the onset of the collapse. In the current model, they are approximately $3\times 10^5$\,km and $27M_\odot$, and hence, $t_\mathrm{ff}\sim 90$\,s. The viscous timescale is much shorter than this timescale assuming that the viscous alpha parameter is of order $10^{-2}$ and the torus radius is smaller than $100M_\mathrm{BH}$ (see, e.g., Ref.~\cite{Fujibayashi2023BH}). Thus, in $\sim 100$\,s, the matter density in the vicinity of the black hole is likely to become low and the opening angle of the poloidal magnetic field becomes wide. Indeed, in some of our present models for which the initial field strength is high, the widening of the poloidal magnetic field configuration is seen (see Fig.~\ref{fig6}). This widening could eventually lead to the magnetic-field configuration similar to the split monopole and to a magnetic reconnection near the equatorial plane~(see, e.g., Refs.~\cite{Komissarov:2004ms,Komissarov:2005wj,Lyutikov:2011tk}). Exploring this possibility for the very late stage of stellar collapses is one of the issues in our future work. 

The other possible mechanism is the reconnection resulting from the interaction between the aligned magnetic fields along the black-hole spin axis and the magnetic loop ejected from the accretion torus that is in a turbulent state. General relativistic magnetohydrodynamics simulations have shown that accretion disks/tori around spinning black holes are in a turbulent state as a result of the MRI (see, e.g., Ref.~\cite{Hayashi:2022cdq} for the latest investigation). From such a turbulent disk/torus, the matter and magnetic loop are ejected, and some of the magnetic loops  move toward the spin axis of the black hole. Here, the polarity of the magnetic loops should be quite random. Hence, if an ejected magnetic loop has a polarity different from that of the aligned magnetic field along the spin axis of the black hole, the magnetic-field strength becomes weaker by the reconnection. If this process continuously occurs, the Poynting luminosity associated with the Blandford-Znajek mechanism may decrease with time. Indeed, this process is often observed in a magnetohydrodynamics simulation with a phenomenological dynamo term~\cite{Shibata2021b}. 

We note that the same problem (overproduction energy problem) is present for the short gamma-ray burst scenario by neutron-star mergers. For the case of binary neutron star mergers, the formed black hole is likely to have mass between $2.5M_\odot$ and $3M_\odot$ with the dimensionless spin of $0.6$--$0.8$~(e.g., Ref.~\cite{Shibata2016a}), while for black hole-neutron star mergers, the black-hole mass and spin are likely to be similar to those in the collapsar scenario. In both cases, the total rotation kinetic energy of the black hole available is larger than $10^{53}$\,erg (see Eq.~(\ref{eq7})) which is much larger than the typical energy of short gamma-ray bursts ($\sim 10^{49}$--$10^{50}$\,erg~\cite{Nakar:2007yr}). The latest neutrino-radiation magnetohydrodynamics simulations have shown that a strong magnetic field is developed by magnetohydrodynamics instabilities such as MRI for the merger remnants irrespective of the binary type~\cite{Kiuchi:2022nin,Hayashi:2021oxy,Hayashi:2022cdq}. As Eq.~(\ref{eq9}) shows, the spin-down timescale of the black hole is of order $10^3$--$10^4$\,s, much longer than the typical timescale of short gamma-ray bursts. This suggests that to explain the short timescale ($\alt 2$\,s) of short gamma-ray bursts, we need a mechanism to stop the emission toward the observer direction associated with the Blandford-Znajek mechanism (e.g., Ref.~\cite{Hayashi:2022cdq}), and in addition, we need a dissipation process of the magnetic field that penetrates a black hole within a timescale much shorter than the spin-down timescale of $10^3$--$10^4$\,s. In other words, it might not be particularly strange that long gamma-ray bursts took place after neutron-star mergers~\cite{2006Natur.444.1044G,2006Natur.444.1050D,Gal-Yam:2006qle,Rastinejad:2022zbg,Troja:2022yya,Levan:2023ssd,Yang:2023mqt} if the dissipation timescale of magnetic fields cannot be very short in a class of neutron-star merger remnants. 

The present work is based on an axisymmetric simulation, and as a result, we cannot follow the turbulence, which should be developed by the MRI in the formed disk/torus. In its presence, the magnetic-field strength is likely to be amplified more in them, the magnetic-flux supply onto the black hole may be more efficient, and a jet may be launched earlier. It is also likely that the turbulence activity develops the effective viscosity in the disk/torus, which can contribute to the explosion of the entire star as found in our viscous hydrodynamics simulation~\cite{Fujibayashi2023BH}. These are the issues to be pursued in the next step. 

\acknowledgements

We thank David Aguilera-Dena for providing their stellar evolution models. Numerical computation was performed on Sakura, Momiji, Cobra, and Raven clusters at Max Planck Computing and Data Facility. This work was in part supported by Grant-in-Aid for Scientific Research (grant Nos.~20H00158, 22H00130, 23H04900, and 23H05430) of Japanese MEXT/JSPS.

\appendix

\section{Definition of $L_\mathrm{BZ}$ and explosion energy}\label{appB}

In the following, the Greek and Latin indices denote the spacetime and spatial components, respectively. 

Our (approximate) definition of the Poynting flux as well as the total energy flux on the horizon is based on the energy equation in the lab frame (see, e.g., Eq.~(4.144) of Ref.~\cite{Shibata2016a}). By combining the continuity equation for the rest-mass density $\rho$, we have 
\beqn
\pa_t \bar S_0 + \pa_i F_0^i=\alpha \sqrt{\gamma} (T_{ij} K^{ij}-\gamma^{ij}
J_i \pa_j \ln\alpha),\label{eqene}
\eeqn
where $\bar S_0=(\alpha^2T^{tt}-\rho \alpha u^t) \sqrt{\gamma}$, $J_i=-\alpha T^t_{~i}$, $\alpha$ is the lapse function, $K_{ij}$ is the extrinsic curvature, $T_{\mu\nu}$ is the energy-momentum tensor, $u^\mu$ is the four velocity of the fluid, and 
\beqn
F_0^i&=& \bar S_0 v^i +\left(P-{E^2 + B^2 \over 8\pi}\right)(v^i + \beta^i)\sqrt{\gamma} \nonumber \\
&& +{\alpha \sqrt{\gamma} \over 4\pi} \epsilon^i_{~jk} E^j B^k\nonumber\\
&=& \sqrt{\gamma} \rho w (h w -1) v^i + \sqrt{\gamma} \bigg(P-\frac{E^2+B^2}{8\pi}\bigg)\beta^i \nonumber\\
&& + \frac{\alpha\sqrt{\gamma}}{4\pi} \epsilon^i{}_{jk} E^j B^k.
\label{F0}
\eeqn
Here, $v^i=u^i/u^t$ is the three velocity, $w=\alpha u^t$, $P$ is the pressure, $h$ is the specific enthalpy, $E^i$ and $B^i$ are an electric field and a magnetic field in the lab frame, $\beta^i$ is the shift vector, and $\epsilon_{ijk}$ is the completely anti-symmetric tensor in three dimension. Note that $E^i$ and $B^i$ are defined from the electromagnetic tensor $F^{\mu\nu}$ by 
\beq
E^i=-\alpha F^{it}~~{\rm and}~~
B^i={1 \over 2} \epsilon^{i\mu\nu}F_{\mu\nu}.
\eeq
The last term of Eq.~(\ref{F0}) denotes the Poynting flux, and hence, the Poynting luminosity on the horizon is defined by
\beq
L_\mathrm{BZ}^\mathrm{full}=\oint_\mathrm{horizon} {\alpha\sqrt{\gamma} \over 4\pi} \epsilon^i_{~jk} E^j B^k dS_i
\eeq
where $dS_i$ denotes an area element on the horizon. Throughout this paper, the surface integral for the Poynting luminosity is performed on apparent horizons. 

For a wide portion of the black-hole horizon, in particular near the equatorial plane, the net extracted energy can be negative if the matter inflow onto the black hole is present. For such a situation, it would not be appropriate to consider that the energy is extracted from the horizon. Thus, in this paper, we practically define a Poynting luminosity by
\beq
L_\mathrm{BZ}=\oint_\mathrm{horizon} {\alpha\sqrt{\gamma} \over 4\pi} \epsilon^i_{~jk} E^j B^k \Theta(F_0^l n_l) dS_i,
\eeq
where $\Theta$ is the Heaviside step function and $n_l$ is the spatial unit vector normal to horizons. With the factor of $\Theta(F_0^l n_l)$, we integrate the Poynting flux only if the total energy flux on the portion of the surface of apparent horizon is positive. Thus, $L_\mathrm{BZ}$ defined in this paper does not give the entire Poynting luminosity extracted from the black hole (i.e., $L_\mathrm{BZ} < L_\mathrm{BZ}^\mathrm{full}$) but a net one that is emitted outward. 

On the other hand, outflow energy (explosion energy) is extracted from the quantities in the outer region. Assuming that $(\del_t)^\mu$ is a timelike Killing vector, with which the energy-momentum tensor satisfies $\del_\mu (\sqrt{-g}T^\mu{}_\nu (\del_t)^\nu) = \del_\mu (\sqrt{-g}T^\mu{}_t)=0$, the outflow energy is defined by
\begin{align}
E_\mathrm{exp} &= \int^t \oint_{r=r_\mathrm{ext}} \sqrt{-g}\left[-T^k{}_t-h_\mathrm{min}\rho u^k\right]\Theta(e_\mathrm{bind}) dS_k dt \notag\\
&+ \int_{r<r_\mathrm{ext}} \sqrt{-g}\left[-T^t{}_t-h_\mathrm{min} \rho u^t \right]\Theta(e_\mathrm{bind}) d^3x,
\end{align}
where $r_\mathrm{ext}$ denotes an extraction radius, which is chosen to be $\approx 10^5$\,km, $e_\mathrm{bind}$ is the specific binding energy of a fluid element defined by
\begin{align}
e_\mathrm{bind} = \frac{-T^t{}_t}{\rho u^t} - h_\mathrm{min},
\end{align}
and $h_\mathrm{min}=c^2+\varepsilon_\mathrm{min}$ is the minimum value of the specific enthalpy for a given equation-of-state table. For the DD2 equation of state~\cite{banik2014a} which we employ in this paper, $\varepsilon_\mathrm{min}\approx-0.0013c^2$.
In electromagnetohydrodynamics, the energy and momentum flux densities conserved in stationary spacetime, $-\sqrt{-g}T^\mu{}_t$, are written as
\begin{align}
-\sqrt{-g}T^t{}_t &= \sqrt{\gamma} \rho u^t \left[ \alpha(hw - P/\rho w ) - \beta^k hu_k \right] \notag \\
                  &+ \sqrt{\gamma} \left[ \alpha\frac{E^2 + B^2}{8\pi} + \beta^k \frac{1}{4\pi}\epsilon_{klm}E^lB^m\right], \label{eq:Ttt}
\end{align}
\begin{alignat}{3}
-\sqrt{-g}T^i{}_t 
&= \sqrt{\gamma} &\rho  & w v^i \bigg(\alpha h w - hu_k\beta^k \bigg) \notag\\
&+ \sqrt{\gamma} &\bigg[&-\alpha\frac{E^2 + B^2}{8\pi}\beta^i+\beta^k(E_kE^i+B_kB^i) \notag\\
&                       &&+\alpha \bigg(\frac{\beta^i\beta^k}{\alpha^2}+\gamma^{ik}\bigg) \frac{1}{4\pi}\epsilon_{klm}E^lB^m\bigg], \label{eq:Tti}
\end{alignat}
where we used $V_t = \beta^k V_k$ for a spatial vector $V_\mu$ that satisfies $V_\mu n^\mu=0$; i.e., for $E_\mu$, $B_\mu$, and $\epsilon_{\mu kl}E^k B^l$. The expression found in the first lines of Eqs.~\eqref{eq:Ttt} and \eqref{eq:Tti} corresponds to  $-\sqrt{-g}T^\mu{}_t$ of the ideal fluid~\cite{fujibayashi2021oct}.

\bibliography{reference}

\end{document}